%% file: main.tex
\documentclass[a4paper,11pt]{article}
\pdfoutput=1 

\usepackage{jheppub} 

\usepackage[T1]{fontenc} 
\usepackage[utf8]{inputenc}
\usepackage{amsmath,amssymb,amstext,amsfonts} 

\usepackage[dvipsnames]{xcolor}
\usepackage{graphicx}
\usepackage{subfig}
\usepackage{hyperref}
\usepackage{cancel} 
\usepackage{float}
\usepackage{bbm}
\usepackage{mathtools}
\usepackage{cleveref}
\usepackage{lscape}
\usepackage{siunitx}
\usepackage{slashed}
\usepackage{physics}

\makeatletter
\renewcommand*\env@matrix[1][\arraystretch]{%
  \edef\arraystretch{#1}%
  \hskip -\arraycolsep
  \let\@ifnextchar\new@ifnextchar
  \array{*\c@MaxMatrixCols c}}
\makeatother

\usepackage{tikz}
    \usetikzlibrary{positioning}
    \usetikzlibrary{arrows}
    \usetikzlibrary{shapes}
    \usepackage{pgfplots}
    \usetikzlibrary{calc}
    \usetikzlibrary{decorations.markings}
     \usetikzlibrary{decorations.pathmorphing}
    \tikzset{snake it/.style={decorate, decoration=snake}}
\def\centerarc[#1](#2)(#3:#4:#5) 
    { \draw[#1] ($(#2)+({#5*cos(#3)},{#5*sin(#3)})$) arc (#3:#4:#5); }
    \usepackage{booktabs}

\DeclareMathAlphabet\mathbfcal{OMS}{cmsy}{b}{n}

\makeatletter
\g@addto@macro\bfseries{\boldmath}
\makeatother

\usepackage{caption}
\usepackage{tabularx}
\usepackage{array}
\usepackage{hhline}

\definecolor{cnblue}{RGB}{7,82,154}

\newcommand{\mt}{m}
\newcommand{\mR}{m_R}

\allowdisplaybreaks[4]

\title{Analytic two-loop amplitudes for di-jet and
$\gamma+$jet production mediated by a heavy-quark loop}

\author[a]{Federico Coro,}
\author[b]{Christoph Nega,}
\author[b]{Lorenzo Tancredi,}
\author[b]{Fabian J. Wagner}
\affiliation[a]{Department of Physics and Astronomy, Ghent University, 9000 Ghent, Belgium}
\affiliation[b]{Technical University of Munich, TUM School of Natural Sciences, Physics Department, James-Franck-Straße 1, 85748 Garching, Germany}
\emailAdd{federico.coro@ugent.be}
\emailAdd{c.nega@tum.de}
\emailAdd{lorenzo.tancredi@tum.de}
\emailAdd{fabianjohannes.wagner@tum.de}

\preprint{{\raggedleft
            TUM-HEP 1570/25 
}}

\abstract{
In this paper, we present analytical results for the two-loop QCD corrections to the production of two partons or a photon and a parton in hadronic collisions, mediated by loops of massive quarks. These amplitudes involve Feynman integrals defined on an elliptic curve. We compute them by generalizing our recent results for the production of two photons to include additional crossings of the corresponding master integrals, which we compute in terms of the same basis of independent iterated integrals. We discuss the analytical properties of the amplitudes, highlighting the cancellations of a large number of elliptic differential forms in their finite remainders. Finally, we elaborate on a strategy for their numerical evaluation based on generalized series expansions at singular points of the physical amplitude, through the introduction of suitable sets of variables that allow us to resolve all singularities.}

\begin{document}

\maketitle

\input{introduction}

\input{setup}
\input{masters}
\input{uvandir}
\input{numerics}
\input{conclusions}

\acknowledgments
We thank Melih A. Ozcelik for pointing us to the triangle relations in~\cite{Abreu:2022vei} and Federico Buccioni for help in dealing with large expressions in Mathematica. We also thank Martin Link for bringing the existence of Shanks transformations to our attention.
Further, we are thankful to Matteo Becchetti and Federico Ripani for many discussions and for their collaboration in the initial stages of this project. 
We are grateful to the Munich Institute for Astro-, Particle and BioPhysics (MIAPbP), funded by the Deutsche Forschungsgemeinschaft (DFG, German Research Foundation) under Germany's Excellence Strategy – EXC-2094 – 390783311, where part of these results have been obtained.
CN, LT, and FW were supported in part by the Excellence Cluster ORIGINS 
funded by the Deutsche Forschungsgemeinschaft (DFG, German Research Foundation) under Germany’s 
Excellence Strategy – EXC-2094-390783311 and in part by the European Research Council (ERC) under the European Union’s research and innovation program grant agreements 949279 (ERC Starting Grant HighPHun). 
The work of FC was supported by the European Research Council (ERC) under the European Union’s Horizon Europe research and innovation program grant agreement 101078449 (ERC Starting Grant MultiScaleAmp).
Views and opinions expressed are, however, those of the authors only and do not necessarily reflect those of the European Union or the European Research Council Executive Agency. Neither the European Union nor the granting authority can be held responsible for them.

\appendix

\input{appendices}

\bibliographystyle{JHEP} 
\bibliography{biblio} 

\end{document}

%% file: introduction.tex
\section{Introduction}
\label{sec:intro}

Photons and jets are among the most abundant and widely studied products of hadronic collisions, as they provide important observables to scrutinize the details of the theory of strong interactions to the highest precision. In fact, scattering processes involving isolated photons, photons produced in association with jets, and jet production have been at the center of both theory and experimental studies for many decades now, and are among the first $2 \to 2$ processes for which complete studies in Next-to-Next-to-Leading-Order (NNLO) QCD have been performed~\cite{Catani:2011qz, Campbell:2016yrh, Campbell:2016lzl, Campbell:2017dqk, Chen:2019zmr, Czakon:2019tmo, Britzger:2022lbf, Chen:2022tpk}.
For example, the production of isolated photons $pp \to \gamma + X$ and photons in association with a jet $pp \to \gamma + j$ can be measured over large ranges of rapidities and transverse momenta of the corresponding photon, and provide a direct way to probe parton distribution functions, including strong sensitivity to the gluon distribution~\cite{dEnterria:2012kvo, Carminati:2013lvm}.
Moreover, precision studies for the production of two jets in hadron collisions and their comparison with experimental results have an important role, among other things, for the precise measurement of the strong coupling constant and the study of its running across large energy scales, see for example~\cite{Ahmadova:2024emn} for a recent analysis.
These, and many other precision studies for related observables, have become possible thanks, on the one hand, to the computation of two-loop virtual amplitudes for the relevant partonic subprocesses~\cite{Anastasiou:2002zn, Bern:2001df} in \emph{massless QCD} and, on the other, to the development of increasingly sophisticated subtraction methods~\cite{Binoth:2004jv, Anastasiou:2003gr, Catani:2007vq, Gehrmann-DeRidder:2005btv, Czakon:2010td, Boughezal:2011jf, Boughezal:2015eha, Gaunt:2015pea, Cacciari:2015jma, DelDuca:2016ily, Caola:2017dug, Magnea:2020trj} which make it possible to remove nonphysical infra-red divergences and compute infra-red finite observables.

While massless NNLO QCD radiative corrections are expected to provide the bulk of the effects in most distributions, the astonishing precision reached by current LHC measurements and the large amount of data that will be produced and analysed by the end of the LHC lifetime, motivate the scrutiny of the impact of smaller effects. Among these are higher-order corrections in massless QCD, and subtle effects induced by heavy virtual particles circulating in the loops. First steps in the former direction have been recently achieved with the calculation of the full set of three-loop virtual amplitudes~\cite{Caola:2020dfu, Bargiela:2021wuy, Caola:2021izf, Caola:2021rqz, Caola:2022dfa, Bargiela:2022lxz} and of the corresponding two-loop real-virtual counterparts~\cite{Abreu:2018jgq, Abreu:2019odu, Badger:2019djh, Agarwal:2021grm, Agarwal:2021vdh, Badger:2021imn, Abreu:2021oya, Agarwal:2023suw, DeLaurentis:2023nss, DeLaurentis:2023izi, Badger:2023mgf}.
The inclusion of the effects of massive particles, instead, requires, among others, the computation of $2 \to 2$ amplitudes in QCD up to two loops, where the production of photons and jets happens through loops of massive (top) quarks.
Compared to their massless counterparts, progress in the computation of massive amplitudes has been slower, especially due to the fact that massive internal states lead to the early appearance of new mathematical structures, as special functions related to elliptic geometries and their generalizations. The appearance of these objects in quantum field theory has a long history, and can be traced back to a calculation by A. Sabry of the two-loop corrections to the QED electron propagator~\cite{Sabry}. 
The initial difficulties in dealing with these new functions using standard analytic methods have been among the motivations for the development of powerful numerical methods, which have made it possible to compute many scattering amplitudes, including the two-loop massive corrections to diphoton production in QCD~\cite{Dugad:2016kdv, Maltoni:2018zvp, Chen:2019fla, Becchetti:2023wev, Becchetti:2023yat}.

Since Sabry's paper, our understanding of the mathematics behind these objects has substantially improved, such that we are finally in the position of addressing some of these calculations fully analytically. The recent progress has been mainly possible due to parallel developments in calculational methods and in our general understanding of wide classes of special functions. In particular, the differential equations method~\cite{Kotikov:1990kg, Remiddi:1997ny, Gehrmann:1999as} has provided us with a unique tool to investigate the analytic properties of multi-loop Feynman integrals. In this context, the concept of a canonical basis~\cite{Arkani-Hamed:2010pyv, Henn:2013pwa} has revealed to be extremely powerful in simplifying the calculation of large classes of Feynman integrals that evaluate to the well understood class of functions of multiple polylogarithms~\cite{Kummer, Goncharov:1998kja, Remiddi:1999ew, Vollinga:2004sn}, and their generalizations to iterated integrals over logarithmic differential forms \emph{à la} Chen~\cite{ChenSymbol}. 

An important part of the recent efforts in extending these techniques beyond integrals of polylogarithmic type has focused, on the one hand, on identifying the correct mathematical language to generalize iterated integrals with logarithmic singularities to elliptic geometries~\cite{Brown:2011wfj, Broedel:2014vla, Broedel:2017kkb, Broedel:2018qkq, Blumlein:2018jgc} and beyond~\cite{DHoker:2023vax, DHoker:2024ozn, DHoker:2025dhv} and, on the other, 
to extend the concept of a canonical bases to arbitrary geometries~\cite{Primo:2016ebd, Primo:2017ipr, Frellesvig:2017aai, Adams:2018yfj, Dlapa:2022wdu, Pogel:2022ken, Frellesvig:2023iwr, Gorges:2023zgv, Frellesvig:2024rea, Chen:2025hzq, Duhr:2025lbz, Chaubey:2025adn, e-collaboration:2025frv}.
While there is not yet a consensus on what the correct generalization of canonical bases beyond polylogarithms should be, it has recently become clear that many of the employed approaches share important common features. In particular, the two apparently distinct procedures outlined in~\cite{Adams:2018yfj, Pogel:2022ken} and~\cite{Gorges:2023zgv, Duhr:2025lbz} have been demonstrated to lead to equivalent results for large classes of problems related to geometries of Calabi-Yau type~\cite{Duhr:2025lbz}. As first highlighted in~\cite{Gorges:2023zgv}, crucial to the success of these strategies is the selection of a suitable basis of pre-canonical master integrals, which can be achieved by a generalization of the integrand analysis commonly used to select polylogarithmic integrals~\cite{Arkani-Hamed:2010pyv, Henn:2020lye}, to integrals of elliptic type in~\cite{Gorges:2023zgv} and beyond~\cite{Duhr:2024uid, Duhr:2025lbz}.

At variance with the polylogarithmic case, the structure of the (co)-homology associated with these non-trivial geometries introduces a new element in this analysis. In particular, the periods of these geometries satisfy higher-order Picard-Fuchs differential equations, and the Frobenius method allows us to classify their transcendental behaviour close to each regular singular point in terms of the powers of logarithmic singularities they develop. In particular, close to a so-called point of Maximal Unipotent Monodromy (MUM)~\cite{Bonisch:2021yfw}, the period solutions will develop a complete set of logarithmic divergences and, in general, the non-holomorphic periods will not have uniform transcendental weight, but will instead have mixed regular and transcendental behaviour. Starting from this observation, a general procedure was devised first for elliptic Feynman integrals in~\cite{Broedel:2018qkq} to split the corresponding period matrix close to a MUM point into a part of transcendental weight zero (referred to as \emph{semi-simple}) and one containing pure and uniform transcendental terms (also called \emph{unipotent}). In~\cite{Gorges:2023zgv, Duhr:2025lbz} it was then shown that leveraging this splitting, one can define a new basis of Feynman integrals by rotating away the \emph{semi-simple} part of their period matrix and rescaling them by appropriate $\epsilon$ powers to realign their transcendental weight. After some minor clean-up procedure, the resulting basis of master integrals would then satisfy $\epsilon$-factorized differential equations with linearly independent differential forms with at most single poles. Moreover, close to those singular points where the non-trivial geometries degenerate to simple polylogarithms, the solution of these differential equations would be pure and of uniform transcendental weight in the usual polylogarithmic sense. 
In~\cite{Duhr:2025lbz},  these properties have been elevated to formal requirements for a conjectural definition of a canonical basis beyond polylogarithms.

This strategy has by now been applied to many non-trivial problems, including the calculation of two-point correlators involving elliptic and K3 geometries~\cite{Duhr:2024bzt, Forner:2024ojj, Duhr:2025kkq}, generalizations to higher genus~\cite{Duhr:2024uid}, $2 \to 1$, $2 \to 2$ and $2 \to 3$ scattering amplitudes of elliptic type~\cite{Delto:2023kqv, Becchetti:2025rrz, Marzucca:2025eak, Becchetti:2025oyb} and Calabi-Yau geometries relevant for observables related to black-hole scattering~\cite{Klemm:2024wtd, Driesse:2024feo, Frellesvig:2023bbf}.
In this paper, we follow this strategy and extend the results we recently obtained for the scattering amplitudes relevant for the production of two photons mediated by loops of heavy quarks\footnote{See~\cite{Ahmed:2025osb} for an independent study.}~\cite{Becchetti:2025rrz}, to include all missing master integrals relevant for the production of two jets or a photon and a jet. 

The rest of the paper is organized as follows. In~\cref{sec:setup} we provide our notation and the general setup of the calculation. We continue in~\cref{sec:ampcomp} with details of the calculation of the amplitude and of the relevant master integrals. There, we also discuss in detail the origin of extra relations among the masters. We discuss UV renormalization, the subtraction of IR poles in~\cref{sec:ren}, and finally provide a discussion of the numerical evaluation of our amplitudes by generalized series expansions in~\cref{sec:numerics}. Finally, we draw our conclusions in~\cref{sec:conc}. We include additional material in various appendices.

%% file: setup.tex
\section{Notation and computational setup}
\label{sec:setup}
We consider the two-loop QCD amplitudes for the production of a single photon plus jet as well as two jets through a heavy-quark loop. In particular, we focus on the following partonic subchannels
\begin{align} 
    q(p_1) + \bar{q}(p_2) \ &\longrightarrow \ g(-p_3) + \gamma(-p_4)  \, , \nonumber \\
    g(p_1) + g(p_2) \ &\longrightarrow \ g(-p_3) + \gamma(-p_4) \, , \nonumber \\
    q(p_1) + \bar{q}(p_2) \ &\longrightarrow \ \bar{Q}(-p_3) + Q(-p_4)  \, , \label{eq:channels} \\
    q(p_1) + \bar{q}(p_2) \ &\longrightarrow \ g(-p_3) + g(-p_4)  \, , \nonumber \\
    g(p_1) + g(p_2) \ &\longrightarrow \ g(-p_3) + g(-p_4) \, . \nonumber 
\end{align}
Here, $q$ and $Q$ denote different quark flavors, but we note that the amplitude for any $2 \rightarrow 2$ massless quark scattering process, including the equal-flavor case,  can be obtained from the above four-quark scattering process by suitable crossings of the external invariants (for example, see sec.~7 of~\cite{Caola:2021rqz}).

The external particles are on-shell, i.e., $p_i^2 = 0$, and the kinematics of the process can be described by the usual Mandelstam variables 
\begin{equation}
    s = (p_1+p_2)^2\,, \hspace{0.3cm} t = (p_1+p_3)^2\,, \hspace{0.3cm} u = (p_2+p_3)^2\,, \hspace{0.3cm} \text{with} \hspace{0.3cm} s+t+u = 0\, .
\end{equation}
In the physical scattering region, one has $s>0$ and $t<0,\; u<0$, which imply $0 < -t < s $. We denote the mass of the heavy quark by the symbol $\mt$.

Working in the 't Hooft-Veltman dimensional regularization scheme~\cite{tHooft:1972tcz}, we decompose the bare scattering amplitudes corresponding to the processes in eq.~\eqref{eq:channels} in a basis of four-dimensional tensors~\cite{Peraro:2019cjj, Peraro:2020sfm} and scalar form factors as follows: 
\begin{align}
    \mathbfcal{A}_{q\bar{q}g\gamma}(s,t,\mt^2) &= \sqrt{4 \pi \alpha} \sqrt{4 \pi \alpha_{s,b}} \, \left[ \, \sum_{n=1}^4 \mathbfcal{F}_n(s,t,\mt^2)\, \bar{u}(p_2) \, \Gamma_n^{\mu \nu} \,u(p_1) \right] \epsilon_{3,\mu} \, \epsilon_{4,\nu}\,,
    \label{eq:ampqqga} \\
    \mathbfcal{A}_{ggg\gamma}(s,t,\mt^2) &= \sqrt{4 \pi \alpha} \sqrt{4 \pi \alpha_{s,b}} \,\left[ \, \sum_{n=1}^8 \mathbfcal{G}_n(s,t,\mt^2) \, T_n^{\mu \nu \rho \sigma} \right] \epsilon_{1,\mu} \, \epsilon_{2,\nu} \, \epsilon_{3,\rho} \, \epsilon_{4,\sigma}\,, \label{eq:ampggga} \\
    \mathbfcal{A}_{q\bar{q}\bar{Q}Q}(s,t,\mt^2) &= \left(4 \pi \alpha_{s,b}\right) \,\left[ \, \sum_{n=1}^2 \mathbfcal{H}_n(s,t,\mt^2)\, \Pi_n \right] \,,
    \label{eq:ampqqQQ} \\
    \mathbfcal{A}_{q\bar{q}gg}(s,t,\mt^2) &= \left(4 \pi \alpha_{s,b}\right) \,\left[ \, \sum_{n=1}^4 \mathbfcal{K}_n(s,t,\mt^2)\, \bar{u}(p_2) \, \Gamma_n^{\mu \nu} \,u(p_1) \right] \epsilon_{3,\mu} \, \epsilon_{4,\nu} \,,
    \label{eq:ampqqgg} \\ 
    \mathbfcal{A}_{gggg}(s,t,\mt^2) &= \left(4 \pi \alpha_{s,b}\right) \,\left[ \, \sum_{n=1}^8 \mathbfcal{J}_n(s,t,\mt^2) \, T_n^{\mu \nu \rho \sigma} \right] \epsilon_{1,\mu} \, \epsilon_{2,\nu} \, \epsilon_{3,\rho} \, \epsilon_{4,\sigma} \,. \label{eq:ampgggg}
 \end{align}

In these equations, we used boldfaced symbols to indicate that the amplitudes and the corresponding form factors are in general vectors in color space, see~\cref{eq:colordecomqqga,eq:colordecomggga,eq:colordecomqqQQ,eq:colordecomqqgg,eq:colordecomgggg}.
Moreover, $\alpha=e^2/4\pi$ denotes the electromagnetic coupling constant and $\alpha_{s,b}=g_{s,b}^2/4\pi$ the \emph{bare} strong coupling constant. The spinors of the massless (anti)quarks with momentum $p_i$ are represented by $u(p_i)$ and $\bar{u}(p_i)$, respectively, while we denote the polarization vector of the 
gauge boson of momentum $p_i$ by $\epsilon_{i,\mu}$. 

To reduce the number of independent tensor structures, we perform the following
gauge choices for the external vector bosons
\begin{align}
    \left\{   
    \begin{array}{llc}
    \epsilon_3 \cdot p_2 &= \epsilon_4 \cdot p_1 = 0\,,  & \quad  \mbox{for} \ \ q\bar{q} \to g \gamma \ \  \mbox{and} \ \ q\bar{q} \to g g   \\
    \epsilon_i \cdot p_{i+1} &= 0\,,  & \quad \mbox{for} \ \ g g \to g \gamma \ \  \mbox{and} \ \ g g \to g g
    \end{array} \right. \label{eq:gaugeeps}
\end{align}
with $p_5 = p_1$. In this way, a convenient four-dimensional basis for the tensor structures appearing on the right-hand sides is given by (see also~\cite{Caola:2021rqz, Caola:2020dfu, Bargiela:2021wuy})
\begin{align}
    \Pi_1 &= \left[ \bar{u}(p_2) \, \gamma_\alpha \,u(p_1) \right] \left[ \bar{u}(p_4) \, \gamma^\alpha \,u(p_3) \right] \,, &&\Pi_2 = \left[ \bar{u}(p_2) \, \slashed{p}_3 \,u(p_1) \right] \left[  \bar{u}(p_4) \, \slashed{p}_2 \,u(p_3) \right] \, ,
    \label{eq:tensqqQQ}
\end{align}
\begin{align}
    \Gamma_1^{\mu \nu} &= \gamma^\mu p_2^{\nu}\,,  &&\Gamma_2^{\mu \nu} = \gamma^\nu p_1^{\mu}\,, \nonumber \\
    \Gamma_3^{\mu \nu} &=  \slashed{p}_3\,p_1^\mu p_2^\nu \,, 
    &&\Gamma_4^{\mu \nu} =  \slashed{p}_3\,g^{\mu \nu}\,, \label{eq:tensqq} 
\end{align}
\begin{align}
    T_1^{\mu \nu \rho \sigma} &= p_3^{\mu}p_1^{\nu}p_1^{\rho}p_2^{\sigma}\,, 
    && T_2^{\mu \nu \rho \sigma} = p_3^{\mu}p_1^{\nu}g^{\rho\sigma}\,, \nonumber \\
    T_3^{\mu \nu \rho \sigma} &= p_3^{\mu}p_1^{\rho}g^{\nu\sigma} \,, 
    && T_4^{\mu \nu \rho \sigma} = p_3^{\mu}p_2^{\sigma}g^{\nu\rho}\,, \label{eq:tensgg} \\
    T_5^{\mu \nu \rho \sigma} &= p_1^{\nu}p_1^{\rho}g^{\mu\sigma} \,, 
    && T_6^{\mu \nu \rho \sigma} = p_1^{\nu}p_2^{\sigma}g^{\mu\rho}\,, \nonumber \\
    T_7^{\mu \nu \rho \sigma} &= p_1^{\rho}p_2^{\sigma}g^{\mu\nu} \,, 
    && T_8^{\mu \nu \rho \sigma} = g^{\mu\nu}g^{\rho\sigma}+g^{\mu\sigma}g^{\nu\rho}+g^{\mu\rho}g^{\nu\sigma}\,. \nonumber
\end{align}
As vectors in color space, each form factor 
can then be decomposed as follows:
\begin{align}
    \mathbfcal{F}_n(s,t,\mt^2) &= \mathcal{F}_n(s,t,\mt^2) \, T_{i_2 i_1}^{a_1} \, , \label{eq:colordecomqqga} \\
    \mathbfcal{G}_n(s,t,\mt^2) &= \mathcal{G}_n(s,t,\mt^2) \, \text{tr} \left( T^{a_1} \left[T^{a_2},T^{a_3} \right]\right) \, , \label{eq:colordecomggga} \\
    \mathbfcal{H}_n(s,t,\mt^2) &= \sum_{i=1}^{2} \mathcal{H}_n^{[i]}(s,t,\mt^2) \ \mathcal{C}_i^{i_1,i_2,i_3,i_4} \, , \label{eq:colordecomqqQQ} \\
    \mathbfcal{K}_n(s,t,\mt^2) &= \sum_{i=1}^{3} \mathcal{K}_n^{[i]}(s,t,\mt^2) \ \mathcal{C}_i^{i_1,i_2,a_3,a_4} \, , \label{eq:colordecomqqgg} \\
    \mathbfcal{J}_n(s,t,\mt^2) &= \sum_{i=1}^{6} \mathcal{J}_n^{[i]}(s,t,\mt^2) \ \mathcal{C}_i^{a_1,a_2,a_3,a_4} \, , \label{eq:colordecomgggg}
\end{align}
with the colour structures $\mathcal{C}_i$ given by (see also~\cite{Caola:2021rqz,Caola:2022dfa,Caola:2021izf})
\begin{equation}
    \mathcal{C}_1^{i_1,i_2,i_3,i_4} = \delta_{i_1 i_4} \delta_{i_2 i_3} \, , \qquad \mathcal{C}_2^{i_1,i_2,i_3,i_4} = \delta_{i_1 i_2} \delta_{i_3 i_4}\, ,
    \label{eq:colqqQQ}
\end{equation}
\begingroup
\vskip -2em
\endgroup
\begin{equation}
    \mathcal{C}_1^{i_1,i_2,a_3,a_4} = \left(T^{a_3} T^{a_4} \right)_{i_2 i_1 } \,, \quad \mathcal{C}_2^{i_1,i_2,a_3,a_4} = \left(T^{a_4} T^{a_3} \right)_{i_2 i_1} \,, \quad
    \mathcal{C}_3^{i_1,i_2,a_3,a_4}= \delta_{i_1 i_2} \delta^{a_3 a_4} \,, 
    \label{eq:colqq}
\end{equation}
\begin{equation}
\begin{aligned}
    \mathcal{C}_1^{a_1,a_2,a_3,a_4} &= \text{tr} \left( T^{a_1} T^{a_2} T^{a_3} T^{a_4} \right) + \text{tr} \left( T^{a_1} T^{a_4} T^{a_3} T^{a_2} \right) \,, \\ 
    \mathcal{C}_2^{a_1,a_2,a_3,a_4} &= \text{tr} \left( T^{a_1} T^{a_2} T^{a_4} T^{a_3} \right) + \text{tr} \left( T^{a_1} T^{a_3} T^{a_4} T^{a_2} \right) \,, \\
    \mathcal{C}_3^{a_1,a_2,a_3,a_4} &= \text{tr} \left( T^{a_1} T^{a_3} T^{a_2} T^{a_4} \right) + \text{tr} \left( T^{a_1} T^{a_4} T^{a_2} T^{a_3} \right) \,, \\
    \mathcal{C}_4^{a_1,a_2,a_3,a_4} &=  \text{tr} \left( T^{a_1} T^{a_2}\right) \, \text{tr} \left( T^{a_3} T^{a_4} \right) \,,\\
    \mathcal{C}_5^{a_1,a_2,a_3,a_4} &= \text{tr} \left( T^{a_1} T^{a_3}\right) \, \text{tr} \left( T^{a_2} T^{a_4} \right) \,, \\
    \mathcal{C}_6^{a_1,a_2,a_3,a_4} &= \text{tr} \left( T^{a_1} T^{a_4}\right) \, \text{tr} \left( T^{a_2} T^{a_3} \right) \,.
    \label{eq:colgg} 
\end{aligned}
\end{equation}
In the formulas above, we used indices $i_j$ for the colour of (anti)quarks with momentum $p_j$ and indices $a_j$ for the colour of gluons with momentum $p_j$.

Instead of computing the scattering amplitudes in~\cref{eq:ampggga,eq:ampgggg,eq:ampqqga,eq:ampqqgg,eq:ampqqQQ} for general helicities, we choose to consider all possible independent helicity configurations individually, i.e., we aim at computing so-called helicity amplitudes. 
In fact, our form factor decomposition is non-physical (not even gauge invariant, due to the gauge fixing in~\cref{eq:gaugeeps}), and it should be thought of only as an intermediate decomposition to obtain a compact representation for physical quantities, such as the helicity amplitudes, or the amplitude squared and summed over helicities.
The polarization vectors $\epsilon_{j,\mu}$ of each gauge boson can have one of two possible physical polarizations, $\lambda= \pm$. However, helicity is conserved along massless fermion lines, so for any external quark-antiquark pair, making a choice for the helicity of one of them also fixes the helicity of the other. 
We denote the helicity of such a  $q\bar{q}$ pair by the helicity of the corresponding quark taken as an incoming particle. We use a superscript $\lambda_q \in \{L,R\}$ for the helicity of a quark and $\lambda_j \in \{ \pm \}$ for the helicity of a gauge boson with momentum $p_j$. In this way, we write 
for the individual helicity amplitudes of our scattering processes
\begin{equation}
\begin{aligned}
    \mathbfcal{A}_{q\bar{q}g\gamma}^{\lambda_q \lambda_3 \lambda_4} &(s,t,\mt^2) \, , \quad
    \mathbfcal{A}_{ggg\gamma}^{\lambda_1 \lambda_2 \lambda_3 \lambda_4}(s,t,\mt^2) \, , \quad
    \mathbfcal{A}_{q\bar{q}\bar{Q}Q}^{\lambda_q\lambda_Q}(s,t,\mt^2) \, , \\
    &\mathbfcal{A}_{q\bar{q}gg}^{\lambda_q \lambda_3 \lambda_4}(s,t,\mt^2) \, , \quad
    \mathbfcal{A}_{gggg}^{\lambda_1 \lambda_2 \lambda_3 \lambda_4}(s,t,\mt^2)\,.
\end{aligned}
\end{equation}
To fix our conventions for the helicity brackets, we  
define \emph{incoming} left-handed spinors as
\begin{equation}
    \bar{u}_L(p_2) = \langle 2 | \qquad \text{and} \qquad u_L(p_1) = | 1 ]\,,
\end{equation}
and  polarization vector associated to \emph{incoming} gauge boson $j$ of momentum $p_j$ and gauge vector $q_j$ (see eq.~\eqref{eq:gaugeeps}) as
\begin{equation}
    \epsilon^\mu_{j,+}(q_j) = \frac{\langle q_j | \gamma^\mu | j ] }{ \sqrt{2} \langle q_j j \rangle} \qquad \text{and} \qquad \epsilon^\mu_{j,-}(q_j) = \frac{\langle j | \gamma^\mu | q_j ] }{ \sqrt{2} [ j q_j  ]}\,.
\end{equation}
Performing the spinor algebra, we can then write the individual helicity amplitudes as an overall spinor factor times a ``helicity coefficient'', which is a ``helicity free'' function of the Mandelstam variables and $\mt^2$ only. 
Specifically, we have 
\begin{equation}
\label{eq:helampqqQQ}
    \mathbfcal{A}_{q\bar{q}\bar{Q}Q}^{LR}(s,t,\mt^2) = \frac{[13]}{[24]} \, \boldsymbol{\eta}_1 (s,t,\mt^2) \, , \quad 
    \mathbfcal{A}_{q\bar{q}\bar{Q}Q}^{LL}(s,t,\mt^2) = \frac{[14]}{[23]} \,\boldsymbol{\eta}_2 (s,t,\mt^2) \, ,
\end{equation}
while for $X\in \{g\gamma,gg \}$,
\begin{equation}
\begin{aligned}
     \mathbfcal{A}_{q\bar{q}X}^{L++}&= \frac{2[34]^2}{\langle 1 3 \rangle [23]} 
    \boldsymbol{\alpha}_{X}(s,t,\mt^2) \,, \quad 
     &\mathbfcal{A}_{q\bar{q}X}^{L+-} &= \frac{2\langle 24 \rangle [13]}{\langle 2 3 \rangle [24]} 
     \boldsymbol{\beta}_{X}(s,t,\mt^2)\,, \\
     \mathbfcal{A}_{q\bar{q}X}^{L-+} &= \frac{2\langle 23 \rangle [41]}{\langle 2 4 \rangle [32]} 
    \boldsymbol{\gamma}_{X}(s,t,\mt^2) \,, \quad
     &\mathbfcal{A}_{q\bar{q}X}^{L--}&= \frac{2\langle 34\rangle^2}{\langle 3 1 \rangle [23]} 
    \boldsymbol{\delta}_{X}(s,t,\mt^2) \, , 
\label{eq:helampqq}
\end{aligned}
\end{equation}
and
\begin{equation}
\begin{aligned}
    \mathbfcal{A}_{ggX}^{++++} &= \frac{[1 2][3 4]}{\langle1 2\rangle\langle3
    4\rangle} \boldsymbol{f}_{X}^{++++}(s,t,\mt^2) \,, & %
    \mathbfcal{A}_{ggX}^{-+++} &= \frac{\langle1 2\rangle\langle1 4\rangle[2 4]}{\langle3
    4\rangle\langle2 3\rangle\langle2 4\rangle} \boldsymbol{f}_{X}^{-+++}(s,t,\mt^2)\,, \\ 
    \mathbfcal{A}_{ggX}^{+-++} &= \frac{\langle2 1\rangle\langle2 4\rangle[14]}{\langle3 4\rangle\langle1 3\rangle\langle1 4\rangle} \boldsymbol{f}_{X}^{+-++}(s,t,\mt^2) \,, &
    \mathbfcal{A}_{ggX}^{++-+} &= \frac{\langle3 2\rangle\langle3 4\rangle[24]}{\langle1 4\rangle\langle2 1\rangle\langle2 4\rangle} \boldsymbol{f}_{X}^{++-+}(s,t,\mt^2) \,,  \\
    \mathbfcal{A}_{ggX}^{+++-} &= \frac{\langle4 2\rangle\langle4 3\rangle[23]}{\langle1 3\rangle\langle2 1\rangle\langle2 3\rangle} \boldsymbol{f}_{X}^{+++-}(s,t,\mt^2) \,,&
    \mathbfcal{A}_{ggX}^{--++} &= \frac{\langle1 2\rangle[3 4]}{[12]\langle3 4\rangle}\, \boldsymbol{f}_{X}^{--++}(s,t,\mt^2) \,, \\
    \mathbfcal{A}_{ggX}^{-+-+} &= \frac{\langle1 3\rangle[2 4]}{[1 3]\langle24\rangle} \boldsymbol{f}_{X}^{-+-+}(s,t,\mt^2) \,,&
    \mathbfcal{A}_{ggX}^{+--+} &= \frac{\langle2 3\rangle[1 4]}{[2 3]\langle1 4\rangle}\, \boldsymbol{f}_{X}^{+--+}(s,t,\mt^2) \,.
\label{eq:helampgg}
\end{aligned}
\end{equation}
Each helicity coefficient is a linear combination of the form factors given in~\cref{eq:ampqqga,eq:ampggga,eq:ampgggg,eq:ampqqgg,eq:ampqqQQ} with coefficients that are rational functions of the Mandelstam variables, which we provide in~\cref{app:helcoeffs}. The notation with respect to their colour structure is analogous to the form factors, see~\cref{eq:colordecomqqga,eq:colordecomggga,eq:colordecomqqQQ,eq:colordecomqqgg,eq:colordecomgggg}.

All the remaining helicity amplitudes can be obtained from those given above by a parity transformation, under which QCD is invariant. This symmetry transformation acts by flipping all the helicities ($L \leftrightarrow R$ and $\pm \leftrightarrow \mp$) and exchanging the spinor brackets according to $\langle ij \rangle \leftrightarrow [ji]$. For instance, we have
\begin{equation}
    \mathbfcal{A}_{X}^{R-+} = \mathbfcal{A}_{X}^{L +-}\vert_{\langle ij \rangle \leftrightarrow [ji]} = \frac{2 [42] \langle 31 \rangle}{[32] \langle 42 \rangle} 
     \boldsymbol{\beta}_{X}(s,t,\mt^2) \,.
\end{equation}
Furthermore, charge-conjugation symmetry and  Bose symmetry under the exchange of two identical gauge bosons imply additional relations between the helicity amplitudes. In terms of the helicity coefficients, these can be written as follows
\begin{align}
    &\boldsymbol{\gamma}_{X}=\boldsymbol{\beta}_{X}\vert_{1 \leftrightarrow 2}\,, &&\boldsymbol{\delta}_{X}=\boldsymbol{\alpha}_{X}\vert_{1 \leftrightarrow 2}\,, \label{eq:symQQ1} \\
    &\boldsymbol{\alpha}_{gg}=-\boldsymbol{\alpha}_{gg}\vert_{3 \leftrightarrow 4}\,, &&\boldsymbol{\delta}_{gg}=-\boldsymbol{\delta}_{gg}\vert_{3 \leftrightarrow 4}\,, \label{eq:symQQ2}
\end{align}
where $j \leftrightarrow k$ means to swap both momenta, $p_j \leftrightarrow p_k$, and colour indices, $i_j \leftrightarrow i_k$ or $a_j \leftrightarrow a_k$, respectively. Similarly, we find for the four-gauge-boson amplitudes 
\begin{equation}
\begin{split}
\boldsymbol{f}_{X}^{\lambda_2 \lambda_1 \lambda_3 \lambda_4} = 
\boldsymbol{f}_{X}^{\lambda_1\lambda_2 \lambda_3 \lambda_4}\vert_{1 \leftrightarrow 2} \,, \qquad 
\boldsymbol{f}_{gg}^{\lambda_1 \lambda_2 \lambda_4 \lambda_3} = 
\boldsymbol{f}_{gg}^{\lambda_1 \lambda_2 \lambda_3 \lambda_4}\vert_{3 \leftrightarrow 4} \,. 
\end{split}
\label{eq:bosesymmgg}
\end{equation}
In the case of four gluons, Bose symmetry under the exchange of any two gluons can even be used to show that the helicity coefficients of the single-minus helicity configurations are all equal
\begin{equation}
    \boldsymbol{f}_{gg}^{-+++}(s,t,\mt^2) = \boldsymbol{f}_{gg}^{+-++}(s,t,\mt^2) = \boldsymbol{f}_{gg}^{++-+}(s,t,\mt^2) = \boldsymbol{f}_{gg}^{+++-}(s,t,\mt^2) \, .
\end{equation}
For the expansion of the bare helicity coefficients in the bare strong coupling constant $\alpha_{s,b}$, we write up to two-loop order 
\begin{align}
    \boldsymbol{\Omega}_{q\bar{q}g\gamma} &=  \sqrt{4 \pi \alpha}\sqrt{4 \pi \alpha_{s,b}}\,\sum_{\ell=0}^{2} \left(\frac{\alpha_{s,b}}{2\pi}\right)^\ell  \boldsymbol{\Omega}_{q\bar{q}g\gamma}^{(\ell,b)}\,, \label{eq:asbexpqqga} \\
    \boldsymbol{\Omega}_{ggg\gamma} &= \sqrt{4 \pi \alpha}\sqrt{4 \pi \alpha_{s,b}}\, \sum_{\ell=1}^{2} \left(\frac{\alpha_{s,b}}{2\pi}\right)^\ell \boldsymbol{\Omega}_{ggg\gamma}^{(\ell,b)}\,, \label{eq:asbexpggga} \\
    \boldsymbol{\Omega}_{q\bar{q}\bar{Q}Q} &=  (4 \pi \alpha_{s,b})\,\sum_{\ell=0}^{2} \left(\frac{\alpha_{s,b}}{2\pi}\right)^\ell  \boldsymbol{\Omega}_{q\bar{q}\bar{Q}Q}^{(\ell,b)}\,, \label{eq:asbexpqqQQ} \\
    \boldsymbol{\Omega}_{q\bar{q}gg} &=  (4 \pi \alpha_{s,b})\,\sum_{\ell=0}^{2} \left(\frac{\alpha_{s,b}}{2\pi}\right)^\ell  \boldsymbol{\Omega}_{q\bar{q}gg}^{(\ell,b)}\,, \label{eq:asbexpqqgg} \\
    \boldsymbol{\Omega}_{gggg} &=  (4 \pi \alpha_{s,b})\, \sum_{\ell=0}^{2} \left(\frac{\alpha_{s,b}}{2\pi}\right)^\ell \boldsymbol{\Omega}_{gggg}^{(\ell,b)}\,,  \label{eq:asbexpgggg}  
\end{align}
where $\ell$ represents the number of loops and $ \boldsymbol{\Omega}_{q\bar{q}\bar{Q}Q} = \{ \boldsymbol{\eta}_1, \boldsymbol{\eta}_2\}$, $ \boldsymbol{\Omega}_{q\bar{q}X} = \{ \boldsymbol{\alpha}_X, \boldsymbol{\beta}_X, \boldsymbol{\gamma}_X, \boldsymbol{\delta}_X\}$ and 
 $ \boldsymbol{\Omega}_{ggX} = \{ \boldsymbol{f}_X^{++++},\cdots ,\boldsymbol{f}_X^{+--+} \}$\, the helicity coefficients defined in \cref{eq:etaqqQQ,eq:alphabetaqq,eq:alphabetagg}.
Note that the gluon-fusion amplitudes for the production of a jet and a photon are one-loop induced, which implies that the sum starts with $\ell = 1$.
To compute the helicity coefficients $\boldsymbol{\Omega}_{q\bar{q}\bar{Q}Q}^{(\ell,b)} $, $\boldsymbol{\Omega}_{q\bar{q}X}^{(\ell,b)}$, $\boldsymbol{\Omega}_{ggX}^{(\ell,b)}$,  we follow the strategy proposed in~\cite{Peraro:2020sfm} and define a suitable set of projector operators which act directly on the Feynman diagrams representing the amplitudes to the corresponding loop order.
Explicitly, the projectors only depend on the spins of the external particles and commute with the color structure. We can therefore define three types of projector operators 
\begin{align}
    &\mathcal{P}_{q\bar{q}\bar{Q}Q}^{(i)} = \sum_{j=1}^2 b_{ij}\, \Pi_j^\dagger\,,\\
    &\mathcal{P}_{q\bar{q}}^{(i)} = \sum_{j=1}^4 c_{ij}\, \bar{u}(p_1) \Gamma_j^{\mu \nu} u(p_2) \epsilon^*_{3, \mu}(p_3)\epsilon^*_{4, \nu}(p_4)\,,\\
    &\mathcal{P}_{gg}^{(i)} = \sum_{j=1}^8 d_{ij}\, T_j^{\mu \nu \rho \sigma} \epsilon^*_{1, \mu}(p_1)\epsilon^*_{2, \nu}(p_2)\epsilon^*_{3, \rho}(p_3)\epsilon^*_{4, \sigma}(p_4)\,,
\end{align}
where the tensors $\Pi_i$, $\Gamma_j^{\mu \nu}$ and $T_j^{\mu \nu \rho \sigma}$ were defined in \cref{eq:tensqqQQ,eq:tensqq,eq:tensgg}.
The projectors are constructed such that they single out the individual form factors implicitly defined in~\cref{eq:ampggga,eq:ampgggg,eq:ampqqga,eq:ampqqgg,eq:ampqqQQ}, when applied to the respective amplitudes,
\begin{align}
    &\mathcal{P}_{q\bar{q}}^{(i)} \cdot \mathbfcal{A}_{q\bar{q}g\gamma} \coloneqq \sum_{pol} \mathcal{P}_{q\bar{q}}^{(i)} \,\mathbfcal{A}_{q\bar{q}g\gamma} =  \sqrt{4 \pi \alpha} \sqrt{4 \pi \alpha_{s,b}}\mathbfcal{F}_i \,,\\
    &\mathcal{P}_{gg}^{(i)} \cdot \mathbfcal{A}_{ggg\gamma} \coloneqq \sum_{pol} \mathcal{P}_{gg}^{(i)} \,\mathbfcal{A}_{ggg\gamma} = \sqrt{4 \pi \alpha} \sqrt{4 \pi \alpha_{s,b}} \mathbfcal{G}_i\,, \\
    &\mathcal{P}_{q\bar{q}\bar{Q}Q}^{(i)} \cdot \mathbfcal{A}_{q\bar{q}\bar{Q}Q} \coloneqq \sum_{pol} \mathcal{P}_{q\bar{q} \bar{Q}Q}^{(i)} \,\mathbfcal{A}_{q\bar{q}\bar{Q}Q} =  (4 \pi \alpha_{s,b})\mathbfcal{H}_i \,,\\
    &\mathcal{P}_{q\bar{q}}^{(i)} \cdot \mathbfcal{A}_{q\bar{q}gg} \coloneqq \sum_{pol} \mathcal{P}_{q\bar{q}}^{(i)} \,\mathbfcal{A}_{q\bar{q}gg} =  (4 \pi \alpha_{s,b})\mathbfcal{K}_i \,,\\
     &\mathcal{P}_{gg}^{(i)} \cdot \mathbfcal{A}_{gggg} \coloneqq \sum_{pol} \mathcal{P}_{gg}^{(i)} \,\mathbfcal{A}_{gggg} = (4 \pi \alpha_{s,b})\mathbfcal{J}_i\,.
\end{align}
As shown explicitly in the formulas above, their action, represented by the dot operator ``$\cdot$'', includes summing over the polarization of all external particles (i.e., quarks, gluons, and photons). Note that the polarization sums have to be performed consistently with the choice of gauge made in \cref{eq:gaugeeps} for the polarization vectors.
The explicit form of the coefficients $b_{ij}, $$c_{ij}$ and $d_{ij}$ can be found in~\cite{Peraro:2020sfm, Caola:2021rqz}. Instead, we combine them with the help of~\cref{app:helcoeffs} in such a way that they single out directly the helicity coefficients $\boldsymbol{\Omega}_{q\bar{q}\bar{Q}Q}$, $\boldsymbol{\Omega}_{q\bar{q}X}$ and $\boldsymbol{\Omega}_{ggX}$. Expressions for the helicity projectors obtained this way can be found in~\cite{Becchetti:2025rrz}, and explicitly in~\cref{app:helproj}.

%% file: masters.tex
\section{Amplitude Computation}
\label{sec:ampcomp}
We compute the scattering amplitudes starting from their representation in terms of Feynman diagrams, following a standard procedure, which we summarize in the following. First, for each of the processes in~\eqref{eq:channels}, we generate all the Feynman diagrams containing at least one massive fermion loop with \texttt{QGRAF}~\cite{Nogueira:1991ex}. We then apply the projector operators on each diagram and perform the relevant color and Dirac algebra using \texttt{FORM}~\cite{Vermaseren:2000nd, Kuipers:2012rf, Ruijl:2017dtg}. In~\cref{fig:numberofdiagrams}, we show the number of contributing two-loop diagrams for each process.

\begin{table}[H]
\begin{center}
\begin{tabular}{| m{2.45cm} | m{2cm} | m{2cm} | m{2cm} | m{2cm} | m{2cm} |} 
 \hline
 \centering Process 
 & \centering\arraybackslash $q\bar{q}\ \longrightarrow \ g \gamma$ 
 & \centering\arraybackslash $gg \ \longrightarrow \ g \gamma$ 
 & \centering\arraybackslash $q\bar{q} \ \longrightarrow \ \bar{Q}Q$ 
 & \centering\arraybackslash $q\bar{q}\ \longrightarrow \ gg$ 
 & \centering\arraybackslash $gg\ \longrightarrow \ gg$  \\ 
 \hline
 \centering \# of diagrams 
 & \centering\arraybackslash $71$ 
 & \centering\arraybackslash $300$ 
 & \centering\arraybackslash $23$ 
 & \centering\arraybackslash $135$ 
 & \centering\arraybackslash $723$ \\ 
 \hline
\end{tabular}
\caption{\label{fig:numberofdiagrams} Number of contributing two-loop diagrams with at least one heavy quark loop. We notice here that in all processes involving four partons, starting at two loops also diagrams with snails have to be included.}
\end{center}
\end{table}
Having performed these manipulations, we can express all two-loop helicity coefficients as linear combinations of scalar Feynman integrals from the five integral families introduced in~\cite{Becchetti:2025rrz} and their crossings. For convenience, we repeat their definitions in~\cref{app:deffamilies}. Using integration-by-parts identities~\cite{Tkachov:1981wb, Chetyrkin:1981qh} and the Laporta algorithm~\cite{Laporta:2000dsw} as implemented and publicly available in \texttt{Reduze2}~\cite{Studerus:2009ye, vonManteuffel:2012np} and \texttt{KIRA2}~\cite{Maierhofer:2017gsa, Klappert:2019emp, Klappert:2020nbg, Klappert:2020aqs}, we express all the integrals appearing in the amplitudes in terms of a set of finitely many master integrals. For the complete set of processes in~\cref{eq:channels}, we find a total of $220$ master integrals, which include all the $165$ master integrals relevant for diphoton production from~\cite{Becchetti:2025rrz} and 55 extra integrals. 
We stress here that the additional $55$ integrals can be chosen as crossings of the $165$ integrals known from diphoton production. A convenient basis of all $220$ master integrals is given in~\cref{app:masters}. Taking into account the 11 extra relations reported in~\cite{Becchetti:2025rrz}, the number of independent master integrals can be reduced further to $209$.

\subsection{Master Integral Computation}
\label{sec:mastercomp}
To compute all $220$ relevant master integrals as a Laurent series in $\epsilon$, we generalized the same strategy outlined in~\cite{Becchetti:2025rrz}. In detail, we derived a system of differential equations for all master integrals and cast it into an $\epsilon$-factorized form utilizing the method of~\cite{Gorges:2023zgv, Duhr:2025lbz}. As demonstrated in many other instances, this method conjecturally guarantees that the resulting system of differential equations fulfills some standard criteria that we expect to constitute the natural generalization of a canonical basis beyond multiple polylogarithms. In particular, the differential equations can be expressed in terms of an independent set of differential forms with at most single poles locally close to each regular singular point. Moreover, the equations degenerate to standard canonical differential equations in dlog-form when expanded close to singular points of the corresponding non-polylogarithmic geometries, in this case, an elliptic curve. 
We recall here that, for many polylogarithmic sectors, we could take canonical candidates from the literature~\cite{Caron-Huot:2014lda, Becchetti:2017abb, Becchetti:2023wev}. Beyond the simple polylogarithmic graphs, and as already discussed in~\cite{Maltoni:2018zvp, Becchetti:2023wev, Ahmed:2024tsg, Becchetti:2025rrz}, in computing the relevant integrals, one encounters two graphs related to a single elliptic curve. As demonstrated in~\cite{Becchetti:2025rrz}, the underlying geometry implies that the differential forms of our $\epsilon$-factorized system of differential equations are expressed not only in terms of rational and algebraic functions, but also two irreducible new transcendental functions: one can be chosen as the holomorphic period of the elliptic curve, which we denote by $\varpi_0 (s,\mt^2)$, while the other can be expressed as integrated integral over this $\varpi_0(s,\mt^2)$ with an algebraic kernel
\begin{align}
    G(s,t,\mt^2)& = \int^{\mt^2} \mathrm{d}x \, \frac{s (s+2 t)\sqrt{P_4(x - t)}}{(t (s+t)-4 s x)^2} \, \varpi_0 (s,x)\,, \label{eq:Gdef} \\
    P_4(X) &= (\mt^2-X) (\mt^2+s-X) \left(\mt^2 (\mt^2-3 s)-X (2 \mt^2+s)+X^2\right) \, . \label{eq:ellcurve}
\end{align}
As a consequence, when writing down a solution for the master integrals as a Laurent series in $\epsilon$, the iterated integrals appearing as the coefficients in the expansion contain a new set of kernels beyond $\mathrm{d}\log$-forms, which depend on $\varpi_0 (s,\mt^2)$ and $G(s,t,\mt^2)$.
We recall here that, as discussed in~\cite{Becchetti:2025rrz}, the extra function $G(s,t,\mt^2)$ can be related to an extra puncture on the elliptic curve and can be expressed analytically in terms of complete elliptic integrals of the third kind. This form is nevertheless not particularly illuminating and of scarce practical utility. In fact, everything one needs in practice is the differential equation it satisfies, from which one can readily obtain its form as a series expansion close to every singular or regular point.

Since the extra $55$ integrals can be chosen as crossings of the $165$ master integrals already computed, their solution can be obtained rather straightforwardly by performing their relevant crossings. As an interesting feature of these partonic channels, while only the elliptic curve parametrized by $s$ was relevant in the case of diphoton production, the additional crossings give rise to the same elliptic curve parameterized by $t$ or $u$. As a consequence, besides $\varpi_0 (s,\mt^2)$ and $G(s,t,\mt^2)$, we also have to consider
\begin{equation}
    \varpi_0 (t,\mt^2) \, , \ G(t,u,\mt^2) \ \ \text{and} \ \ \varpi_0 (u,\mt^2) \, ,  \ G(u,s,\mt^2)
\end{equation}
as part of our differential forms. Note that not all of the amplitudes considered in this work actually receive contributions from the elliptic functions. In particular, the coefficients of different colour structures might contain contributions from different crossings of the elliptic curve, as summarized in~\cref{fig:ellcurves}. However, we note that the presence of the elliptic curve parametrized by a particular Mandelstam variable is, in our case, independent of the specific helicity configuration.
\begin{table}[H]
\begin{center}
\begin{tabular}{| m{2.8cm} | m{1.5cm} | m{1.5cm} | m{1.5cm} | m{1.5cm} | m{1.5cm} | m{1.5cm} |} 
 \hline
 \centering  
 & \centering\arraybackslash $\mathcal{C}_1$
 & \centering\arraybackslash $\mathcal{C}_2$
 & \centering\arraybackslash $\mathcal{C}_3$
 & \centering\arraybackslash $\mathcal{C}_4$
 & \centering\arraybackslash $\mathcal{C}_5$
 & \centering\arraybackslash $\mathcal{C}_6$ \\ 
 \hline
 \centering $q\bar{q}\ \longrightarrow \ g \gamma$
 & \centering\arraybackslash $s$
 & \centering\arraybackslash $-$
 & \centering\arraybackslash $-$
 & \centering\arraybackslash $-$
 & \centering\arraybackslash $-$
 & \centering\arraybackslash $-$ \\ 
 \hline
 \centering $gg \ \longrightarrow \ g \gamma$
 & \centering\arraybackslash $s,t,u$
 & \centering\arraybackslash $-$
 & \centering\arraybackslash $-$
 & \centering\arraybackslash $-$
 & \centering\arraybackslash $-$
 & \centering\arraybackslash $-$ \\ 
 \hline
 \centering $q\bar{q} \ \longrightarrow \ \bar{Q}Q$
 &  
 &  
 & \centering\arraybackslash $-$
 & \centering\arraybackslash $-$
 & \centering\arraybackslash $-$
 & \centering\arraybackslash $-$ \\ 
 \hline
 \centering $q\bar{q}\ \longrightarrow \ gg$
 & \centering\arraybackslash $s$
 & \centering\arraybackslash $s$
 & \centering\arraybackslash $s$
 & \centering\arraybackslash $-$
 & \centering\arraybackslash $-$
 & \centering\arraybackslash $-$ \\ 
 \hline
 \centering $gg\ \longrightarrow \ gg$
 &  
 &  
 &  
 & \centering\arraybackslash $t,u$
 & \centering\arraybackslash $s,u$
 & \centering\arraybackslash $s,t$ \\ 
 \hline
\end{tabular}
\caption{\label{fig:ellcurves} This table gives the Mandelstam variables by which the elliptic curves relevant to the coefficient of a particular colour structure in the helicity amplitudes of a given process are parametrized. While an empty cell means that there are no elliptic contributions, a $-$ signifies that the given process does not have that many colour structures, see~\cref{eq:colordecomqqga,eq:colordecomggga,eq:colordecomqqQQ,eq:colordecomqqgg,eq:colordecomgggg}.}
\end{center}
\end{table}
We provide the entire $220$-integral canonical basis as well as the differential equation it satisfies in the ancillary files found in a \href{https://doi.org/10.5281/zenodo.17141555}{\tt zenodo.org} repository submission~\cite{zenodo} accompanying this article. In summary, the differential equations contain a total of $118$ differential forms, which are those listed in appendix C of~\cite{Becchetti:2025rrz} as well as additional crossings thereof as required by the extra 55 master integrals.
In particular, $67$ of the differential forms are $\mathrm{d}\log$-forms, where $16$ of the letters are purely rational (\emph{even}), while $51$ are algebraic (\emph{odd}). The remaining $51$ kernels are of elliptic type.

Focusing on the elliptic letters, we observe interesting patterns of cancellations at the level of the $\mathcal{O}(\epsilon^0)$ part of our amplitudes. These align completely with previous reports of similar cancellations in other calculations of QFT correlators related to non-trivial geometries~\cite{Duhr:2024bzt, Forner:2024ojj, Becchetti:2025rrz}.
Concretely, we find that 18 out of the 51 elliptic differential forms drop from the $O(\epsilon^0)$ of all our amplitudes, in particular all those that are related to the period squared, $\omega_0^2$.

\subsection{Revisiting the extra relations among the master integrals}
\label{sec:extrarel}
In this section, we elaborate on 11 extra relations among the master integrals which we first identified in~\cite{Becchetti:2025rrz}.
With our choice of integral families, we could not obtain these identities from integration-by-parts identities, even including higher sectors, as is often the case. 
In terms of the canonical integrals used in this work, the relations read as follows:
\begin{equation}
\begin{aligned}
    &\mathcal{I}_{106}=\mathcal{I}_{63} \, , \quad &&\mathcal{I}_{109}=
   \mathcal{I}_{65} \, , \quad &&\mathcal{I}_{112}=
   \mathcal{I}_{67} \, ,\quad  \label{eq:extrarel1} \\
       &\mathcal{I}_{105}= \mathcal{I}_{104}-\mathcal{I}_{62} \, , \quad 
    &&  \mathcal{I}_{108}= \mathcal{I}_{107}-\mathcal{I}_{64} \, ,
    \quad &&  \mathcal{I}_{111}= \mathcal{I}_{110}-\mathcal{I}_{66} \, , 
\end{aligned}
\end{equation}
\begin{align}
    \mathcal{I}_{141}&= \mathcal{I}_{139} = \mathcal{I}_{137}\,, \nonumber \\
    \mathcal{I}_{138}&= -\frac{\mathcal{I}_8}{8}-\frac{5
   \mathcal{I}_{10}}{8}-\frac{\mathcal{I}_{12}}{8}
   +\frac{\mathcal{I}_{32}}{2}-\frac{3\mathcal{I}_{33}}{2}+\frac{\mathcal{I}_{34}}{2}-\mathcal{I}_{62}+\mathcal{I}_{64}
   -\mathcal{I}_{66} \nonumber \\ 
   &+\frac{\mathcal{I}_{95}}{2}-\frac{\mathcal{I}_{96}}{2}+\frac{\mathcal{I}_{97}}{2}-\frac{\mathcal{I}_{104}}{2}+\frac{\mathcal{I}_{107}}{2}-\frac{\mathcal{I}_{110}}{2} \, , \nonumber \\
   \mathcal{I}_{140}&=
   -\frac{\mathcal{I}_8}{8}-\frac{\mathcal{I}_{10}}{8}-\frac{5
   \mathcal{I}_{12}}{8}+\frac{\mathcal{I}_{32}}{2}
   +\frac{\mathcal{I}_{33}}{2}-\frac{3\mathcal{I}_{34}}{2}
   -\mathcal{I}_{62}-\mathcal{I}_{64}+\mathcal{I}_{66} \nonumber \\ 
   &+\frac{\mathcal{I}_{95}}{2}+\frac{\mathcal{I}_{96}}{2}-\frac{\mathcal{I}_{97}}{2}-\frac{\mathcal{I}_{104}}{2}-\frac{\mathcal{I}_{107}}{2}+\frac{\mathcal{I}_{110}}{2} \, , \nonumber \\
   \mathcal{I}_{142}&= -\frac{5\mathcal{I}_{8}}{8}-\frac{\mathcal{I}_{10}}{8}-\frac{\mathcal{I}_{12}}{8}-\frac{3\mathcal{I}_{32}}{2}+\frac{\mathcal{I}_{33}}{2}+\frac{\mathcal{I}_{34}}{2}+\mathcal{I}_{62}-\mathcal{I}_{64}-\mathcal{I}_{66} \nonumber \\ 
   &-\frac{\mathcal{I}_{95}}{2}+\frac{\mathcal{I}_{96}}{2}+\frac{\mathcal{I}_{97}}{2}+\frac{\mathcal{I}_{104}}{2}-\frac{\mathcal{I}_{107}}{2}-\frac{\mathcal{I}_{110
   }}{2} \, . \label{eq:extrarel2} 
\end{align}
As described in~\cite{Becchetti:2025rrz}, they can be proven straightforwardly through the system of differential equations satisfied by the integrals. In fact, starting from the system of differential equations in $\epsilon$-factorized (conjecturally canonical) form, one can easily see that all these combinations of masters satisfy purely homogeneous differential equations and that their boundary conditions are identically zero. We stress that having at disposal a canonical form is what allows us to search for these relations systematically, under the assumption that the coefficients in the linear combinations are simple rational numbers.

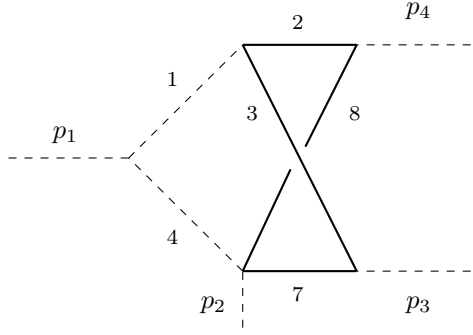
\begin{figure}[t]
\centering
\begin{tikzpicture}[scale=1.5]
\coordinate (links) at (-0.6,0);
\coordinate (mitte) at (0.5,0);
\coordinate (mmitte) at (1.92,-0.1); 
\coordinate (mmmitte) at (2.05,0.1); 
\coordinate (oben) at (1.5,1);
\coordinate (unten) at (1.5,-1);
\coordinate (obenr) at (2.5,1);
\coordinate (untenr) at (2.5,-1);
\coordinate (obenrr) at (3.5,1);
\coordinate (untenrr) at (3.5,-1);
\coordinate (untenl) at (1.5,-1.5);
\begin{scope}
\draw [-, thick,postaction={decorate}] (oben) to [bend right=0]  (obenr);
\draw [-, thick,postaction={decorate}] (untenr) to [bend right=0]  (unten);
\draw [-, thick,postaction={decorate}] (oben) to [bend right=0]  (untenr);
\draw [-, thick,postaction={decorate}] (obenr) to [bend right=0]  (mmmitte);
\draw [-, thick,postaction={decorate}] (unten) to [bend right=0]  (mmitte);
\draw [-, dashed,postaction={decorate}] (obenrr) to [bend right=0]  (obenr);
\draw [-, dashed,postaction={decorate}] (untenrr) to [bend right=0]  (untenr);
\draw [-, dashed,postaction={decorate}] (mitte) to [bend right=0]  (oben);
\draw [-, dashed,postaction={decorate}] (unten) to [bend right=0]  (mitte);
\draw [-, dashed,postaction={decorate}] (mitte) to [bend right=0]  (links);
\draw [-, dashed,postaction={decorate}] (untenl) to [bend right=0]  (unten);
\end{scope}
\node (d1) at (0.9,0.7) [font=\scriptsize, text width=.2 cm]{1};
\node (d2) at (0.9,-0.7) [font=\scriptsize, text width=.2 cm]{4};
\node (d3) at (2,1.2) [font=\scriptsize, text width=.2 cm]{2};
\node (d4) at (2,-1.2) [font=\scriptsize, text width=.2 cm]{7};
\node (d5) at (1.6,0.4) [font=\scriptsize, text width=.2 cm]{3};
\node (d6) at (2.5,0.4) [font=\scriptsize, text width=.2 cm]{8};
\node (d7) at (-0.1,0.2) [font=\small, text width=.2 cm]{$p_{1}$};
\node (d7) at (1.2,-1.3) [font=\small, text width=.2 cm]{$p_{2}$};
\node (d8) at (3,1.3) [font=\small, text width=.2 cm]{$p_4$};
\node (d9) at (3,-1.3) [font=\small, text width=.2 cm]{$p_3$};
\end{tikzpicture}
\caption{Feynman graph associated with the integrals given in~\eqref{eq:familyextrarels}. Dashed edges are massless, while the thick internal edges have mass $\mt^2$. The edge labels correspond to the propagator numbering in family $\text{NPA}$.}
\label{fig:GraphExtraRelations}
\end{figure}

Moreover, it was shown that the two relations in the first line of eq.~\eqref{eq:extrarel2} can also be proven by performing non-linear changes of variables in the Feynman parameter representation, following observations made in~\cite{Buccioni:2023okz}. It turns out that the remaining relations in eq.~\eqref{eq:extrarel2} admit a similar proof by generalizing the argument from~\cite{Becchetti:2025rrz} to higher propagator powers, which we present in the following. All the relations in eq.~\eqref{eq:extrarel2} originate from the integrals given by,\footnote{See~\cref{app:deffamilies} for the definition of the inverse propagators.}
\begin{equation}
\mathcal{I}_{\text{NPA}}(n_1,n_2,n_3,n_4,0,0,n_7,n_8,0) = 
\int \prod_{\ell=1}^2 \left[\frac{\mu_0^{2\epsilon}}{ C_\epsilon}\frac{\mathrm d^d k_\ell}{(2 \pi)^d}\right]
\frac{1}{D_{1}^{n_1} D_{2}^{n_2} D_{3}^{n_3} D_{4}^{n_4} D_{7}^{n_7} D_{8}^{n_8}}  \label{eq:familyextrarels}
\end{equation}
with $n_i \in \mathbb{N} \ \forall i \in \{1,2,3,4,7,8\}$. The associated graph is depicted in~\cref{fig:GraphExtraRelations}. 
Studying its symmetries, we can immediately see that under exchanging $p_3 \leftrightarrow p_4$, the integrals satisfy the following symmetry relations 
\begin{equation}
     \label{eq:graphsymmetryrel}
    \mathcal{I}_{\text{NPA}}(n_1,n_2,n_3,n_4,0,0,n_7,n_8,0) = \mathcal{I}_{\text{NPAx34}}(n_1,n_3,n_2,n_4,0,0,n_8,n_7,0) \, ,
\end{equation}
where we used the notation for crossings defined in~\cref{app:masters}. Let us now consider the Feynman parameter representation of the integrals~\eqref{eq:familyextrarels}. Its first and second Symanzik polynomials $\mathcal{U}$ and $\mathcal{F}$ are given by
\begin{align}
    \mathcal{U}(\underline x) =& \, \left(x_3+x_7\right) \left(x_2+x_8\right)+x_1 \left(x_2+x_3+x_7+x_8\right)+x_4 \left(x_2+x_3+x_7+x_8\right) \, , \\
    \mathcal{F}(s,u,\underline x) =& \, s \, x_7 \left(x_2 x_4-x_1 x_8\right)+u \, x_4 \left(x_2 x_7-x_3 x_8\right) +\mt^2 \left[x_1 \left(x_2+x_3+x_7+x_8\right)^2 \right. \nonumber \\ & \ \left.+x_4 \left(x_2+x_3+x_7+x_8\right)^2+\left(x_3+x_7\right) \left(x_2+x_8\right)
   \left(x_2+x_3+x_7+x_8\right)\right] \, ,
\end{align}
where we denote the Feynman parameter associated with edge $i$ by $x_i$ and we abbreviate the set of all Feynman parameters by $\underline x =(x_1,x_2,x_3,x_4,x_7,x_8)$. First, notice that the graph symmetry under $p_3 \leftrightarrow p_4$, which corresponds to $u\leftrightarrow t=-s-u$, can also be seen in Feynman parameter representation by simply exchanging $x_2 \leftrightarrow x_3$ and $x_7 \leftrightarrow x_8$. However, consider now the following change of variables:
\begin{equation}
\label{eq:quadcov}
\begin{aligned}
    x_1 &\to \frac{\left(x_1+x_4\right)x_3 }{x_3+x_7} \, , \quad &x_2 &\to x_2 \, , \quad &x_3 &\to \frac{
   \left(x_3+x_7\right) x_1}{x_1+x_4} \, , \\
   x_4 &\to \frac{\left(x_1+x_4\right) x_7}{x_3+x_7} \, , \quad &x_7 &\to \frac{\left(x_3+x_7\right)x_4}{x_1+x_4} \, , \quad &x_8&\to x_8 \, .
\end{aligned}
\end{equation}
The determinant of the corresponding Jacobian is equal to $-1$, and it is easy to see that the transformation maps 
\begin{align}
    x_1 + x_2 + x_3 + x_4 + x_7 + x_8 \quad &\rightarrow \quad x_1 + x_2 + x_3 + x_4 + x_7 + x_8 \, , \\
    \mathcal{U}(x_1,x_2,x_3,x_4,x_7,x_8) \quad &\rightarrow \quad \mathcal{U}(x_1,x_2,x_3,x_4,x_7,x_8) \, , \\
    \mathcal{F}(s,u,x_1,x_2,x_3,x_4,x_7,x_8) \quad &\rightarrow \quad \mathcal{F}(u,s,x_1,x_2,x_3,x_4,x_7,x_8) \, .
\end{align}
In this way, using that $s \leftrightarrow u$ corresponds to the crossing x1234, i.e., $p_1 \rightarrow p_2 \rightarrow p_3 \rightarrow p_4 \rightarrow p_1$, we obtain the relation
\begin{equation}
    \mathcal{I}_{\text{NPA}}(1,n_2,1,1,0,0,1,n_8,0) = \mathcal{I}_{\text{NPAx1234}}(1,n_2,1,1,0,0,1,n_8,0) \, .
\end{equation}
Combining this with the symmetry relation from eq.~\eqref{eq:graphsymmetryrel}, we may also write this as
\begin{equation}
    \mathcal{I}_{\text{NPA}}(1,n_2,1,1,0,0,1,n_8,0) = \mathcal{I}_{\text{NPAx123}}(1,1,n_2,1,0,0,n_8,1,0) \, .
\end{equation}
We can now take additional crossings of this identity until we have found all relations of this type among all the crossings of the family. Explicitly relevant to our case is to consider the additional crossings $\text{x123}$. Using that $\text{x123x123}=\text{x124}$ and $\text{x124x123}=\text{id}$, we obtain two more such relations
\begin{align}
    \mathcal{I}_{\text{NPAx123}}(1,n_2,1,1,0,0,1,n_8,0) &= \mathcal{I}_{\text{NPAx124}}(1,1,n_2,1,0,0,n_8,1,0) \, , \\
    \mathcal{I}_{\text{NPAx124}}(1,n_2,1,1,0,0,1,n_8,0) &= \mathcal{I}_{\text{NPA}}(1,1,n_2,1,0,0,n_8,1,0) \, .
\end{align}
Specializing these identities to the cases $(n_2,n_8)=(1,1)$ and $(n_2,n_8)=(2,1)$ and inserting integration-by-parts identities to rewrite them in terms of our canonical basis yields all the relations in~\eqref{eq:extrarel2}.

\begin{figure}[t]
\centering
\begin{tikzpicture}[scale=1.5]
\coordinate (links) at (-0.6,0);
\coordinate (mitte) at (0.5,0);
\coordinate (oben) at (2,1);
\coordinate (unten) at (2,-1);
\coordinate (obenr) at (3.1,1);
\coordinate (untenr) at (3.1,-1);
\begin{scope}
\draw [-, thick,postaction={decorate}] (oben) to [bend right=0]  (obenr);
\draw [-, thick,postaction={decorate}] (untenr) to [bend right=0]  (unten);
\draw [-, ultra thick,postaction={decorate}] (oben) to [bend right=0]  (unten);
\draw [-, very thin,postaction={decorate}] (mitte) to [bend right=0]  (oben);
\draw [-, very thin,postaction={decorate}] (unten) to [bend right=0]  (mitte);
\draw [-, dashed,postaction={decorate}] (mitte) to [bend right=0]  (links);
\end{scope}
\node (d1) at (1,0.7) [font=\small, text width=.35 cm]{$m^2$};
\node (d2) at (1,-0.7) [font=\small, text width=.35 cm]{$m^2$};
\node (d3) at (2.3,0.2) [font=\small, text width=.35 cm]{$M^2$};
\node (d4) at (0,0.2) [font=\small, text width=.35 cm]{$l$};
\node (d5) at (2.7,1.3) [font=\small, text width=.35 cm]{$q_2$};
\node (d6) at (2.7,-1.3) [font=\small, text width=.35 cm]{$q_1$};
\end{tikzpicture}
\caption{One-loop triangle graph associated with the integrals given in~\eqref{eq:triangle}. The dashed line is massless, $l^2=0$.}
\label{fig:trianglegraph}
\end{figure}
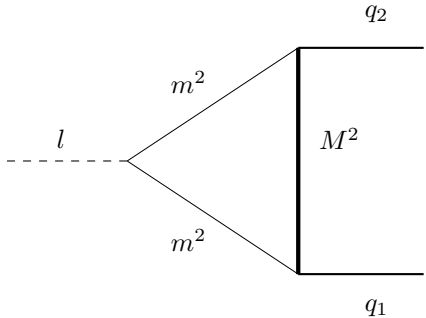

The relations in eq.~\eqref{eq:extrarel1} can also be proven from the Feynman parameter representation, although in a very different fashion. At the core of this proof lies the one-loop triangle relation discussed in~\cite{Abreu:2022vei} of which we present a generalization here. Consider the one-loop triangle integral family given by 
\begin{equation}
\mathcal{I}^{\text{tri}}_{a_1,a_2,b}(m^2,M^2) = 
\frac{\mu_0^{2\epsilon}}{ C_\epsilon} \int \frac{\mathrm d^d k}{(2 \pi)^d}
\frac{1}{\left[k^2-m^2\right]^{a_1}\left[(k-l)^2-m^2\right]^{a_2} \left[(k-l+q_2)^2-M^2\right]^{b}} \, ,\label{eq:triangle}
\end{equation}
where $l^2=0$, and we suppressed the depence of the integral on $q_1^2 \neq 0$ and $q_2^2 \neq 0$. The graph associated with this family is shown in~\cref{fig:trianglegraph}. 
The Feynman parameter representation of the integral~\eqref{eq:triangle} reads
\begin{align}
	\mathcal{I}^{\text{tri}}_{a_1,a_2,b}(m^2,M^2) &= 
    \mathcal{N} \int \left( \prod\limits_{i=1}^3 \mathrm{d}x_i \right) x_1^{a_1-1} x_2^{a_2-1} x_3^{b-1}   \dfrac{\mathcal{U}^{\, n-d}}{ \mathcal{F}^{\, n - \frac{d}{2}}} \, , \label{eq:Feynmantriangle} 
\end{align}
where
\begin{align}
    n&=a_1+a_2+b \quad \text{and} \quad \mathcal{N} = \frac{\mu_0^{2\epsilon}}{ C_\epsilon} \frac{ (-1)^n \, \Gamma (n - \frac{d}{2})}{\Gamma(a_1) \Gamma(a_2) \Gamma(b)} \, .
\end{align}
The first and second Symanzik polynomials of the graph are given by 
\begin{align}
    \mathcal{U} &= x_1+x_2+x_3 \, , \\
    \mathcal{F} &= m^2 (x_1+x_2) (x_1+x_2+x_3)+x_3 (M^2 (x_1+x_2+x_3)-q_1^2 x_1-q_2^2 x_2) \, ,
\end{align}
respectively. The Feynman parameter integral is a projective integral, and we may, for instance, evaluate it over the contour given by $x_1+x_2=1$ and $x_3 \in (0,\infty)$, yielding
\begin{align}
	\mathcal{I}^{\text{tri}}_{a_1,a_2,b}(m^2,M^2) &= 
    \mathcal{N} \int\limits_{0}^{\infty} \mathrm{d}x_3 \, x_3^{b-1} (1+x_3)^{n-d} \, \mathcal{I}_{a_1,a_2,b}(x_3,m^2,M^2) \, , \label{eq:triproj} \\
    \mathcal{I}_{a_1,a_2,b}(x_3,m^2,M^2) &= \int\limits_{0}^{1} \mathrm{d}x_1 \, \dfrac{x_1^{a_1-1} (1-x_1)^{a_2-1}}{ \left(m^2 (1+x_3)+x_3 (M^2 (1+x_3)-q_1^2 x_1-q_2^2 (1-x_1))\right)^{n - \frac{d}{2}}} \, .
\end{align}
Clearly, the $x_1$-integral could be performed in terms of a ${}_2F_1$, but this
is not necessary for the argument below.
In fact, the $x_1$-integral has an interesting property. Upon setting $x_3=1/y$, we find
\begin{equation}
    \mathcal{I}_{a_1,a_2,b}(1/y,m^2,M^2) = y^{2n-d} \ \mathcal{I}_{a_1,a_2,b}(y,M^2,m^2) \, .
\end{equation}
As a consequence, if we change variables in~\eqref{eq:triproj} according to $x_3=1/y$, we obtain
\begin{equation}
    \mathcal{I}^{\text{tri}}_{a_1,a_2,b}(m^2,M^2) = 
    \mathcal{N} \int\limits_{0}^{\infty} \mathrm{d}y \, y^{n-b-1} (1+y)^{n-d} \, \mathcal{I}_{a_1,a_2,b}(y,M^2,m^2) \, ,
\end{equation}
so for $2b=n=a_1+a_2+b$ (or, equivalently, $b=a_1+a_2$), we find that 
\begin{equation}
    \mathcal{I}^{\text{tri}}_{a_1,a_2,a_1+a_2}(m^2,M^2) = \mathcal{I}^{\text{tri}}_{a_1,a_2,a_1+a_2}(M^2,m^2)\,.
\end{equation}
Additionally, the triangle integral is symmetric under the exchange of the two internal masses. For the special case $a_1=a_2=1$ and $b=2$, we recover the relation presented in~\cite{Abreu:2022vei}. Graphically, this relation can be represented as 
\begin{equation}
\begin{tikzpicture}[baseline=-0.5ex, scale=1]
\coordinate (links) at (-0.5,0);
\coordinate (mitte) at (0.5,0);
\coordinate (oben) at (2,1);
\coordinate (unten) at (2,-1);
\coordinate (obenr) at (3.1,1);
\coordinate (untenr) at (3.1,-1);
\begin{scope}
\draw [-, thick,postaction={decorate}] (oben) to [bend right=0]  (obenr);
\draw [-, thick,postaction={decorate}] (untenr) to [bend right=0]  (unten);
\draw [-, ultra thick,postaction={decorate}] (oben) to [bend right=0]  (unten);
\draw [-, very thin,postaction={decorate}] (mitte) to [bend right=0]  (oben);
\draw [-, very thin,postaction={decorate}] (unten) to [bend right=0]  (mitte);
\draw [-, dashed,postaction={decorate}] (mitte) to [bend right=0]  (links);
\end{scope}
\node (d1) at (0.68,0.54) [font=\small, text width=.35 cm]{$m^2$};
\node (d2) at (0.68,-0.54) [font=\small, text width=.35 cm]{$m^2$};
\node (d3) at (2.3,0.5) [font=\small, text width=.35 cm]{$M^2$};
\node (d4) at (0,0.2) [font=\small, text width=.35 cm]{$l$};
\node (d5) at (2.7,1.3) [font=\small, text width=.35 cm]{$q_2$};
\node (d6) at (2.7,-1.3) [font=\small, text width=.35 cm]{$q_1$};
\node (d7) at (2.7,0) []{\tiny $(a_1+a_2)$};
\node[draw=black, fill=black, rectangle, inner sep=2.3pt] at (2,0) {};
\node (d8) at (1.25,-0.8) []{\tiny $(a_1)$};
\node[draw=black, fill=black, regular polygon, regular polygon sides=3, inner sep=1.3pt] at (1.25,-0.5) {};
\node (d9) at (1.25,0.8) []{\tiny $(a_2)$};
\node[draw=black, thick, circle, inner sep=2pt] at (1.25,0.5) {};
\end{tikzpicture}
\quad = \quad 
\begin{tikzpicture}[baseline=-0.5ex, scale=1]
\coordinate (links) at (-0.5,0);
\coordinate (mitte) at (0.5,0);
\coordinate (oben) at (2,1);
\coordinate (unten) at (2,-1);
\coordinate (obenr) at (3.1,1);
\coordinate (untenr) at (3.1,-1);
\begin{scope}
\draw [-, thick,postaction={decorate}] (oben) to [bend right=0]  (obenr);
\draw [-, thick,postaction={decorate}] (untenr) to [bend right=0]  (unten);
\draw [-, very thin,postaction={decorate}] (oben) to [bend right=0]  (unten);
\draw [-, ultra thick,postaction={decorate}] (mitte) to [bend right=0]  (oben);
\draw [-, ultra thick,postaction={decorate}] (unten) to [bend right=0]  (mitte);
\draw [-, dashed,postaction={decorate}] (mitte) to [bend right=0]  (links);
\end{scope}
\node (d1) at (0.68,0.54) [font=\small, text width=.35 cm]{$M^2$};
\node (d2) at (0.68,-0.54) [font=\small, text width=.35 cm]{$M^2$};
\node (d3) at (2.3,0.5) [font=\small, text width=.35 cm]{$m^2$};
\node (d4) at (0,0.2) [font=\small, text width=.35 cm]{$l$};
\node (d5) at (2.7,1.3) [font=\small, text width=.35 cm]{$q_2$};
\node (d6) at (2.7,-1.3) [font=\small, text width=.35 cm]{$q_1$};
\node (d7) at (2.7,0) []{\tiny $(a_1+a_2)$};
\node[draw=black, fill=black, rectangle, inner sep=2.3pt] at (2,0) {};
\node (d8) at (1.25,-0.8) []{\tiny $(a_1)$};
\node[draw=black, fill=black, regular polygon, regular polygon sides=3, inner sep=1.3pt] at (1.25,-0.5) {};
\node (d9) at (1.25,0.8) []{\tiny $(a_2)$};
\node[draw=black, thick, circle, inner sep=2pt] at (1.25,0.5) {};
\end{tikzpicture},
\quad 
\label{eq:trianglerel}
\end{equation}
where we used different symbols to indicate the symbolic different powers of
the corresponding propagators.
The fact that $q_1^2 \neq 0$ and $q_2^2 \neq 0$ are generic allows us to embed this identity into higher-loop graphs to obtain relations between higher-loop integrals. This way, combined with integration-by-parts identities and specifying explicit integer values for $a_1$ and $a_2$, we can obtain the relations in eq.~\eqref{eq:extrarel1}, which graphically correspond to
\begin{align}
\begin{tikzpicture}[baseline=-0.5ex, scale=1.5]
\coordinate (oben) at (0,1);
\coordinate (moben) at (0,0.5);
\coordinate (links) at (-1,0);
\coordinate (mlinks) at (-0.5,0);
\coordinate (rechts) at (1,0);
\coordinate (mrechts) at (0.5,0);
\coordinate (unten) at (0,-1);
\coordinate (munten) at (0,-0.5);
\begin{scope}
\draw [-, dashed,postaction={decorate}] (oben) to [bend right=0]  (moben);
\draw [-, dashed,postaction={decorate}] (links) to [bend right=0]  (mlinks);
\draw [-, dashed,postaction={decorate}] (rechts) to [bend right=0]  (mrechts);
\draw [-, dashed,postaction={decorate}] (unten) to [bend right=0]  (munten);
\draw [-, dashed,postaction={decorate}] (moben) to [bend right=0]  (mrechts);
\draw [-, very thick,postaction={decorate}] (mrechts) to [bend right=0]  (munten);
\draw [-, very thick,postaction={decorate}] (munten) to [bend right=0]  (mlinks);
\draw [-, dashed,postaction={decorate}] (mlinks) to [bend right=0]  (moben);
\draw [-, very thick,postaction={decorate}] (mlinks) to [bend right=0]  (mrechts);
\end{scope}
\node (d1) at (0.2,0.7) [font=\small, text width=.35 cm]{$p_1$};
\node (d2) at (-0.7,0.15) [font=\small, text width=.35 cm]{$p_2$};
\node (d3) at (0.2,-0.7) [font=\small, text width=.35 cm]{$p_3$};
\node (d4) at (0.7,0.15) [font=\small, text width=.35 cm]{$p_4$};
\node[draw=black, fill=black, circle, inner sep=2pt] at (0,0) {};
\end{tikzpicture}
\quad &= \quad 
\begin{tikzpicture}[baseline=-0.5ex, scale=1.5]
\coordinate (oben) at (0,1);
\coordinate (moben) at (0,0.5);
\coordinate (links) at (-1,0);
\coordinate (mlinks) at (-0.5,0);
\coordinate (rechts) at (1,0);
\coordinate (mrechts) at (0.5,0);
\coordinate (unten) at (0,-1);
\coordinate (munten) at (0,-0.5);
\begin{scope}
\draw [-, dashed,postaction={decorate}] (oben) to [bend right=0]  (moben);
\draw [-, dashed,postaction={decorate}] (links) to [bend right=0]  (mlinks);
\draw [-, dashed,postaction={decorate}] (rechts) to [bend right=0]  (mrechts);
\draw [-, dashed,postaction={decorate}] (unten) to [bend right=0]  (munten);
\draw [-, very thick,postaction={decorate}] (moben) to [bend right=0]  (mrechts);
\draw [-, very thick,postaction={decorate}] (mrechts) to [bend right=0]  (munten);
\draw [-, very thick,postaction={decorate}] (munten) to [bend right=0]  (mlinks);
\draw [-, very thick,postaction={decorate}] (mlinks) to [bend right=0]  (moben);
\draw [-, dashed,postaction={decorate}] (mlinks) to [bend right=0]  (mrechts);
\end{scope}
\node (d1) at (0.2,0.7) [font=\small, text width=.35 cm]{$p_1$};
\node (d2) at (-0.7,0.15) [font=\small, text width=.35 cm]{$p_2$};
\node (d3) at (0.2,-0.7) [font=\small, text width=.35 cm]{$p_3$};
\node (d4) at (0.7,0.15) [font=\small, text width=.35 cm]{$p_4$};
\node[draw=black, fill=black, circle, inner sep=2pt] at (0,0) {};
\end{tikzpicture}
\quad , \\
& \nonumber \\
\begin{tikzpicture}[baseline=-0.5ex, scale=1.5]
\coordinate (oben) at (0,1);
\coordinate (moben) at (0,0.5);
\coordinate (links) at (-1,0);
\coordinate (mlinks) at (-0.5,0);
\coordinate (rechts) at (1,0);
\coordinate (mrechts) at (0.5,0);
\coordinate (unten) at (0,-1);
\coordinate (munten) at (0,-0.5);
\begin{scope}
\draw [-, dashed,postaction={decorate}] (oben) to [bend right=0]  (moben);
\draw [-, dashed,postaction={decorate}] (links) to [bend right=0]  (mlinks);
\draw [-, dashed,postaction={decorate}] (rechts) to [bend right=0]  (mrechts);
\draw [-, dashed,postaction={decorate}] (unten) to [bend right=0]  (munten);
\draw [-, dashed,postaction={decorate}] (moben) to [bend right=0]  (mrechts);
\draw [-, very thick,postaction={decorate}] (mrechts) to [bend right=0]  (munten);
\draw [-, very thick,postaction={decorate}] (munten) to [bend right=0]  (mlinks);
\draw [-, dashed,postaction={decorate}] (mlinks) to [bend right=0]  (moben);
\draw [-, very thick,postaction={decorate}] (mlinks) to [bend right=0]  (mrechts);
\end{scope}
\node (d1) at (0.2,0.7) [font=\small, text width=.35 cm]{$p_1$};
\node (d2) at (-0.7,0.15) [font=\small, text width=.35 cm]{$p_2$};
\node (d3) at (0.2,-0.7) [font=\small, text width=.35 cm]{$p_3$};
\node (d4) at (0.7,0.15) [font=\small, text width=.35 cm]{$p_4$};
\node[draw=black, fill=black, circle, inner sep=2pt] at (-0.15,0) {};
\node[draw=black, fill=black, circle, inner sep=2pt] at (0.15,0) {};
\node[draw=black, fill=black, circle, inner sep=2pt] at (0.25,0.25) {};
\end{tikzpicture}
\quad &= \quad 
\begin{tikzpicture}[baseline=-0.5ex, scale=1.5]
\coordinate (oben) at (0,1);
\coordinate (moben) at (0,0.5);
\coordinate (links) at (-1,0);
\coordinate (mlinks) at (-0.5,0);
\coordinate (rechts) at (1,0);
\coordinate (mrechts) at (0.5,0);
\coordinate (unten) at (0,-1);
\coordinate (munten) at (0,-0.5);
\begin{scope}
\draw [-, dashed,postaction={decorate}] (oben) to [bend right=0]  (moben);
\draw [-, dashed,postaction={decorate}] (links) to [bend right=0]  (mlinks);
\draw [-, dashed,postaction={decorate}] (rechts) to [bend right=0]  (mrechts);
\draw [-, dashed,postaction={decorate}] (unten) to [bend right=0]  (munten);
\draw [-, very thick,postaction={decorate}] (moben) to [bend right=0]  (mrechts);
\draw [-, very thick,postaction={decorate}] (mrechts) to [bend right=0]  (munten);
\draw [-, very thick,postaction={decorate}] (munten) to [bend right=0]  (mlinks);
\draw [-, very thick,postaction={decorate}] (mlinks) to [bend right=0]  (moben);
\draw [-, dashed,postaction={decorate}] (mlinks) to [bend right=0]  (mrechts);
\end{scope}
\node (d1) at (0.2,0.7) [font=\small, text width=.35 cm]{$p_1$};
\node (d2) at (-0.7,0.15) [font=\small, text width=.35 cm]{$p_2$};
\node (d3) at (0.2,-0.7) [font=\small, text width=.35 cm]{$p_3$};
\node (d4) at (0.7,0.15) [font=\small, text width=.35 cm]{$p_4$};
\node[draw=black, fill=black, circle, inner sep=2pt] at (-0.15,0) {};
\node[draw=black, fill=black, circle, inner sep=2pt] at (0.15,0) {};
\node[draw=black, fill=black, circle, inner sep=2pt] at (0.25,0.25) {};
\end{tikzpicture}
\quad 
\end{align}
and crossings thereof. We stress that here, at variance with~\cref{eq:trianglerel}, we use a filled dot as standard representation for a squared propagator.

%% file: uvandir.tex
\section{UV renormalization and IR structure}
\label{sec:ren}
The bare helicity amplitudes defined in~\cref{sec:setup} exhibit both IR and UV divergences. It is well-known that the latter can be removed consistently by the renormalization procedure. To this end, we choose to work in the $\overline{ \rm MS}$ scheme for the massless quarks and in the on-shell scheme for the massive quarks. The resulting relations between the bare and the renormalized QCD parameters, up to the order in $\epsilon$ and $\alpha_s$ needed for this calculation, are given by
\begin{align}
\mu_0^{2 \epsilon} S_{\epsilon}\alpha_{s,b} &
= \mu^{2\epsilon} \alpha_s
\left\{ 1- \left(\frac{\alpha_s}{2\pi}\right) \left( \frac{\beta_0 }{\epsilon}  + \delta_w \right) +\left(\frac{\alpha_s}{2\pi}\right)^2 \left[\left( \frac{\beta_0 }{\epsilon}  + \delta_w \right)^2 - \frac{\beta_1}{2\epsilon} + \Delta \right] + \mathcal O(\alpha_s^3) \right\} \,,  \\ 
\mt &= \mR \left[ 1+ \delta_m\, \left(\frac{\alpha_s}{2\pi}\right)+ \mathcal O(\alpha_s^2) \right]\,,
\end{align}
where
\begin{equation}
    S_{\epsilon} = (4\pi)^{\epsilon}e^{-\gamma_E \epsilon} \, ,
\end{equation}
and $\alpha_s = \alpha_s(\mu)$ denotes the renormalized strong coupling evaluated at the renormalization scale $\mu$. From here on, we set $\mu_0 = \mu$ and we have defined
\begin{align}
    \beta_0 &= \frac{11}{6}C_A - \frac{2}{3}T_F\, n_f\,, \\
    \beta_1 &= \frac{17}{6}C_A^2 - T_F\, n_f \left( \frac{5}{3} C_A + C_F \right) \, , \\
    \delta_w &=  T_F\, n_h\, \left( \frac{\mR^2}{\mu^2}\right)^{-\epsilon} \left(-\frac{2}{3 \epsilon} - \frac{\pi^2}{18} \epsilon + \frac{2}{9} \zeta_3 \epsilon^2 + \mathcal{O}(\epsilon^3)\right)\,, \label{eq:renconst} \\ 
    \Delta &=T_F\, n_h \left( \frac{\mR^2}{\mu^2}\right)^{-2\epsilon} \left[
     C_A  \left(\frac{5}{6 \epsilon}-\frac{8}{9}
     \right) + C_F  \left(\frac{1}{2 \epsilon} + \frac{15}{4} \right) + \mathcal{O}(\epsilon) \right]\,, \\ 
    \delta_m &= - C_F \left( \frac{\mR^2}{\mu^2}\right)^{-\epsilon} \left[ \frac{3}{2 \epsilon} + 2 + \epsilon \left(4  + \frac{\pi^2}{8} \right) + \epsilon^2 \left(8 + \frac{\pi^2}{6} - \frac{\zeta_3}{2} \right)+ \mathcal{O}(\epsilon^3)\right] \, .
\end{align}
In these formulas, $n_f$ denotes the number of massless active fermions and $n_h$ the number of massive flavours with (renormalized) mass $\mR$. In practice, $n_f = 5$ and $n_h = 1$, but we keep the symbols explicit for convenience. Further, we use the notation $C_A = N_c$  for the Casimir of the adjoint representation and $C_F =T_F (N_c^2-1)/N_c$ for the Casimir of the fundamental representation with $N_c = 3$ and $T_F = 1/2$.

Finally, we give the wave-function renormalization constants for gluons and massless quarks
\begin{align}
    Z_A &= 1 + \left(\frac{\alpha_s}{2\pi}\right) \delta_w  - \left(\frac{\alpha_s}{2\pi}\right)^2 \left[ \frac{\beta_0}{\epsilon} \, \delta_w + \Delta +   \delta_A + \mathcal{O}(\epsilon) \right]+ \mathcal O(\alpha_s^3)\,, \nonumber \\ 
    Z_q &= 1 + \left(\frac{\alpha_s}{2\pi}\right)^2 \delta_q + \mathcal O(\alpha_s^3)\, , \\
    &\text{where} \quad \delta_A = C_F\, T_F\, n_h \left(\frac{\mR^2}{\mu^2}\right)^{-2\epsilon} \left[\frac{1}{ 4\epsilon^2} - \frac{5}{24\epsilon} +\frac{89}{144} +\frac{\pi^2}{24} +\mathcal{O}(\epsilon) \right] \, , \\
    & \quad \text{and} \quad \delta_q = C_F\, T_F\, n_h \left(\frac{\mR^2}{\mu^2}\right)^{-2\epsilon} \left[\frac{1}{ 4\epsilon} - \frac{5}{24} +\mathcal{O}(\epsilon) \right] \, .
\end{align}
Analogously to~\cref{eq:asbexpqqga,eq:asbexpggga,eq:asbexpqqQQ,eq:asbexpqqgg,eq:asbexpgggg}, we expand the renormalized helicity coefficients in the renormalized strong coupling constant,
\begin{align}
    \boldsymbol{\Omega}^{\rm UV}_{q\bar{q}g\gamma} &=  \sqrt{4 \pi \alpha}\sqrt{4 \pi \alpha_{s}}\,\sum_{\ell=0}^{2} \left(\frac{\alpha_{s}}{2\pi}\right)^\ell  \boldsymbol{\Omega}_{q\bar{q}g\gamma}^{(\ell,\rm UV)}\,, \\
    \boldsymbol{\Omega}^{\rm UV}_{ggg\gamma} &= \sqrt{4 \pi \alpha}\sqrt{4 \pi \alpha_{s}}\, \sum_{\ell=1}^{2} \left(\frac{\alpha_{s}}{2\pi}\right)^\ell \boldsymbol{\Omega}_{ggg\gamma}^{(\ell,\rm UV)}\,, \\
    \boldsymbol{\Omega}^{\rm UV}_{q\bar{q}\bar{Q}Q} &=  (4 \pi \alpha_{s})\,\sum_{\ell=0}^{2} \left(\frac{\alpha_{s}}{2\pi}\right)^\ell  \boldsymbol{\Omega}_{q\bar{q}\bar{Q}Q}^{(\ell,\rm UV)}\,, \\
    \boldsymbol{\Omega}^{\rm UV}_{q\bar{q}gg} &=  (4 \pi \alpha_{s})\,\sum_{\ell=0}^{2} \left(\frac{\alpha_{s}}{2\pi}\right)^\ell  \boldsymbol{\Omega}_{q\bar{q}gg}^{(\ell,\rm UV)}\,, \\
    \boldsymbol{\Omega}^{\rm UV}_{gggg} &=  (4 \pi \alpha_{s})\, \sum_{\ell=0}^{2} \left(\frac{\alpha_{s}}{2\pi}\right)^\ell \boldsymbol{\Omega}_{gggg}^{(\ell,\rm UV)}\,.  
\end{align}
The relation to the bare helicity coefficients is given by 
\begin{equation}
    \boldsymbol{\Omega}^{\rm UV}_{Z} = \left( \sqrt{Z_A} \right)^{N_{Z}^{(g)}} \left( \sqrt{Z_q} \right)^{N_{Z}^{(q)}}  \boldsymbol{\Omega}_{Z}\,,
    \label{eq:rengen}
\end{equation}
with $N_{Z}^{(g)}$ and $N_{Z}^{(q)}$ the number of external gluons and external (massless) quarks plus antiquarks involved in process $Z$, respectively. We stress here that, since the processes considered do not involve any on-shell massive external quarks, we do not need to compute explicitly any mass counter-term, but we can instead implement mass renormalization by simply re-expressing $\mt$ in terms of $\mR$ and re-expanding in the renormalized coupling $\alpha_s$.
The explicit relations of the $\boldsymbol{\Omega}_{Z}^{(\ell,\rm UV)}$ to the bare $\boldsymbol{\Omega}_{Z}^{(\ell,b)}$ are shown in~\cref{app:renhelcoeffs}.

The renormalized helicity coefficients $\boldsymbol{\Omega}^{\rm UV}_{Z}$ still contain further poles in $\epsilon$, which are, at this point, exclusively of IR origin. In gauge theories, their form is known to be universal, and it can be predicted order by order in the strong coupling constant, only from the knowledge of the previous orders in the perturbative expansion. 
In particular, up to NNLO, it has been proven that only dipole-like terms contribute, and the form of the IR poles is especially simple~\cite{Catani:1998bh, Becher:2009cu, Becher:2009kw, Becher:2009qa, Gardi:2009qi, Ferroglia:2009ii}. While we discuss this matter in more detail in~\cref{app:IRstructure}, the upshot is that we can define finite remainders for our heavy-quark induced two-loop amplitudes as
\begin{equation}
    \boldsymbol{\Omega}^{(2,\rm fin)}_{Z} = \lim_{\epsilon \rightarrow 0} \left( \boldsymbol{\Omega}^{(2,\rm UV)}_{Z} - \mathbfcal{I}^{(1)}_Z \boldsymbol{\Omega}^{(1,\rm UV)}_{Z} \right) \, . \label{eq:IRsubtraction}
\end{equation}
The operator $\mathbfcal{I}^{(1)}_Z$ is, in general, a non-diagonal operator in color space. However, it is easy to see that, as long as at most three colored partons are involved, the operator becomes diagonal in color space and we can write $\mathbfcal{I}^{(1)} \rightarrow \mathcal{I}^{(1)} $~\cite{Catani:1998bh}. This is clearly the case for those partonic sub-channels considered in this paper, which involve the production of one photon and a jet. 
Concretely, for the two cases that are relevant here, we find explicitly
\begin{align}
    \mathcal{I}_{ggga}^{(1)} = &- \frac{1}{\epsilon^2}\frac{3C_A}{2} - \frac{1}{2\epsilon}\, \left[3\beta_0-N_c\left( \log{\left( \frac{-s-i 0^+}{\mu^2} \right)} + \log{\left( \frac{-t}{\mu^2} \right)}  + \log{\left( \frac{-u}{\mu^2} \right)} \right)\right] \, , \\
    \mathcal{I}_{q\bar{q}ga}^{(1)} = &- \frac{1}{\epsilon^2}\frac{2C_F + C_A}{2} \nonumber \\
    &+ \frac{1}{2 \epsilon} \left[ -\frac{1}{N_c}\log{\left(\frac{-s-i 0^+}{\mu^2} \right)} + N_c \left(  \log{\left(\frac{-t}{\mu^2} \right)}  + \log{\left(\frac{-u}{\mu^2} \right)} \right)-3C_F -\beta_0\right] \, , \nonumber
\end{align}
where the $i0^+$ provides the prescription to analytically continue the logarithm to the physical region $s>0$. For the other three processes involving four external partons, the corresponding color space is multi-dimensional. The helicity coefficients can be decomposed in their respective color basis  in analogy to~\cref{eq:colordecomqqQQ,eq:colordecomqqgg,eq:colordecomgggg}, i.e., 
\begin{equation}
    \boldsymbol{\Omega}_{Z} = \sum_{i=1}^{n_Z} \boldsymbol{\Omega}_{Z}^{[i]} \ \mathcal{C}_{i}^{[Z]} \label{eq:coulourdecompgeneral}
\end{equation}
with $n_{q\bar{q}\bar{Q}Q}=2$, $n_{q\bar{q}gg}=3$, $n_{gggg}=6$ and the $\mathcal{C}_{i}^{[Z]}$ the respective colour structures given in~\cref{eq:colqqQQ,eq:colqq,eq:colgg}. We can interpret the coefficients $\boldsymbol{\Omega}_{Z}^{[i]}$ as the components of an $n$-dimensional vector and represent the operator $\mathbfcal{I}^{(1)}_Z$ as an $(n \times n)$-dimensional matrix acting on that vector. In this way, the operators can be written as 
\begin{align}
    \mathbfcal{I}^{(1)}_{q\bar{q}\bar{Q}Q} &= - C_F\left(\frac{2}{\epsilon^2} + \frac{3}{\epsilon}\right)\, \mathbf{I}_2 + \frac{1}{\epsilon} \sum_{x=s,t,u} \mathbf{A}_x \log{\left( \frac{-x-i 0^+}{\mu^2} \right)} \, , \\
    \mathbfcal{I}^{(1)}_{q\bar{q}gg} &= - \left(\frac{C_F+C_A}{\epsilon^2} + \frac{3C_F+2\beta_0}{2\epsilon}\right)\, \mathbf{I}_3 + \frac{1}{\epsilon} \sum_{x=s,t,u} \mathbf{B}_x \log{\left( \frac{-x-i 0^+}{\mu^2} \right)} \, , \\
    \mathbfcal{I}^{(1)}_{gggg} &= - \left(\frac{2C_A}{\epsilon^2} + \frac{2\beta_0}{\epsilon}\right)\, \mathbf{I}_6 + \frac{1}{\epsilon} \sum_{x=s,t,u} \mathbf{C}_x \log{\left( \frac{-x-i 0^+}{\mu^2} \right)} \, ,
\end{align}
where $\mathbf{I}_n$ denotes the $(n \times n)$-dimensional identity matrix and the remaining matrices read explicitly
\begin{align}
    \mathbf{A}_s = \left(
    \begin{array}{cc}
        -\frac{1}{N_c} & 0 \\
        1 & \frac{N_c^2-1}{N_c} \\
    \end{array} \right) \, , \quad 
    \mathbf{A}_t = \left(
    \begin{array}{cc}
        \frac{1}{N_c} & -1 \\
        -1 & \frac{1}{N_c} \\
    \end{array} \right) \, , \quad 
    \mathbf{A}_u = \left(
    \begin{array}{cc}
        \frac{N_c^2-1}{N_c} & 1 \\
        0 & -\frac{1}{N_c} \\
    \end{array} \right) \, , \\
\end{align}
\begin{equation}
    \mathbf{B}_s = \left(
    \begin{array}{ccc}
        \frac{N_c^2-1}{2 N_c} & 0 & 0 \\
        0 & \frac{N_c^2-1}{2 N_c} & 0 \\
        \frac{1}{2} & \frac{1}{2} & \frac{3 N_c^2-1}{2 N_c} \\
    \end{array} \right) \, , \quad 
    \mathbf{B}_t = \left(
    \begin{array}{ccc}
        0 & 0 & -2 \\
        0 & N_c & 2 \\
        -\frac{1}{2} & 0 & 0 \\
    \end{array} \right) \, , \quad 
    \mathbf{B}_u = \left(
    \begin{array}{ccc}
        N_c & 0 & 2 \\
        0 & 0 & -2 \\
        0 & -\frac{1}{2} & 0 \\
    \end{array} \right) \, , 
\end{equation}
\begin{equation}
    \mathbf{C}_s = \left(
    \begin{array}{cccccc}
        N_c & 0 & 0 & 0 & 0 & 1 \\
        0 & N_c & 0 & 0 & 1 & 0 \\
        0 & 0 & 0 & 0 & -1 & -1 \\
        2 & 2 & 0 & 2 N_c & 0 & 0 \\
        0 & 0 & -2 & 0 & 0 & 0 \\
        0 & 0 & -2 & 0 & 0 & 0 \\
    \end{array} \right) \, , \quad 
    \mathbf{C}_t = \left(
    \begin{array}{cccccc}
        0 & 0 & 0 & -1 & 0 & -1 \\
        0 & N_c & 0 & 1 & 0 & 0 \\
        0 & 0 & N_c & 0 & 0 & 1 \\
        -2 & 0 & 0 & 0 & 0 & 0 \\
        0 & 2 & 2 & 0 & 2 N_c & 0 \\
        -2 & 0 & 0 & 0 & 0 & 0 \\
    \end{array} \right) \, , 
\end{equation}
\begin{equation}
    \mathbf{C}_u = \left(
    \begin{array}{cccccc}
        N_c & 0 & 0 & 1 & 0 & 0 \\
        0 & 0 & 0 & -1 & -1 & 0 \\
        0 & 0 & N_c & 0 & 1 & 0 \\
        0 & -2 & 0 & 0 & 0 & 0 \\
        0 & -2 & 0 & 0 & 0 & 0 \\
        2 & 0 & 2 & 0 & 0 & 2 N_c \\
    \end{array} \right) \, . 
\end{equation}
Note that in the formulas above, we have used the explicit values for $C_A$, $C_F$, and $T_F$ for $SU(N_c)$.
While the cancellation of all poles after UV renormalization and subtraction of the IR poles according to~\cref{eq:coulourdecompgeneral} served as a strong consistency check on the helicity amplitudes, we choose to provide instead the results for the bare helicity coefficients in a \href{https://doi.org/10.5281/zenodo.17141555}{\tt zenodo.org} repository submission~\cite{zenodo} accompanying this article. In this way, UV renormalization and IR subtraction can be implemented in any scheme of choice.

%% file: numerics.tex
\section{Numerical evaluation}
\label{sec:numerics}
In order to obtain a fast and reliable numerical implementation of our amplitudes, we resort to the same strategy already used in~\cite{Becchetti:2025rrz} and produce generalized series expansions for the whole amplitudes at different singular points. On the one hand, this allows for a drastic simplification of the numerical evaluation of the individual pieces, since one ultimately only has to evaluate polynomials, algebraic functions, and logarithms. On the other hand, by producing a series expansion for the whole amplitude, we can easily avoid strong numerical cancellations among master integrals or special functions, especially when spurious singularities are present in the rational functions that multiply them. This strategy makes it possible to work with low-precision numerical samples at all steps of the calculation, and is therefore particularly well-suited for an implementation with floating-point precision.

Compared to our previous study for $\gamma \gamma$ production in~\cite{Becchetti:2025rrz}, we consider now all massless partonic channels to produce $\gamma \gamma$, $\gamma j$, and $jj$, and we extend our analysis by considering one more expansion point, i.e., the so-called \emph{small-mass expansion}. Below, we briefly discuss how we obtain these expansions and then illustrate their performances across the physical phase space.

\subsubsection*{Large-mass expansion}
\begin{figure}[h!]
    \centering
\includegraphics[width=0.50\textwidth]{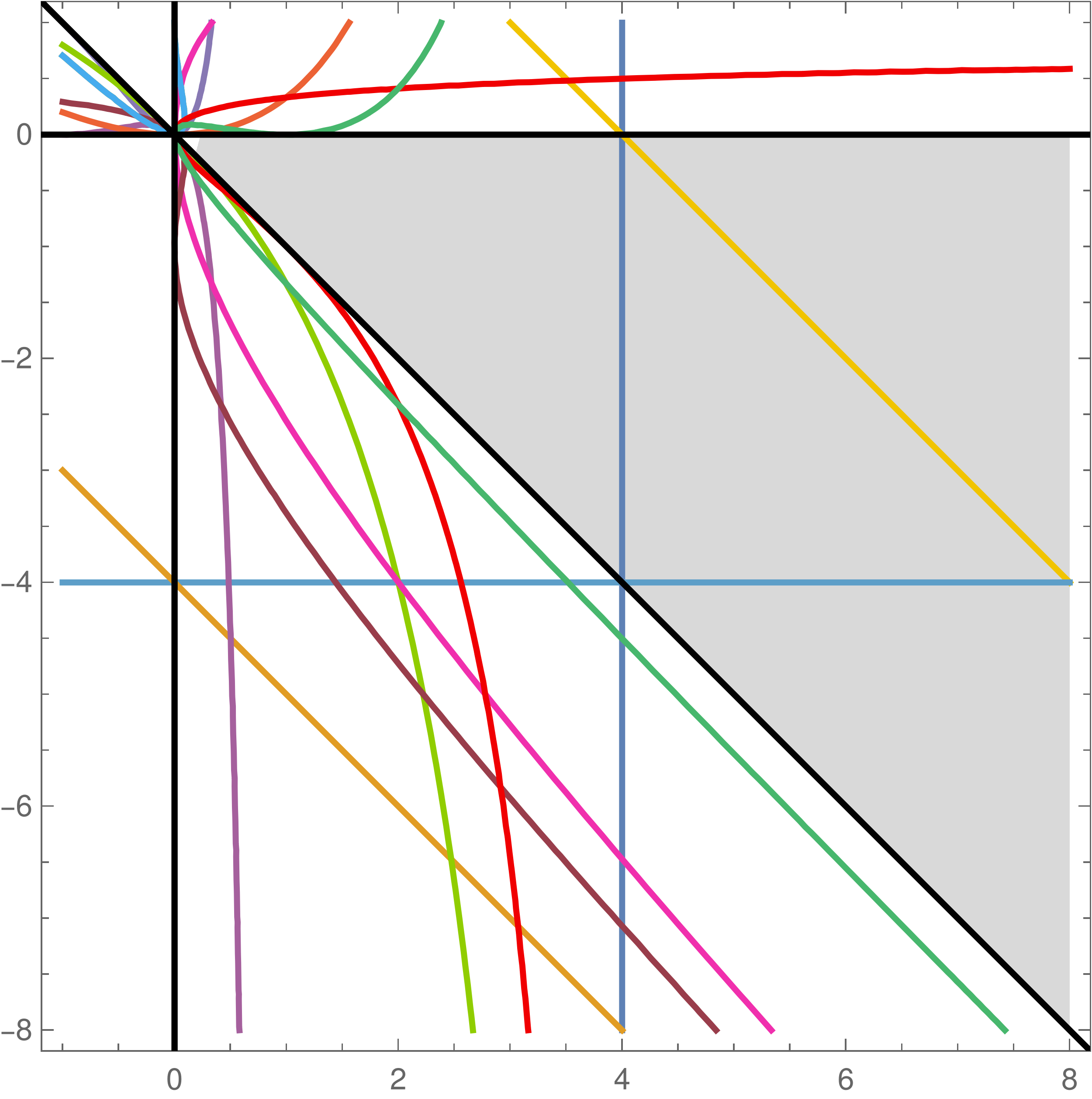}
    \caption{Plot of all singular curves of the system of differential equations for all master integrals. The horizontal axis corresponds to $s$ and the vertical axis to $t$ in units of $\mt^2$. The physical region is shaded in grey.}
    \label{fig:SingPlot}
\end{figure}

The large mass expansion was already studied in~\cite{Becchetti:2025rrz}. As discussed there, we derive the expansion directly from our canonical differential equations. To achieve that, we need to define an appropriate set of variables that allows us to resolve all singularities crossing at the point $s=0, \ t=0$, see~\cref{fig:SingPlot}. 
Compared to~\cite{Becchetti:2025rrz}, we generalize our work to include the missing crossings of the master integrals considered there. This does not require us to change our strategy, so we refer to~\cite{Becchetti:2025rrz} for more details.

\subsubsection*{Threshold expansion}
The second expansion we consider is the threshold expansion, which we had also already studied in~\cite{Becchetti:2025rrz}.
To be precise, we consider two-variable expansions at the two symmetric points $\{t=0,\, s=4\mt^2 \}$ and $\{u=0,\, s=4\mt^2 \}$. We focus here on the first of the two, but the same discussion applies to the second by crossing symmetry. In fact, the various color and helicity amplitudes for all the channels involving at least two gauge bosons have fixed transformation properties under $t \leftrightarrow u$ (see~\cref{eq:symQQ1,eq:symQQ2,eq:bosesymmgg}), which allows us to get all expansions at $u=0$ from the ones at $t=0$. For the four-quark channel, we rely instead on our canonical basis being closed under crossings of the external momenta.

We recall here that, if we limit ourselves to the $\gamma \gamma$ amplitudes, there are only two singular lines crossing normally at $\{t=0,s=4\mt^2\}$, which means that singularities are already resolved and we can perform the expansion in the standard variables
\begin{equation}
    y_1 = \frac{4\mt^2-s}{\mt^2} \, , \qquad y_2 = \frac{-t}{\mt^2} \, ,
    \label{eq:oldTexpVars}
\end{equation}
see eq.(7.4) in~\cite{Becchetti:2025rrz}. As a matter of fact, once we consider all partonic subchannels, more singularities arise and, in particular, a third singular line, $u=-4\mt^2$, crosses the other two at the expansion point considered, see~\cref{fig:SingPlot}. 
For this reason, we define a new set of variables that allows us to produce consistently all expansions for all relevant master integrals at once, resolving all singularities. In detail, we use
\begin{equation}
    v = \frac{4\mt^2-s}{-t} = \frac{y_1}{y_2}\, , \qquad w = \frac{-t}{\mt^2} = y_2 \, , \label{eq:varthreshold}
\end{equation}
while treating $w \sim \mathcal{O}(v^2)$.
Given the asymmetric definition of the two variables, we expand all master integrals to up to $45$ orders in $w$, which allows us to get a series expansion for the amplitude correct up to at least order $\mathcal{O}(w^{39})$ and $\mathcal{O}(w^{51/2} v^{53/2})$, due to the coefficients multiplying the special functions in the amplitudes. 
This behavior can be easily understood from the fact that all the amplitudes have at most a simple pole in $t$. As a consequence, each power of $v$ needs to come together with sufficiently many powers of $w$ to exclude higher-power divergences in the limit $t \to 0$, see~\cref{eq:varthreshold}.
This also implies that we can rewrite the expansions for the amplitudes in terms of our original variables from~\cref{eq:oldTexpVars}, which allows us to get good convergence also in the region $t/(s-4 \mt^2) \to 0$. In fact, while the additional singularity at $u=-4\mt^2$ (which requires the introduction of the blow-up variables in~\cref{eq:varthreshold}) is relevant to individual master integrals, one can easily see from the analytic expression of the amplitudes in terms of iterated integrals, that the associated letter drops from all amplitudes to $\mathcal{O}(\epsilon^0)$.

\subsubsection*{Small-mass expansion}
In addition to the two expansion points considered above, in this paper we have also derived the small-mass (or equivalently high-energy) expansion for our amplitudes, i.e., $\mt^2 \to 0$.
As it is well known, in this limit these amplitudes are expected to be expressed in terms of Harmonic Polylogarithms~\cite{Remiddi:1999ew} only (for similar results see~\cite{Davies:2023vmj, Delto:2023kqv, Lee:2024dbp}). 
While the small-mass expansion is typically considered as a very good approximation, we will see below that, depending on the helicity amplitudes and color factors considered, the performance of this expansion can change drastically: in some (simple) cases, it indeed allows to cover the full phase space once matched to the large-mass and threshold expansions. In most non-trivial cases, though, the convergence can become very poor, such that it correctly reproduces the amplitude only for extremely large values of $s,t$.

Compared to the large-mass and threshold expansions, the small-mass expansion does not require any variable redefinition to resolve non-normal crossing singular lines. 
We therefore use the rather standard variables
\begin{align}
    y= \frac{\mt^2}{-t}\,, \qquad x = \frac{-t}{s} \, ,
\end{align}
and perform an expansion in $y$ up to $\mathcal{O}(y^{30})$.
Given the fact that all elliptic curves degenerate at this point, we decided to keep the dependence on $x$ exact, i.e., we perform effectively a single variable expansion in $y$ and express the coefficients in terms of Harmonic Polylogarithms of $x$. All boundary values are rational linear combinations of uniform weight of $\{ 1,\pi,\zeta(3) \}$ and suitable products thereof.

\subsubsection*{Numerical results}
\begin{figure}[htbp!]
     \centering
     \subfloat[Real part of the coefficient of $\mathcal{C}_3$ in the $\mathcal{O}(\epsilon^0)$ part of $\boldsymbol{f}_{gg}^{--++}(s,t,\mt^2)$]{\includegraphics[width=0.41\textwidth]{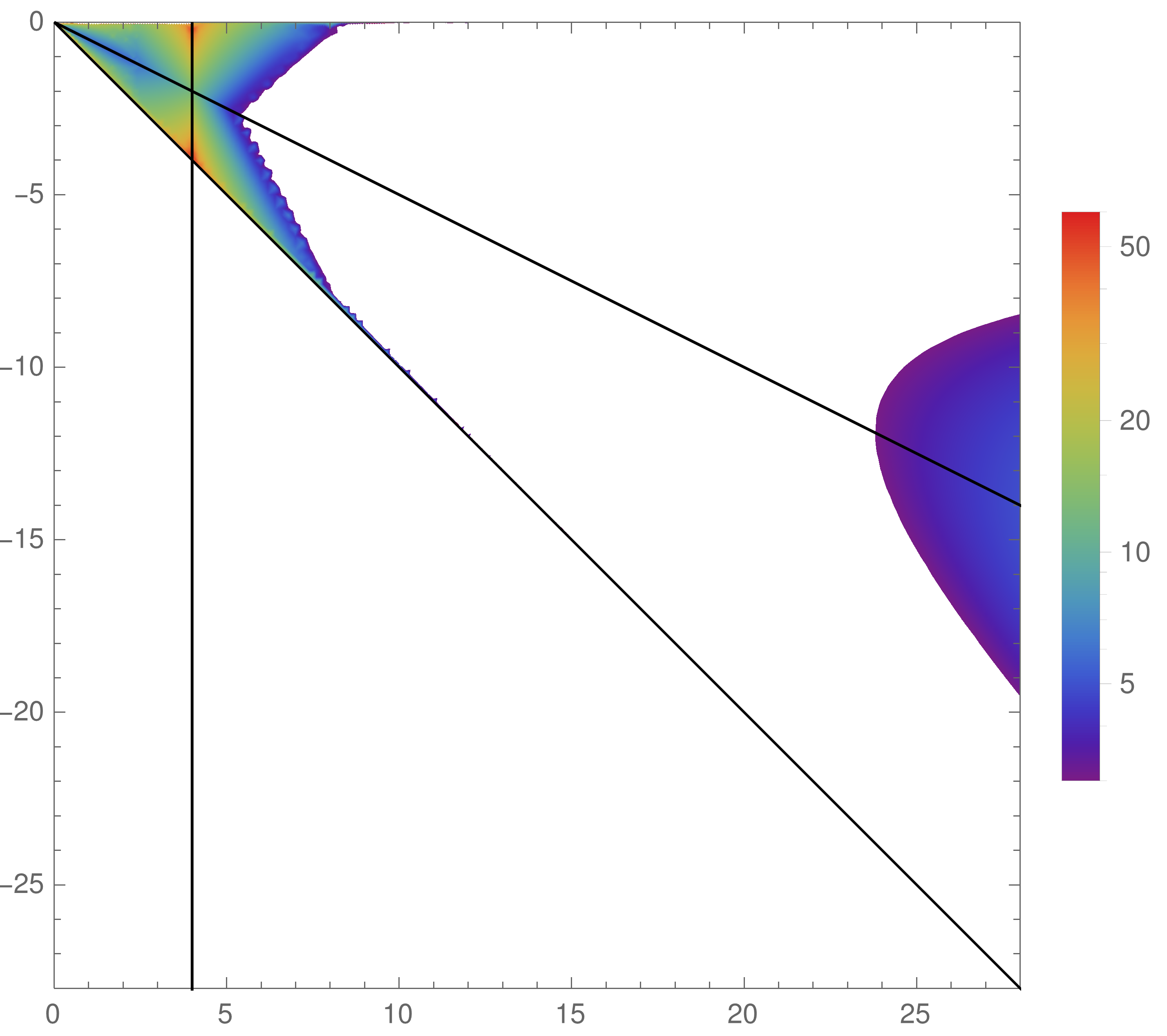}\label{subfig1}}
     \hspace{0.03\textwidth}
     \subfloat[Real part of the $\mathcal{O}(\epsilon^0)$ part of $f_{g\gamma}^{+--+}(s,t,\mt^2)$]{\includegraphics[width=0.41\textwidth]{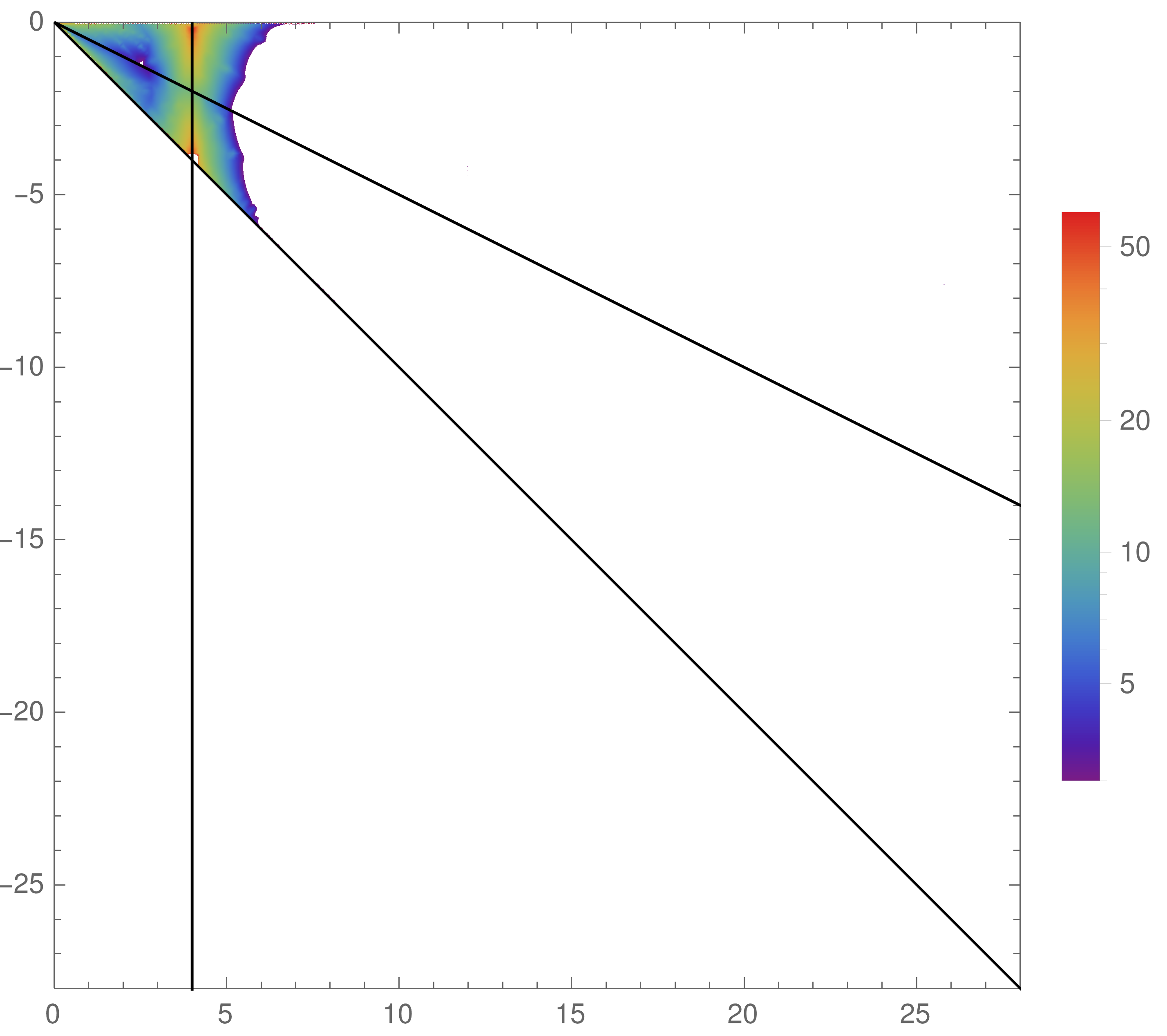}\label{subfig2}} 
     \\
    
    \subfloat[Real part of the coefficient of $\mathcal{C}_3$ of the $\mathcal{O}(\epsilon^0)$ part of  $\boldsymbol{\alpha}_{gg}(s,t,\mt^2)$]{\includegraphics[width=0.41\textwidth]{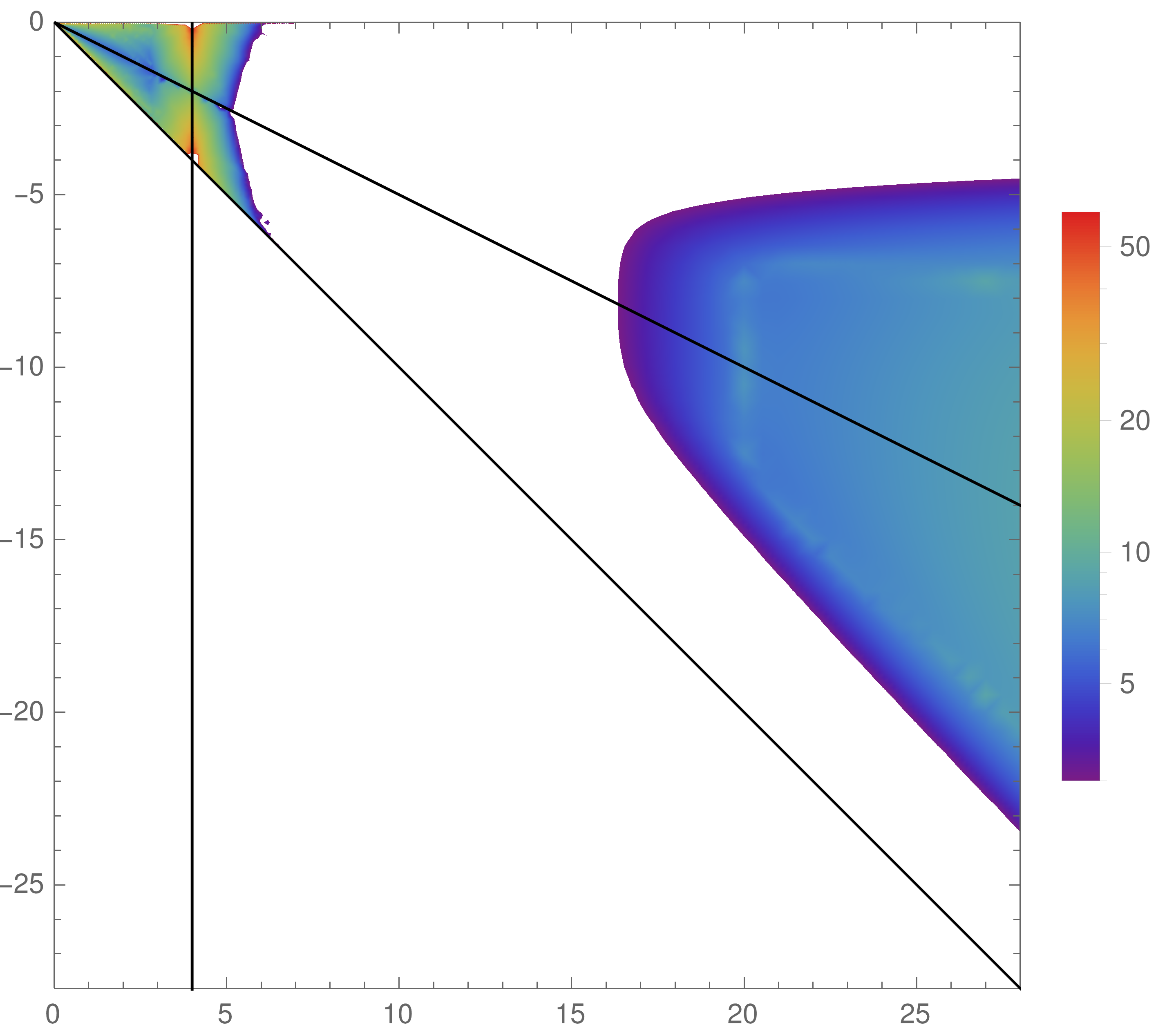}\label{subfig3}}
     \caption{Some relative precision estimates for the series expansions of the amplitudes, picking the best performing expansion in every phase-space point. The horizontal axis corresponds to $s/\mt^2$, the vertical axis to $t/\mt^2$, and the physical region is given by $0<-t<s$. To estimate the error, we calculate a suitable difference of partial sums $\Delta$, which is chosen based on comparisons in a few phase-space points to numerical evaluations using the method of auxiliary mass flow to determine the master integrals. We thereby rely on the \textsc{Mathematica} package \texttt{AMFlow}~\cite{Liu:2017jxz,Liu:2022chg}. This difference of partial sums $\Delta$ is then divided by the result for the full expansion $\Omega$. To turn this into an estimate for the number of digits of relative precision, we plot $-\log_{10}{(\vert\Delta/\Omega\vert})$. Note that the logarithmic colour scale starts at $3$ and ends at $60$. White spots bordering red colour, like in the proximity of the expansion point $(s,u)=(4\mt^2,0)$ in the plot to the right, signify more than $60$ digits of estimated relative precision, while white spots/regions bordering purple colour amount to an estimated less than $3$ digits of relative precision.}
    \label{fig:plotsnoacceleration}
 \end{figure}

An interesting question at this point is, how far do these expansions converge, and so how much of the physical phase space can we cover by using all of them together?
To investigate this question, we studied for each helicity configuration and color factor in each partonic subchannel how the partial sums of the various series representations behave. For those cases where the analysis of the partial sums indicates that two (or in principle more) series converge in overlapping regions, we have then compared the different expressions to check that the various expansions agree. Following~\cite{Becchetti:2025rrz}, this analysis can be most easily described using heat plots which indicate the number of digits of relative precision computed from the corresponding partial sums, see~\cref{fig:plotsnoacceleration}.

Similar to our findings for $\gamma \gamma$ production in~\cite{Becchetti:2025rrz}, we observe that also for $jj$ and $\gamma j$ production, the large-mass and threshold expansion always overlap at least slightly in the region $s \leq 4\mt^2$, yielding agreement of the leading digit at a minimum, often better. For values $s > 4\mt^2$, all the amplitudes are reproduced by the threshold expansions to a few digits up to values around $s \lesssim 7\mt^2$.

On the other hand, we notice that the small-mass expansion has a much less obvious pattern of convergence, as demonstrated in~\cref{fig:plotsnoacceleration}, where the large white regions indicate no convergence of any of the series up to the order considered.
However, we observe that the small mass expansion converges in relatively large patches of the phase space for simple color factors and helicity amplitudes. On the other hand, when we consider the most complex color factors in the less symmetric helicity configurations, we observe no convergence whatsoever for values of $s < 28\mt^2$, see for example the plot to the upper-right in~\cref{fig:plotsnoacceleration}.

This indicates that these series expansions are definitely not enough to cover the whole phase space, especially for medium-high values of the energy $7\mt^2 < s < 28\mt^2$.
A possible strategy to address this problem is to produce extra series expansions at different singular or non-singular points. Having at disposal differential equations in canonical form, producing such expansions is very simple.
In the rest of this section, we investigate instead a more interesting possibility, namely that of using accelerating techniques to improve the convergence of our series even beyond their naive radius of convergence.

\subsection{Acceleration techniques}
The theory behind the acceleration of the convergence of generalized series expansions has a long history. In particular, various techniques have been employed with different levels of success in the study of Feynman integrals and Feynman amplitudes. Among the most notable examples are Padé approximants, Shanks transformations, and the so-called \emph{Bernoulli-like variables}. While the literature on applying  Shanks transformations in the context of Feynman integrals is, to the best of our knowledge, relatively scarce, Padé approximants have been used thoroughly in many non-trivial realistic examples of physically relevant scattering amplitudes, see for example~\cite{Davies:2019roy, Davies:2023vmj}.
On the other hand, Bernoulli-like variables have demonstrated to be particularly effective in the numerical evaluation of multiple polylogarithms (for recent applications see~\cite{Gehrmann:2001pz, Gehrmann:2001jv, Vollinga:2004sn}), and also of functions of elliptic~\cite{Pozzorini:2005ff, Caffo:2008aw} type and even beyond~\cite{Duhr:2024bzt, Forner:2024ojj}, but their applicability for the evaluation of entire amplitudes has not been studied thoroughly. We stress that most of the literature on the use of Bernoulli-like variables is focused on one-variable problems, with the notable exception of~\cite{Gehrmann:2001jv}.

In this paper, we have focused mostly on investigating how the use of Bernoulli-like variables improves the convergence properties of the threshold and small-mass expansions, which would otherwise not overlap, see~\cref{fig:plotsnoacceleration}. While it was not possible to find a set of transformations that allows us to cover the full phase-space starting only from the three expansions described above, the result of our analysis is that an appropriate use of Bernoulli-like changes of variables can not only accelerate the convergence of our series within the original radii of convergence significantly, but, in most cases, also expand the regions where our series converge. For certain helicity amplitudes and color factors, this allowed us to fully map the physically relevant phase space without introducing another expansion. This analysis was supplemented by applying Shanks transformations to the resulting threshold expansions, yielding in some cases a further increase in the speed of convergence, depending on the specific color factor and helicity configuration. However, most of the improvement in the convergence seems to be provided by Bernoulli-like variables. 

\subsection{Bernoulli-like variables}
In the following, we provide more details on the definitions of the Bernoulli-like variables we used and discuss the improvements on the convergence of the relevant series.

Let us start considering the threshold expansion. Being a two-variable problem, we follow a strategy similar to that of~\cite{Gehrmann:2001jv} while, of course, in our case the singularity structure is less obvious than for individual multiple polylogarithms, see~\cref{fig:SingPlot}.
Following~\cite{Gehrmann:2001jv}, we proceed with iterated changes of variables. Focusing first on the behavior in $v$ for fixed $w$, we start with substituting $v$ by
\begin{equation}
    z = - \log{\left(1-\frac{v w}{4} \right)} \, ,
\end{equation}
that has, loosely speaking, the effect of pushing the singular line $s=0$ to infinite distance from the expansion point. As a second transformation for $w$, we find that there are two different possible changes of variables, which correspond to moving different singular lines. The first option corresponds to pushing away the next physical singularity of the scattering amplitude, $u=0$, and is defined as follows
\begin{equation}
    z_u = - \log{\left(1-\frac{w}{4 e^{-z}} \right)} \, .
    \label{eq:BernoulliVariable1}
\end{equation}
On the one hand, this has the effect of improving the convergence for small values of $-t$ to larger values of $s$, i.e., it makes it possible to evaluate our series for $s\gg 4\mt^2$ as long as $-t$ is small enough. On the other hand, the convergence gets worse for larger values of $-t$ and $s \approx 4 \mt^2$ and we observe that the series does not converge at all anymore if $s+2t<0$.

The other choice of Bernoulli-like variable corresponds to moving the singular line $s(t-\mt^2)^2 -4t^2 \mt^2 = 0$ to infinity, which is the closest \emph{pseudo-threshold} to the expansion point, see the red-coloured line in~\cref{fig:SingPlot} (the line coming from the right above the horizontal axis, turning at the origin and continuing downwards, initially close to the $u=0$ line and then diverging from it more and more). In this case, the relevant Bernoulli variable is given by
\begin{equation}
    z_p = \log{\left(1+(1+e^{z/2})w\right)} \, .
    \label{eq:BernoulliVariable2}
\end{equation}
Also for this choice of variable, we find that the convergence improves for small enough values of $-t$ to larger values of $s\gg 4\mt^2$, very similarly to the choice of variable from~\cref{eq:BernoulliVariable1}. However, we observe that it also improves the convergence for larger values of $-t$, making this choice of variable overall superior to~\cref{eq:BernoulliVariable1}.

Let us now briefly also discuss the small-mass expansion. Also in this case, we can introduce a Bernoulli-like variable. Since the expansion here is effectively a single-variable expansion in the variable $y$, it is enough to consider only a single Bernoulli-like transformation. Interestingly, we find that the best performances are obtained using
\begin{equation}
    z_B = - \log(1+16xy) \,,
\end{equation}
which is not what one would expect from a naive analysis of the position of the singular lines.
Nevertheless, we find that this variable can substantially improve the convergence properties of the small mass expansion.

\subsection{Shanks transformation}
Given a sequence $(A_n)_{n \in \mathbb{N}}$ (corresponding in our case to the sequence of partial sums), its Shanks transform is defined as~\cite{Shanks}
\begin{equation}
    S(A_n) = \frac{A_{n+1}A_{n-1}-A_n^2}{A_{n+1}+A_{n-1}-2A_n} \, .
    \label{eq:defShanksTransform}
\end{equation}
This transformation eliminates the largest \emph{transient}, slowing down the convergence of the series, i.e., the contribution behaving as $q^n$ with maximal $\vert q \vert$. If this transient is what is primarily responsible for the speed of convergence of the sequence, a much faster converging series with the same limit is obtained by considering its Shanks transform instead. This is guaranteed by the fact that if $A_n=A+b \, q^n + B_n$ with $b \in \mathbb{C}$, $A \in \mathbb{C}$ the actual limit of the sequence, $\vert q \vert <1$ (or the sequence would not be convergent), and $B_n \xrightarrow{n\rightarrow \infty}0$ faster than $q^n$, then 
\begin{equation}
    S(A_n) = A + \frac{B_{n+1}-2q B_n +q^2 B_{n-1}}{(1-q)^2} + \mathcal{O}\left( \left(\frac{B_m}{q^m} \right)^2\right) \, , 
\end{equation}
which will converge to $A$ faster than $(A_n)_{n \in \mathbb{N}}$ for large enough $n$. 

However, a caveat to Shanks transformations is that if the sequence $(A_n)_{n \in \mathbb{N}}$ already converges relatively fast, the denominator of~\cref{eq:defShanksTransform} is close to zero, which can cause numerical instabilities and requires working with high-digit precision.

In the context of this work, we found it useful to apply Shanks transformations to the threshold expansions in the Bernoulli-like variables, yielding in some cases a significant increase in the number of digits of precision obtained within its region of convergence.

\subsection{Numerical performances}
\begin{figure}[htbp!]
    \centering
    \subfloat[Real part of the coefficient of $\mathcal{C}_3$ of the $\mathcal{O}(\epsilon^0)$ part of $\boldsymbol{f}_{gg}^{--++}(s,t,\mt^2)$]{\includegraphics[width=0.41\textwidth]{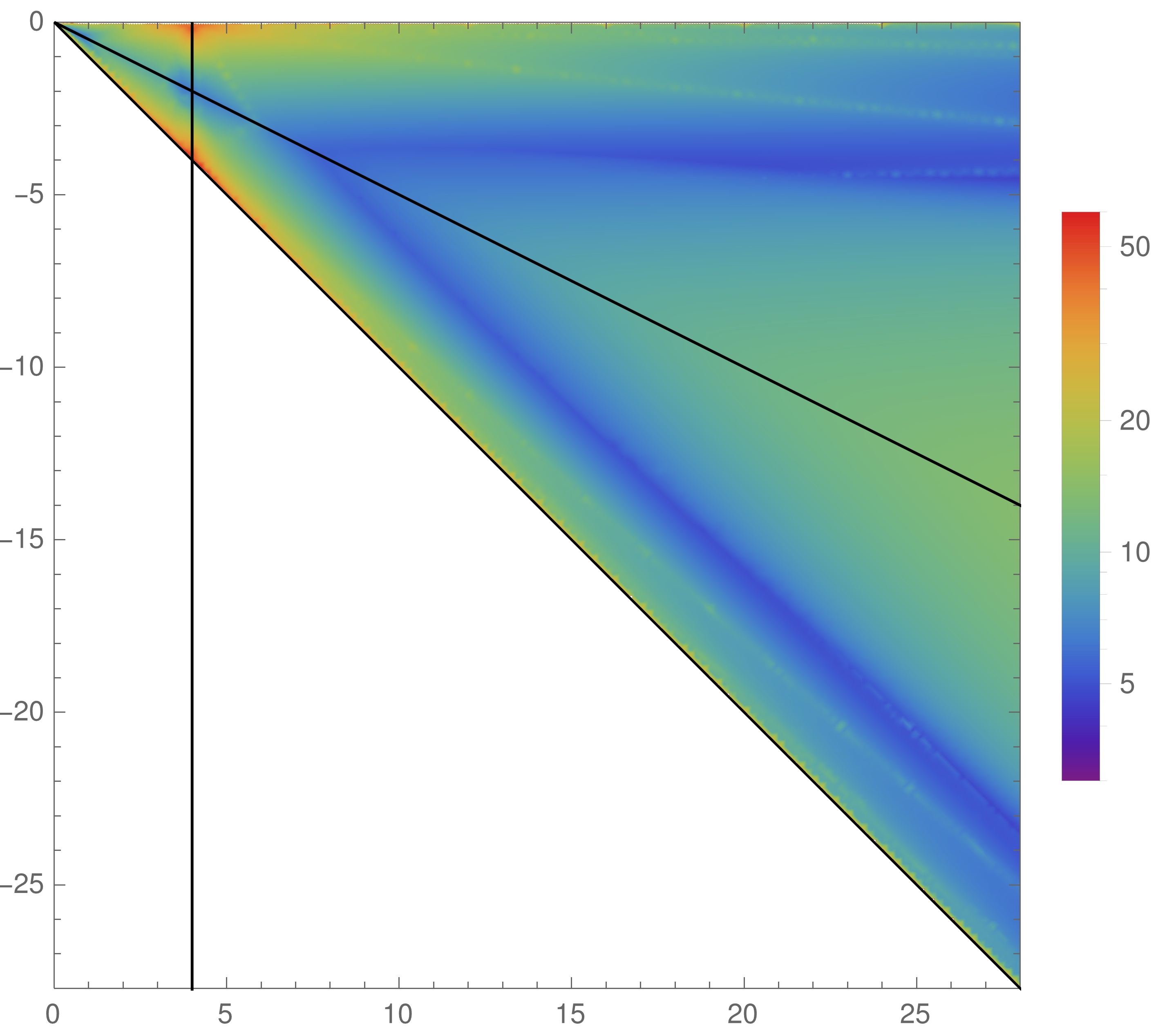}\label{subfig1b}} 
    \hspace{0.03\textwidth}
    \subfloat[Real part of the $\mathcal{O}(\epsilon^0)$ part of  $f_{g\gamma}^{+--+}(s,t,\mt^2)$]{\includegraphics[width=0.41\textwidth]{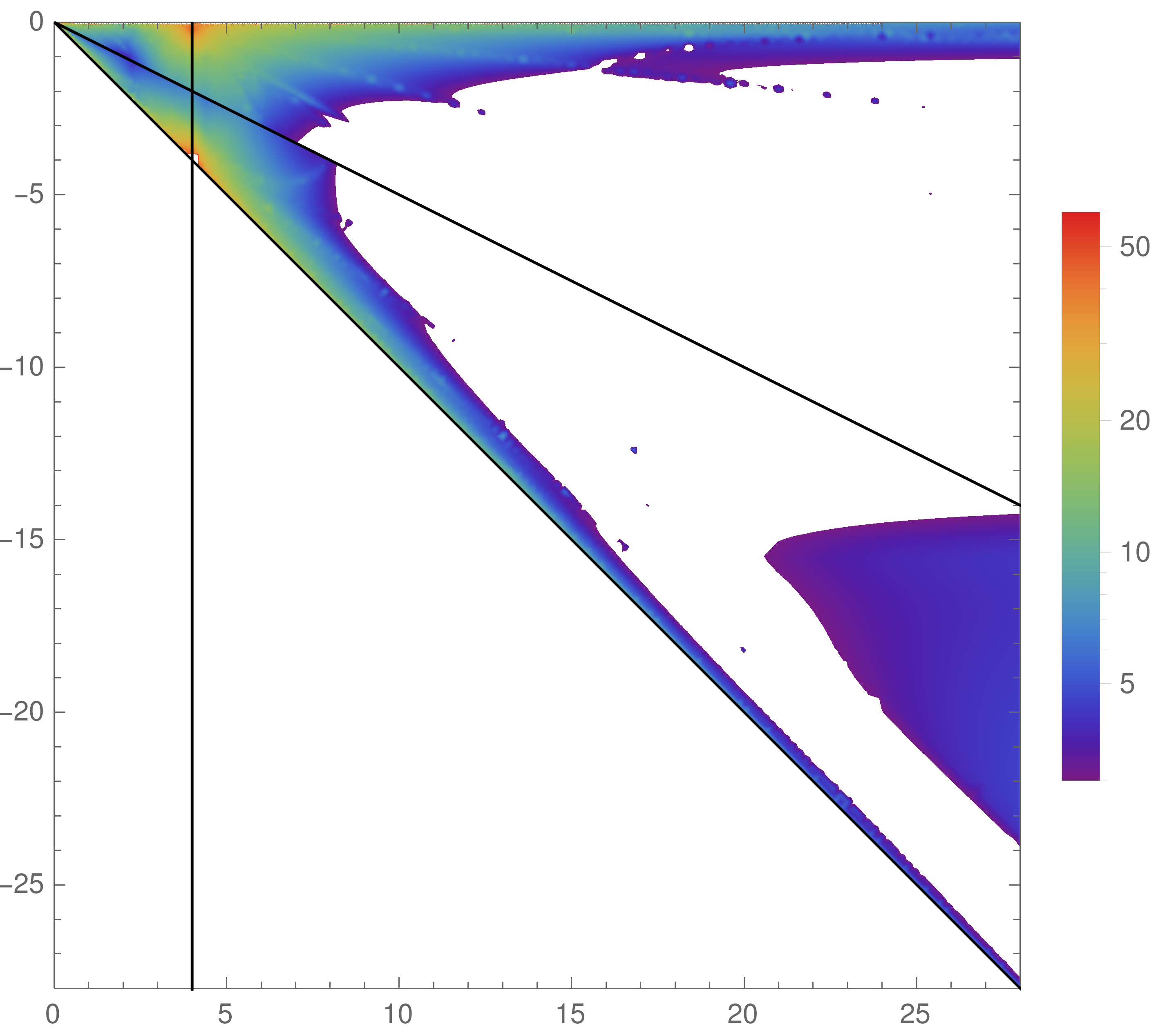}\label{subfig2b}}
    \\
    
    \subfloat[Real part of the coefficient of $\mathcal{C}_3$ of the $\mathcal{O}(\epsilon^0)$ part of  $\boldsymbol{\alpha}_{gg}(s,t,\mt^2)$]{\includegraphics[width=0.41\textwidth]{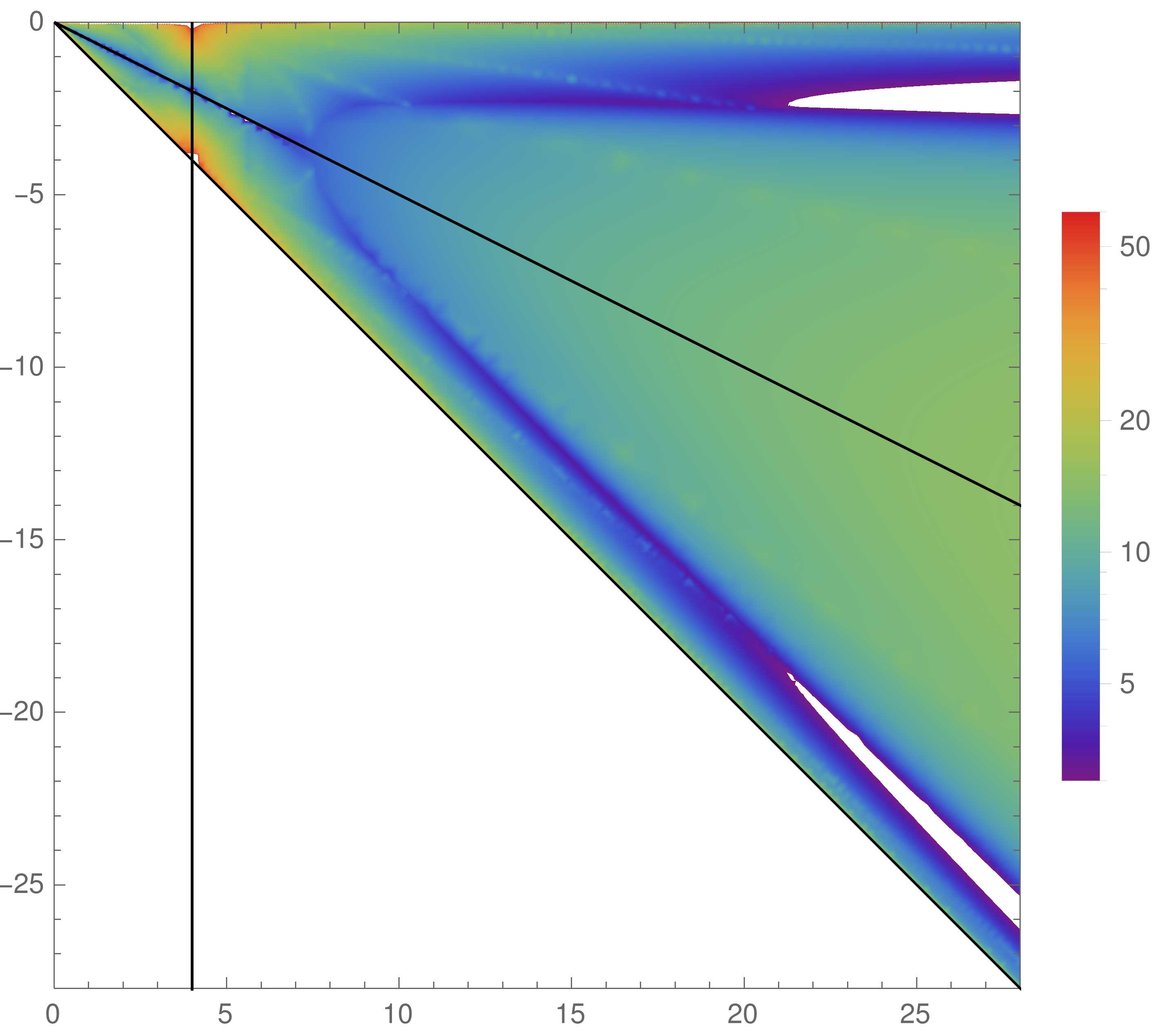}\label{subfig3b}} 
    \hspace{0.03\textwidth}
    \subfloat[Same partial amplitude as in~\cref{subfig1b}, but \emph{without} the use of Shanks transformation]{\includegraphics[width=0.41\textwidth]{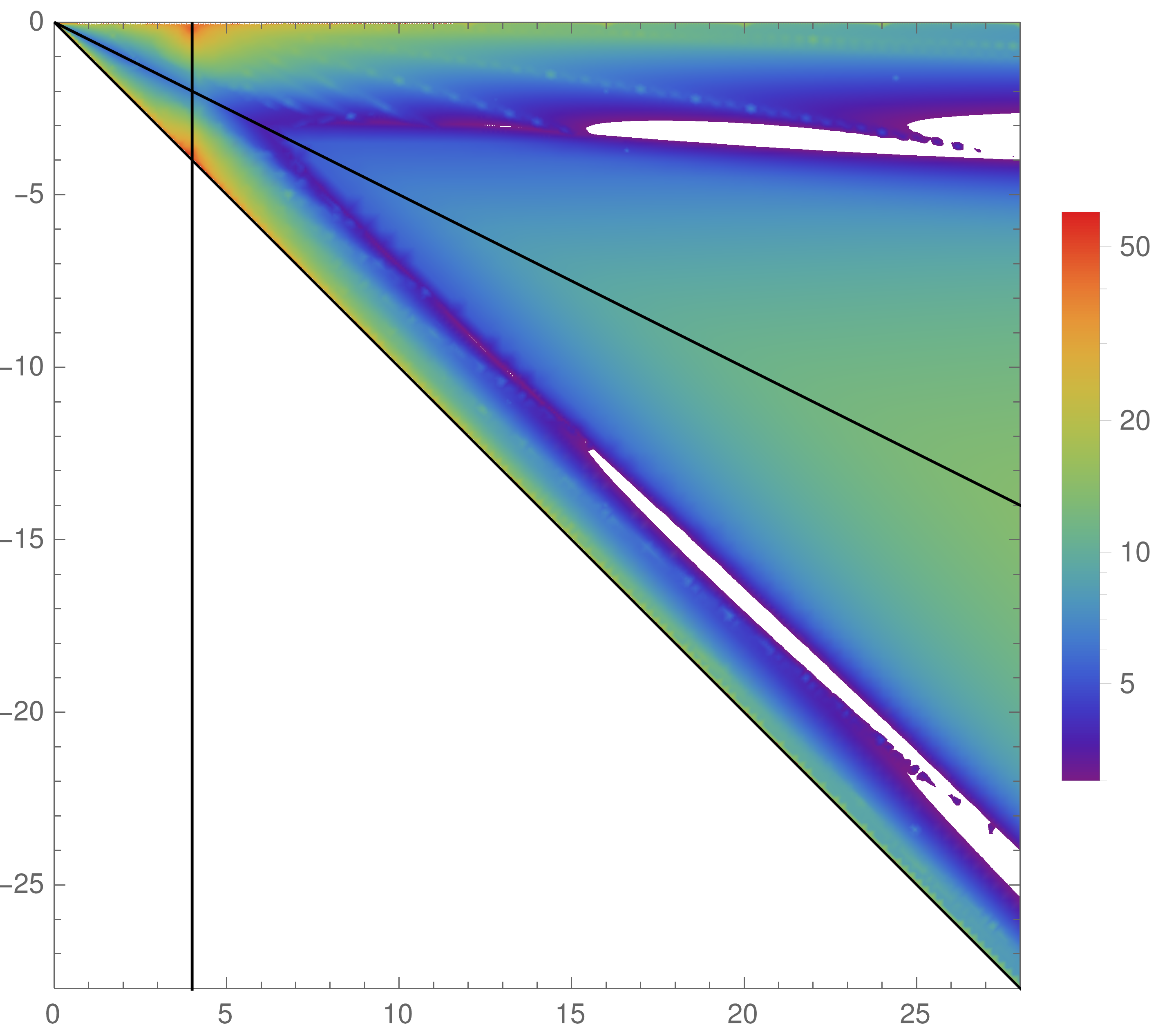}\label{subfig4b}} 
    \\
    
    \caption{Some relative precision estimates for the series expansions of the amplitudes after implementation of acceleration strategies (Bernoulli variables and Shanks transformation). The plots are obtained as described in the caption of~\cref{fig:plotsnoacceleration}.}
    \label{fig:plotsrelerror}
\end{figure}

Combining the techniques of Bernoulli-like variables and Shanks transformation, we can cover much larger regions of the phase space. As an exemplification, in~\cref{fig:plotsrelerror} we show updated versions of the heat plots from~\cref{fig:plotsnoacceleration}, which covered the three scenarios we typically encounter for all our partial amplitudes.

The upper-left plot $(a)$ provides a good example of how the Bernoulli-like variables substantially extend the convergence, especially of the threshold expansion, compared to the same plot in~\cref{fig:plotsnoacceleration}. Comparing to the lower-right plot $(d)$ shows how the usage of a Shanks transform can help to fill the remaining white strips left after using Bernoulli-like variables, such that both techniques combined allow us to cover the full phase-space up to values of $s \sim 28 \mt^2$.

The other two plots $(b)$ and $(c)$ exemplify instead two cases where Bernoulli and Shanks transforms, even combined, do not allow covering the entire region shown. The upper-right one $(b)$, in particular,  corresponds to a two plus and two minus helicity amplitude for $gg \to g\gamma$ for which all three crossings of the elliptic curve contribute to the same (and only) partial amplitude. We stress that this is not the case for any of the di-jet helicity amplitudes, where, thanks to the fact that all four particles are colored, each partial amplitude receives at most a contribution by two crossings of the same elliptic curve, see~\cref{fig:ellcurves}. 
In this case, one can see that the small-mass expansion converges particularly badly, providing a reliable result only for values of $s \gtrsim 20\mt^2$ and for very limited values of $t$. 
As is easy to see, the use of a Bernoulli-like variable at threshold nevertheless still allows us to substantially extend the convergence of the threshold expansion for very large values of $s$ and small values of $-t$ and $-u$, but we are still left with a very large white area which we cannot cover for $s \gtrsim 7\mt^2$.
Finally, the lower-left plot $(c)$ provides an example of an intermediate case, where Bernoulli-like variables and the Shanks transformation are used together to cover almost the full phase space. At variance with $(b)$, some white stripes remain only for large values of $s$ and intermediate values of $t$, where neither the small-mass nor the threshold expansion converges to three leading digits of precision up to the order to which we have derived the respective expansions. While filling the white region in plot $(b)$ would be most easily achieved by adding a further expansion, in the case of $(c)$, it would be likely enough to calculate a few more orders in the expansions already derived.

%% file: conclusions.tex
\section{Conclusions}
\label{sec:conc}
In this paper, we have considered the analytic calculation of all partonic subchannels relevant for the two-loop QCD corrections to the production of two jets or a photon and a jet in hadron collisions. In variance with results already available in the literature, we have kept full dependence on the mass of one heavy quark circulating in the loops. Allowing massive particles in the loops gives rise to two-loop integrals that are related to an elliptic curve, which we have computed leveraging the same techniques introduced by some of us in previous publications. 
In particular, we employed a generalization of the well-known integrand analysis typically performed to identify canonical polylogarithmic integrals to select a precanonical basis of elliptic Feynman integrals that can be matched one-to-one onto the differential forms of first, second, and third kind defined on the corresponding elliptic curve. Starting from this pre-canonical basis, we used the algorithm introduced in~\cite{Gorges:2023zgv} to rotate our basis to a fully $\epsilon$-factorized basis, which we conjecture to be the natural generalization of a canonical basis beyond the polylogarithmic case. In fact, our basis satisfies three fundamental requirements:
\begin{enumerate}
    \item All differential forms appearing in our differential equations are independent up to total derivatives.

    \item All differential forms have at most single poles.

    \item Close to all regular singular points, which are cusps of the corresponding elliptic curve (points where the elliptic curve degenerates to a sphere), our differential equations can be shown to reduce to standard canonical differential equations in $\rm{d}$$\log$-form.
\end{enumerate}

Having differential equations in this form allows us to obtain analytic results in terms of an independent set of iterated integrals, facilitating the study of their analytic structure. In particular, we can use this representation to showcase analytic pole cancellation after UV renormalization and IR subtraction and to demonstrate the cancellation of a large number of elliptic differential forms, following a pattern already observed in other calculations. Moreover, we can use our differential equations to efficiently generate generalized series expansions close to arbitrary regular and singular points. Specifically, in this case, we produce three independent series expansions: the first two corresponding to the large mass and threshold expansion, similarly to what was done in~\cite{Becchetti:2025rrz}. In addition, in this paper, we also consider the small mass expansion, which, together with the two expansions above, allows us to cover a larger region of the physical phase space. To improve the convergence of these three expansions, we introduce suitable Bernoulli-like variables and also investigate the use of Shanks transformations on the corresponding series. These allow us to extend the convergence of the series and, for some helicity amplitudes, to cover the entire two-dimensional phase space. For other configurations, especially those involving all elliptic curves at once, these three expansions are not enough to cover the full phase space, and new expansions produced for larger values of $s$ would be necessary. Depending on the region of phase space one is interested in for phenomenological applications, producing such extra expansions, starting from our differential equations, is entirely straightforward.

%% file: appendices.tex
\section{Helicity coefficients in terms of the form factors}
\label{app:helcoeffs}
In this appendix, we give the relations between the helicity coefficients defined in eqs.~\eqref{eq:helampqqQQ}--\eqref{eq:helampgg} and the form factors (eqs.~\eqref{eq:ampqqga}--\eqref{eq:ampgggg}). For four-quark scattering, we have
\begin{equation}
\label{eq:etaqqQQ}
    \boldsymbol{\eta}_1 (s,t,\mt^2) = 2t \mathbfcal{H}_1 - t u \mathbfcal{H}_2 \, , \qquad \boldsymbol{\eta}_2 (s,t,\mt^2) = 2u \mathbfcal{H}_1 + t u \mathbfcal{H}_2\, .
\end{equation}
For the scattering of a quark-antiquark pair and two gauge bosons, we have
\begin{equation}
\begin{aligned}
\boldsymbol{\alpha}_X(s,t,\mt^2) &=  \frac{t}{2}\,\left( \mathbfcal{X}_2 - \frac{t}{2} \mathbfcal{X}_3 +  \mathbfcal{X}_4 \right) ,
&& \boldsymbol{\beta}_X(s,t,\mt^2) = \frac{t}{2}\, \left( \frac{s}{2} \mathbfcal{X}_3  +  \mathbfcal{X}_4 \right) ,
\\
\boldsymbol{\gamma}_X(s,t,\mt^2) &=  \frac{s\,t}{2u}\left( \mathbfcal{X}_2 - \mathbfcal{X}_1  - \frac{t}{2} \mathbfcal{X}_3 - \frac{t}{s}\mathbfcal{X}_4 \right)\,,
&&\boldsymbol{\delta}_X(s,t,\mt^2) =\frac{t}{2}\left( \mathbfcal{X}_1 + \frac{t}{2} \mathbfcal{X}_3 - \mathbfcal{X}_4 \right)\,,
\label{eq:alphabetaqq}
\end{aligned}
\end{equation}
with $X\in \{g\gamma,gg \}$ and $\mathbfcal{X}\rightarrow\mathbfcal{F},\mathbfcal{K}$, while for four-gauge boson scattering with $\mathbfcal{Y}\rightarrow\mathbfcal{G},\mathbfcal{J}$, we obtain
\begin{equation}
\begin{aligned}
 \boldsymbol{f}_{X}^{++++}(s,t,\mt^2) &=  \frac{t^2}{4}\left(\frac{2\mathbfcal{Y}_{6}}{u}-\frac{2\mathbfcal{Y}_{3}}{s}-\mathbfcal{Y}_{1}\right)+\left(\frac{s}{u}+\frac{u}{s}+4\right)\mathbfcal{Y}_{8}+\frac{t}{2}(\mathbfcal{Y}_{2}-\mathbfcal{Y}_{4}+\mathbfcal{Y}_{5}-\mathbfcal{Y}_{7})\,, \\ 
 \boldsymbol{f}_{X}^{-+++}(s,t,\mt^2) &=  \,\,\,\, \frac{t^2}{4}\left(\frac{2\mathbfcal{Y}_{3}}{s}+\mathbfcal{Y}_{1}\right)+t\left(\frac{\mathbfcal{Y}_{8}}{s}+\frac{1}{2}(\mathbfcal{Y}_{4}+\mathbfcal{Y}_{6}-\mathbfcal{Y}_{2})\right)\,, \\ 
 \boldsymbol{f}_{X}^{+-++}(s,t,\mt^2) &=  -\frac{t^2}{4}\left(\frac{2\mathbfcal{Y}_{6}}{u}-\mathbfcal{Y}_{1}\right)+t\left(\frac{\mathbfcal{Y}_{8}}{u}-\frac{1}{2}(\mathbfcal{Y}_{2}+\mathbfcal{Y}_{3}+\mathbfcal{Y}_{5})\right)\,, \\ 
 \boldsymbol{f}_{X}^{++-+}(s,t,\mt^2) &= \,\,\,\, \frac{t^2}{4}\left(\frac{2\mathbfcal{Y}_{3}}{s}+\mathbfcal{Y}_{1}\right)+t\left(\frac{\mathbfcal{Y}_{8}}{s}+\frac{1}{2}(\mathbfcal{Y}_{6}+\mathbfcal{Y}_{7}-\mathbfcal{Y}_{5})\right)\,, \\ 
 \boldsymbol{f}_{X}^{+++-}(s,t,\mt^2) &=  -\frac{t^2}{4}\left(\frac{2\mathbfcal{Y}_{6}}{u}-\mathbfcal{Y}_{1}\right)+t\left(\frac{\mathbfcal{Y}_{8}}{u}+\frac{1}{2}(\mathbfcal{Y}_{4}+\mathbfcal{Y}_{7}-\mathbfcal{Y}_{3})\right)\,, \\ 
 \boldsymbol{f}_{X}^{--++}(s,t,\mt^2) &=  -\frac{t^2}{4}\mathbfcal{Y}_{1}+\frac{1}{2}t(\mathbfcal{Y}_{2}+\mathbfcal{Y}_{3}-\mathbfcal{Y}_{6}-\mathbfcal{Y}_{7})+2\mathbfcal{Y}_{8}\,, \\ 
 \boldsymbol{f}_{X}^{-+-+}(s,t,\mt^2) &=  t^2\left(\frac{\mathbfcal{Y}_{8}}{su}-\frac{\mathbfcal{Y}_{3}}{2s}+\frac{\mathbfcal{Y}_{6}}{2u}-\frac{\mathbfcal{Y}_{1}}{4}\right)\,, \\ 
 \boldsymbol{f}_{X}^{+--+}(s,t,\mt^2) &=  -\frac{t^2}{4}\mathbfcal{Y}_{1}+\frac{1}{2}t(\mathbfcal{Y}_{3}-\mathbfcal{Y}_{4}+\mathbfcal{Y}_{5}-\mathbfcal{Y}_{6})+2\mathbfcal{Y}_{8}
\,.
\label{eq:alphabetagg}
\end{aligned}
\end{equation}

\section{Helicity Projectors}
\label{app:helproj}
In this appendix, we give explicit expressions for the helicity projectors that single out directly the individual helicity amplitudes in~\cref{eq:helampqqQQ,eq:helampqq,eq:helampgg} when applied to the full amplitude.
We denote the projectors for the helicity amplitudes involving a quark-antiquark pair and two gauge bosons by $\mathcal{P}_{\omega}$ with $\omega=\{\alpha,\beta,\gamma,\delta\}$, i.e.~we have for  $X\in \{g\gamma,gg \}$,
\begin{equation}
    \mathcal{P}_{\omega} \cdot \mathbfcal{A}_{q\bar{q}X} = \boldsymbol{\omega}(s,t,\mt^2)\,.
\end{equation}
Similarly, $\mathcal{P}_{\lambda_1\lambda_2\lambda_3\lambda_4}$ denotes the helicity projectors for the helicity amplitudes involving four gauge bosons, 
\begin{equation}
\mathcal{P}_{\lambda_1\lambda_2\lambda_3\lambda_4} \cdot \mathbfcal{A}_{ggX} = \boldsymbol{f}_{\lambda_1\lambda_2\lambda_3\lambda_4}(s,t, \mt^2)\,, 
\end{equation}
and $\mathcal{P}_{\eta}$ with $\eta=\{\eta_1,\eta_2\}$ the projectors on the four-quark helicity amplitudes
\begin{equation}
    \mathcal{P}_{\eta} \cdot \mathbfcal{A}_{q\bar{q}\bar{Q}Q} = \boldsymbol{\eta}(s,t,\mt^2)\, .
\end{equation}
Decomposed onto the basis of tensors, the projectors can be written as
\begin{align}
    \mathcal{P}_{\alpha}& = \frac{1}{(d-3)} \left[ \frac{u}{8 s^2} \left( \tau_1^{\dag} + \tau_2^{\dag}\right) + \frac{2t + (4-d)u}{8s^2t} \tau_3^{\dag} + \frac{t^2 + t u}{8s^2t} \tau_4^{\dag} \right]\, , \\
    \mathcal{P}_{\beta}& = \frac{1}{(d-3)} \left[ \frac{u}{8st} \left(\tau_2^{\dag} - \tau_1^{\dag}\right) + \frac{2t + du}{8s^2t} \tau_3^{\dag} - \frac{1}{8t} \tau_4^{\dag}\right]\, , \\
    \mathcal{P}_{\gamma}& = \frac{1}{(d-3)} \left[ \frac{1}{8s} \left(\tau_2^{\dag} - \tau_1^{\dag}\right) + \frac{1}{8u} \tau_4^{\dag} - \frac{2t + (d-4)u}{8stu} \tau_3^{\dag}\right]\, , \\
    \mathcal{P}_{\delta}& = \frac{1}{(d-3)} \left[ \frac{u}{8s^2} \left(\tau_1^{\dag} + \tau_2^{\dag}\right) + \frac{1}{s} \tau_4^{\dag} + \frac{(d-4)u - 2t}{8s^2t} \tau_3^{\dag}\right] \, , 
\end{align}
\begin{align}
\mathcal{P}_{++++} &= \frac{1}{3(d-1)(d-3)t^2}\biggl[
    \frac{3 \bigl((4 - d)(d - 2) s^2 + (4 - d)(d - 2) s t - 2 (d - 2) t^2\bigr)}{4 s (s+t)} \mathcal{T}_1^{\dag} \nonumber \\
    &\quad+ \frac{3 \bigl((d - 2) s^2 t + (d - 2) s t^2\bigr)}{4 s (s+t)} \mathcal{T}_2^{\dag} 
    + \frac{3\bigl(-d s^2 t - 2 s t^2\bigr)}{4 s (s+t)}  \mathcal{T}_3^{\dag} \nonumber\\
    &\quad- \frac{3 \bigl(d s^2 t + d s t^2 - 2 s^2 t - 2 s t^2\bigr)}{4 s (s+t)} \left( \mathcal{T}_4^{\dag} - \mathcal{T}_5^{\dag} + \mathcal{T}_7^{\dag}\right) \nonumber \\
    &\quad+ \frac{3 \bigl(d s^2 t + 2 d s t^2 + (d - 2) t^3 - 2 s t^2\bigr)}{4 s (s+t)} \mathcal{T}_6^{\dag}
    + \frac{3 \bigl(s^2 t^2 + s t^3\bigr)}{4 s (s+t)}\mathcal{T}_8^{\dag}
\biggr], \\[6pt]
\mathcal{P}_{-+++} &= \frac{1}{3(d-1)(d-3)t^2}\biggl[
    \frac{3 \bigl(d^2 s (s+t) - 4 d s^2 - 8 d s t - 2 d t^2 + 4 t (s+t)\bigr)}{4 s (s+t)} \mathcal{T}_1^{\dag} \nonumber \\
    &\quad- \frac{3 \bigl(d s^2 t + d s t^2\bigr)}{4 s (s+t)} \left( \mathcal{T}_2^{\dag} - \mathcal{T}_4^{\dag}\right) 
    + \frac{3 \bigl(d s^2 t + d s t^2 - 2 s t (s+t)\bigr)}{4 s (s+t)}\left( \mathcal{T}_5^{\dag} - \mathcal{T}_7^{\dag} \right) \nonumber \\
    &\quad+ \frac{3 \bigl(d s^2 t + 2 d s t^2 + d t^3 - 2 t^2 (s+t)\bigr)}{4 s (s+t)}\mathcal{T}_6^{\dag} + \frac{3 \bigl(d s^2 t - 2 s t (s+t)\bigr)}{4 s (s+t)}\mathcal{T}_3^{\dag} + \frac{3 t^2 }{4}\mathcal{T}_8^{\dag}
\biggr], \\[6pt]
\mathcal{P}_{+-++} &= \frac{1}{3(d-1)(d-3)t^2}\biggl[
    \frac{3 \bigl(d^2 s (s+t) - 4 d s^2 + 2 d t^2 - 4 s t\bigr)}{4 s (s+t)} \mathcal{T}_1^{\dag} \nonumber \\
    &\quad- \frac{3 \bigl(d s^2 t + d s t^2\bigr)}{4 s (s+t)}
    \left( \mathcal{T}_2^{\dag} + \mathcal{T}_5^{\dag} \right) + \frac{3 \bigl(-d s^2 t - 2 s t^2\bigr)}{4 s (s+t)} \mathcal{T}_3^{\dag} \nonumber \\
    &\quad+ \frac{3 \bigl(-d s^2 t - d s t^2 + 2 s^2 t + 2 s t^2\bigr)}{4 s (s+t)} \left( \mathcal{T}_4^{\dag} + \mathcal{T}_7^{\dag} \right) \nonumber \\
    &\quad+ \frac{3 \bigl(-d s^2 t - 2 d s t^2 - d t^3 + 2 s^2 t + 2 s t^2\bigr)}{4 s (s+t)} \mathcal{T}_6^{\dag} + \frac{3 t^2}{4} \mathcal{T}_8^{\dag}
\biggr], \\[6pt]
\mathcal{P}_{++-+} &= \frac{1}{3(d-1)(d-3)t^2}\biggl[
    \frac{3 \mathcal{T}_1^{\dag}\bigl(d^2 s (s+t) - 4 d s^2 - 8 d s t - 2 d t^2 + 4 t (s+t)\bigr)}{4 s (s+t)} \nonumber \\
    &\quad+ \frac{3 \bigl(d s^2 t + d s t^2 - 2 s t (s+t)\bigr)}{4 s (s+t)} \left(  \mathcal{T}_2^{\dag} -  \mathcal{T}_4^{\dag}\right) 
    + \frac{3 \bigl(d s^2 t - 2 s t (s+t)\bigr)}{4 s (s+t)} \mathcal{T}_3^{\dag} \nonumber \\
    &\quad- \frac{3 \bigl(d s^2 t + d s t^2\bigr)}{4 s (s+t)} \left( \mathcal{T}_5^{\dag} - \mathcal{T}_7^{\dag}\right) + \frac{3 t^2 }{4}\mathcal{T}_8^{\dag} \nonumber \\
    &\quad+ \frac{3 \bigl(d s^2 t + 2 d s t^2 + d t^3 - 2 t^2 (s+t)\bigr)}{4 s (s+t)}  \mathcal{T}_6^{\dag}   
\biggr], \\[6pt]
\mathcal{P}_{+++-} &= \frac{1}{3(d-1)(d-3)t^2}\biggl[
    \frac{3 \bigl(d^2 s (s+t) - 4 d s^2 + 2 d t^2 - 4 s t\bigr)}{4 s (s+t)} \mathcal{T}_1^{\dag} \nonumber \\
    &\quad+ \frac{3 \bigl(d s^2 t + d s t^2 - 2 s^2 t - 2 s t^2\bigr)}{4 s (s+t)} \left( \mathcal{T}_2^{\dag} + \mathcal{T}_5^{\dag}\right) \nonumber \\
    &\quad+ \frac{3 \bigl(-d s^2 t - 2 s t^2\bigr)}{4 s (s+t)}\mathcal{T}_3^{\dag} 
    + \frac{3 \bigl(d s^2 t + d s t^2\bigr)}{4 s (s+t)} \left( \mathcal{T}_4^{\dag} + \mathcal{T}_7^{\dag}\right) \nonumber \\
    &\quad+ \frac{3 \bigl(-d s^2 t - 2 d s t^2 - d t^3 + 2 s t (s+t)\bigr)}{4 s (s+t)}\mathcal{T}_6^{\dag} + \frac{3 t^2 }{4}\mathcal{T}_8^{\dag}
\biggr], \\[6pt]
\mathcal{P}_{--++} &= \frac{1}{3(d-1)(d-3)t^2}\biggl[
    -\frac{3 \bigl((d - 4)(d - 2) s^2 + (d - 4)(d - 2) s t - 2 d t^2\bigr)}{4 s (s+t)}\mathcal{T}_1^{\dag} \nonumber \\
    &\quad- \frac{3 \bigl(-d s^2 t + (2 - d) s t^2 + 2 s^2 t\bigr)}{4 s (s+t)} \mathcal{T}_2^{\dag} - \frac{3 \bigl(-d s^2 t + 2 s^2 t + 2 s t^2\bigr)}{4 s (s+t)}\mathcal{T}_3^{\dag} \nonumber \\
    &\quad+ \frac{3 \bigl(d s^2 t + d s t^2\bigr)}{4 s (s+t)}\left( \mathcal{T}_4^{\dag} - \mathcal{T}_5^{\dag}\right) 
    - \frac{3 \bigl(d s^2 t + d s t^2 - 2 s^2 t - 2 s t^2\bigr)}{4 s (s+t)} \mathcal{T}_7^{\dag} \nonumber \\
    &\quad- \frac{3\bigl(d s^2 t + 2 d s t^2 + d t^3 - 2 s^2 t - 2 s t^2\bigr)}{4 s (s+t)}  \mathcal{T}_6^{\dag}- \frac{3 \bigl(-s^2 t^2 - s t^3\bigr)}{4 s (s+t)} \mathcal{T}_8^{\dag}
\biggr], \\[6pt]
\mathcal{P}_{-+-+} &= \frac{1}{3(d-1)(d-3)t^2}\biggl[
    -\frac{3 \bigl(d^2 s (s+t) + 2 d s^2 + 2 d s t + 2 d t^2 - 4 t^2\bigr)}{4 s (s+t)} \mathcal{T}_1^{\dag} \nonumber \\
    &\quad- \frac{3 \bigl(d s^2 t + d s t^2\bigr)}{4 s (s+t)} \left( \mathcal{T}_2^{\dag} - \mathcal{T}_4^{\dag} + \mathcal{T}_5^{\dag} - \mathcal{T}_7^{\dag}\right)
    - \frac{3 \bigl(d s^2 t + 2 s t^2\bigr)}{4 s (s+t)} \mathcal{T}_3^{\dag} \nonumber \\
    &\quad- \frac{3 \bigl(-d s^2 t - 2 d s t^2 - d t^3 + 2 t^2 (s+t)\bigr)}{4 s (s+t)} \mathcal{T}_6^{\dag} + \frac{3 t^2 }{4}\mathcal{T}_8^{\dag} 
\biggr], \\[6pt]
\mathcal{P}_{+--+} &= \frac{1}{3(d-1)(d-3)t^2}\biggl[
    -\frac{3 \bigl((d - 4)(d - 2) s^2 + (d - 4)(d - 2) s t - 2 d t^2\bigr)}{4 s (s+t)}\mathcal{T}_1^{\dag} \nonumber \\
    &\quad- \frac{3 \bigl(d s^2 t + d s t^2\bigr)}{4 s (s+t)} 
    \left(\mathcal{T}_2^{\dag} - \mathcal{T}_7^{\dag} \right) - \frac{3 \bigl(-d s^2 t + 2 s^2 t + 2 s t^2\bigr)}{4 s (s+t)} \mathcal{T}_3^{\dag} \nonumber \\
    &\quad- \frac{3 \bigl(d s^2 t + d s t^2 - 2 s^2 t - 2 s t^2\bigr)}{4 s (s+t)} \left( \mathcal{T}_4^{\dag} - \mathcal{T}_5^{\dag}\right)\nonumber \\
    &\quad- \frac{3 \bigl(d s^2 t + 2 d s t^2 + d t^3 - 2 s^2 t - 2 s t^2\bigr)}{4 s (s+t)} \mathcal{T}_6^{\dag}
    + \frac{3\bigl(s^2 t^2 + s t^3\bigr)}{4 s (s+t)} \mathcal{T}_8^{\dag}
\biggr],
\end{align}
\begin{align}
    \mathcal{P}_{\eta_1} &= \frac{1}{(d-3)4s^2}\Pi_1^{\dagger} + \frac{t-u}{(d-3)s^2tu} \Pi_2^{\dagger}\, ,\\
    \mathcal{P}_{\eta_2} &= \frac{t-u}{(d-3)s^2tu} \Pi_1^{\dagger} + \frac{(d-4)s^2+2t^2+2u^2}{(d-3)4s^2t^2u^2}\Pi_2^{\dagger}\, ,
\end{align}
where 
\begin{equation}
    \tau_i^\dagger = \bar{u}(p_1) \Gamma_i^{\mu \nu} u(p_2) \epsilon_{3,\mu}^*(p_3) \epsilon_{4,\nu}^*(p_4)\,,
\end{equation}
\begin{equation}
\mathcal{T}_i^\dagger = T_i^{\mu \nu \rho \sigma} 
\epsilon_{1,\mu}^*(p_1) \epsilon_{2,\nu}^*(p_2)
\epsilon_{3,\rho}^*(p_3) \epsilon_{4,\sigma}^*(p_4)\,,
\end{equation}
and the $\Pi_i$, $\Gamma_i^{\mu\nu}$ and $T_i^{\mu \nu \rho \sigma}$ are provided in \cref{eq:tensqq,eq:tensgg,eq:tensqqQQ}.

\section{Integral family definitions}
\label{app:deffamilies}
\begin{figure}[h!]
    \centering
\includegraphics[width=0.70\textwidth]{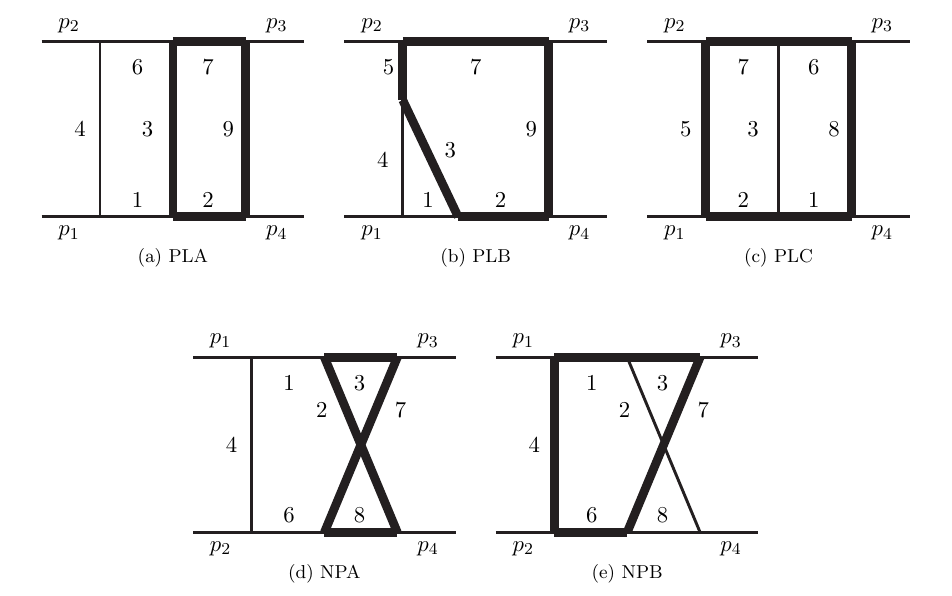}
    \caption{Representative set of two-loop graphs with internal heavy-quark loops. Thick and thin lines represent massive and massless propagators, respectively. The edge labels correspond to the
propagator numbering in~\cref{fig:toposPL} and~\cref{fig:toposNP}.}
    \label{fig:gg_aa_diagrams}
\end{figure}

In this appendix, we repeat the definition of the five integral families introduced in~\cite{Becchetti:2025rrz}. Each of the topology is defined as follows:
\begin{equation}
\mathcal{I}_{\texttt{topo}}(n_1,...,n_9) = 
\int \prod_{\ell=1}^2 \left[\frac{\mu_0^{2\epsilon}}{ C_\epsilon}\frac{\mathrm d^d k_\ell}{(2 \pi)^d}\right]
\frac{1}{D_{1}^{n_1} \cdots D_{9}^{n_9}}\,,  \label{eq:measure}
\end{equation}
where $\mu_0$ is the dimensional regularization scale,
\begin{equation}
C_\epsilon = i\,(4 \pi)^\epsilon \Gamma(1+\epsilon)\,,\label{eq:norm}
\end{equation}
and
\texttt{topo} labels the different integral families. In~\cref{fig:toposPL} and~\cref{fig:toposNP} we give the corresponding propagators $D_1,\hdots,D_9$ for the three planar and two non-planar families, respectively. A graphical representation of the respective top sectors is shown in~\cref{fig:gg_aa_diagrams}. 
\begin{table}[H]
\begin{center}
\begin{tabular}{| m{2.1cm} || m{3.7cm} | m{3.7cm} | m{3.7cm} |} 
 \hline
 Denominator & Integral family PLA & Integral family PLB & Integral family PLC \\[5pt]
 \hline
$D_1$ & $k_1^2$ & $k_1^2$ & $k_1^2 - \mt^2$ \\
$D_2$ & $k_2^2 - \mt^2$ & $k_2^2 - \mt^2$ & $k_2^2 - \mt^2$ \\
$D_3$ & $(k_1-k_2)^2 - \mt^2$ & $(k_1-k_2)^2 - \mt^2$ & $(k_1-k_2)^2$ \\
$D_4$ & $(k_1-p_1)^2$ & $(k_1-p_1)^2$ & $(k_1-p_1)^2 - \mt^2$ \\
$D_5$ & $(k_2-p_1)^2$ & $(k_2-p_1)^2 - \mt^2$ & $(k_2-p_1)^2 - \mt^2$ \\
$D_6$ & $(k_1-p_1-p_2)^2$ & $(k_1-p_1-p_2)^2$ & $(k_1-p_1-p_2)^2 - \mt^2$ \\
$D_7$ & $(k_2-p_1-p_2)^2 - \mt^2$ & $(k_2-p_1-p_2)^2 - \mt^2$ & $(k_2-p_1-p_2)^2 - \mt^2$ \\
$D_8$ & $(k_1-p_1-p_2-p_3)^2$ & $(k_1-p_1-p_2-p_3)^2$ & $(k_1-p_1-p_2-p_3)^2 - \mt^2$ \\
$D_9$ & $(k_2-p_1-p_2-p_3)^2 - \mt^2$ & $(k_2-p_1-p_2-p_3)^2 - \mt^2$ & $(k_2-p_1-p_2-p_3)^2 - \mt^2$ \\
 \hline
\end{tabular}
\caption{\label{fig:toposPL} Propagator definitions for the three planar scalar integrals families PLA, PLB and PLC.}
\end{center}
\end{table}
\begin{table}[H]
\begin{center}
\begin{tabular}{| m{2.1cm} || m{4.5cm} | m{4.5cm} |} 
 \hline
 Denominator & Integral family NPA & Integral family NPB \\[5pt]
 \hline
$D_1$ & $k_1^2$ & $k_1^2 - \mt^2$  \\
$D_2$ & $k_2^2 - \mt^2$ & $k_2^2$  \\
$D_3$ & $(k_1-k_2)^2 - \mt^2$ & $(k_1-k_2)^2 - \mt^2$  \\
$D_4$ & $(k_1-p_1)^2$ & $(k_1-p_1)^2 - \mt^2$  \\
$D_5$ & $(k_2-p_1)^2$ & $(k_2-p_1)^2$  \\
$D_6$ & $(k_1-p_1-p_2)^2$ & $(k_1-p_1-p_2)^2 - \mt^2$  \\
$D_7$ & $(k_1-k_2+p_3)^2 - \mt^2$ & $(k_1-k_2+p_3)^2 - \mt^2$ \\
$D_8$ & $(k_2-p_1-p_2-p_3)^2 - \mt^2$ & $(k_2-p_1-p_2-p_3)^2$ \\
$D_9$ & $(k_1-k_2-p_1-p_2)^2$ & $(k_1-k_2-p_1-p_2)^2$ \\
 \hline
\end{tabular}
\caption{\label{fig:toposNP} Propagator definitions for the two non-planar scalar integrals families NPA and NPB.}
\end{center}
\end{table}


\section{Master Integrals}
\label{app:masters}
In this appendix, we list our choice of initial basis of master integrals. Thereby, we also use crossings of the original integral families, following the notation introduced by \texttt{Reduze2}~\cite{vonManteuffel:2012np}. In particular for crossings of up to four external momenta,
\begin{equation}
    \left\{ 
    \begin{array}{ll} 
    \text{xij} : & \qquad p_i \leftrightarrow p_j   \\
    \text{xijk}: & \qquad p_i \to p_j \to p_k \to p_i  \\
    \text{xijkl}: & \qquad p_i \to p_j \to p_k \to p_l \to p_i \,.
    \end{array}
    \right.
\end{equation}
The list of master integrals then reads
\begin{align*}
&\mathcal{J}_1 = \mathcal{I}_{\text{PLA}}(0,2,2,0,0,0,0,0,0)\, , \quad
&&\mathcal{J}_2 = \mathcal{I}_{\text{PLA}}(2,2,0,0,0,1,0,0,0)\, , \\
&\mathcal{J}_3 = \mathcal{I}_{\text{PLA}}(0,2,2,0,0,1,0,0,0)\, , \quad
&&\mathcal{J}_4 = \mathcal{I}_{\text{PLA}}(0,2,1,0,0,2,0,0,0)\, , \\
&\mathcal{J}_5 = \mathcal{I}_{\text{PLA}}(0,2,2,0,0,0,1,0,0)\, , \quad
&&\mathcal{J}_6 = \mathcal{I}_{\text{PLAx123}}(0,2,2,0,0,0,1,0,0)\, , \\
&\mathcal{J}_7 = \mathcal{I}_{\text{PLAx124}}(0,2,2,0,0,0,1,0,0)\, , \quad
&&\mathcal{J}_8 = \mathcal{I}_{\text{PLA}}(0,2,0,2,0,0,0,1,0)\, , \\
&\mathcal{J}_9 = \mathcal{I}_{\text{PLAx12}}(0,2,0,2,0,0,0,1,0)\, , \quad
&&\mathcal{J}_{10} = \mathcal{I}_{\text{PLA}}(0,0,2,2,0,0,0,0,1)\, , \\
&\mathcal{J}_{11} = \mathcal{I}_{\text{PLA}}(0,0,2,1,0,0,0,0,2)\, , \quad
&&\mathcal{J}_{12} = \mathcal{I}_{\text{PLAx12}}(0,0,2,2,0,0,0,0,1)\, , \\
&\mathcal{J}_{13} = \mathcal{I}_{\text{PLAx12}}(0,0,2,1,0,0,0,0,2)\, , \quad
&&\mathcal{J}_{14} = \mathcal{I}_{\text{PLA}}(0,1,2,1,0,0,1,0,0)\, , \\
&\mathcal{J}_{15} = \mathcal{I}_{\text{PLA}}(0,1,1,2,0,0,1,0,0)\, , \quad
&&\mathcal{J}_{16} = \mathcal{I}_{\text{PLA}}(0,2,2,1,0,0,1,0,0)\, , \\
&\mathcal{J}_{17} = \mathcal{I}_{\text{PLAx123}}(0,1,2,1,0,0,1,0,0)\, , \quad
&&\mathcal{J}_{18} = \mathcal{I}_{\text{PLAx123}}(0,1,1,2,0,0,1,0,0)\, , \\
&\mathcal{J}_{19} = \mathcal{I}_{\text{PLAx123}}(0,2,2,1,0,0,1,0,0)\, , \quad
&&\mathcal{J}_{20} = \mathcal{I}_{\text{PLAx124}}(0,1,2,1,0,0,1,0,0)\, , \\
&\mathcal{J}_{21} = \mathcal{I}_{\text{PLAx124}}(0,1,1,2,0,0,1,0,0)\, , \quad
&&\mathcal{J}_{22} = \mathcal{I}_{\text{PLAx124}}(0,2,2,1,0,0,1,0,0)\, , \\
&\mathcal{J}_{23} = \mathcal{I}_{\text{PLA}}(2,2,0,0,0,1,1,0,0)\, , \quad
&&\mathcal{J}_{24} = \mathcal{I}_{\text{PLAx123}}(2,2,0,0,0,1,1,0,0)\, , \\
&\mathcal{J}_{25} = \mathcal{I}_{\text{PLAx124}}(2,2,0,0,0,1,1,0,0)\, , \quad
&&\mathcal{J}_{26} = \mathcal{I}_{\text{PLA}}(0,2,1,1,0,0,0,1,0)\, , \\
&\mathcal{J}_{27} = \mathcal{I}_{\text{PLA}}(0,2,1,2,0,0,0,1,0)\, , \quad
&&\mathcal{J}_{28} = \mathcal{I}_{\text{PLAx12}}(0,2,1,1,0,0,0,1,0)\, , \\
&\mathcal{J}_{29} = \mathcal{I}_{\text{PLAx12}}(0,2,1,2,0,0,0,1,0)\, , \quad
&&\mathcal{J}_{30} = \mathcal{I}_{\text{PLA}}(0,1,3,1,0,0,0,0,1)\, , \\
&\mathcal{J}_{31} = \mathcal{I}_{\text{PLAx12}}(0,1,3,1,0,0,0,0,1)\, , \quad
&&\mathcal{J}_{32} = \mathcal{I}_{\text{PLA}}(1,0,2,0,0,1,0,0,1)\, , \\
&\mathcal{J}_{33} = \mathcal{I}_{\text{PLA}}(2,0,2,0,0,1,0,0,1)\, , \quad
&&\mathcal{J}_{34} = \mathcal{I}_{\text{PLA}}(0,1,3,0,0,1,0,0,1)\, , \\
&\mathcal{J}_{35} = \mathcal{I}_{\text{PLA}}(0,1,2,0,0,0,1,0,1)\, , \quad
&&\mathcal{J}_{36} = \mathcal{I}_{\text{PLAx123}}(0,1,2,0,0,0,1,0,1)\, , \\
&\mathcal{J}_{37} = \mathcal{I}_{\text{PLAx124}}(0,1,2,0,0,0,1,0,1)\, , \quad
&&\mathcal{J}_{38} = \mathcal{I}_{\text{PLC}}(2,2,0,0,0,1,1,0,0)\, , \\
&\mathcal{J}_{39} = \mathcal{I}_{\text{PLC}}(0,0,0,2,2,0,0,1,1)\, , \quad
&&\mathcal{J}_{40} = \mathcal{I}_{\text{PLCx12}}(0,0,0,2,2,0,0,1,1)\, , \\
&\mathcal{J}_{41} = \mathcal{I}_{\text{PLA}}(1,1,1,0,0,1,1,0,0)\, , \quad
&&\mathcal{J}_{42} = \mathcal{I}_{\text{PLAx123}}(1,1,1,0,0,1,1,0,0)\, , \\
&\mathcal{J}_{43} = \mathcal{I}_{\text{PLAx124}}(1,1,1,0,0,1,1,0,0)\, , \quad
&&\mathcal{J}_{44} = \mathcal{I}_{\text{PLA}}(1,2,0,1,0,1,0,1,0)\, , \\
&\mathcal{J}_{45} = \mathcal{I}_{\text{PLAx123}}(1,2,0,1,0,1,0,1,0)\, , \quad
&&\mathcal{J}_{46} = \mathcal{I}_{\text{PLAx12}}(1,2,0,1,0,1,0,1,0)\, , \\
&\mathcal{J}_{47} = \mathcal{I}_{\text{PLA}}(0,2,1,1,0,1,0,1,0)\, , \quad
&&\mathcal{J}_{48} = \mathcal{I}_{\text{PLA}}(0,2,1,1,0,1,0,1,-1)\, , \\
&\mathcal{J}_{49} = \mathcal{I}_{\text{PLAx12}}(0,2,1,1,0,1,0,1,0)\, , \quad
&&\mathcal{J}_{50} = \mathcal{I}_{\text{PLAx12}}(0,2,1,1,0,1,0,1,-1)\, , \\
&\mathcal{J}_{51} = \mathcal{I}_{\text{PLAx123}}(0,2,1,1,0,1,0,1,0)\, , \quad
&&\mathcal{J}_{52} = \mathcal{I}_{\text{PLAx123}}(0,2,1,1,0,1,0,1,-1)\, , \\
&\mathcal{J}_{53} = \mathcal{I}_{\text{PLA}}(1,1,1,0,0,1,0,0,1)\, , \quad
&&\mathcal{J}_{54} = \mathcal{I}_{\text{PLA}}(1,0,2,1,0,1,0,0,1)\, , \\
&\mathcal{J}_{55} = \mathcal{I}_{\text{PLA}}(1,0,2,1,0,1,0,-1,1)\, , \quad
&&\mathcal{J}_{56} = \mathcal{I}_{\text{PLAx12}}(1,0,2,1,0,1,0,0,1)\, , \\
&\mathcal{J}_{57} = \mathcal{I}_{\text{PLAx12}}(1,0,2,1,0,1,0,-1,1)\, , \quad
&&\mathcal{J}_{58} = \mathcal{I}_{\text{PLAx123}}(1,0,2,1,0,1,0,0,1)\, , \\
&\mathcal{J}_{59} = \mathcal{I}_{\text{PLAx123}}(1,0,2,1,0,1,0,-1,1)\, , \quad
&&\mathcal{J}_{60} = \mathcal{I}_{\text{PLA}}(0,1,1,1,0,1,0,0,1)\, , \\
&\mathcal{J}_{61} = \mathcal{I}_{\text{PLA}}(0,1,2,1,0,1,0,0,1)\, , \quad
&&\mathcal{J}_{62} = \mathcal{I}_{\text{PLAx12}}(0,1,1,1,0,1,0,0,1)\, , \\
&\mathcal{J}_{63} = \mathcal{I}_{\text{PLAx12}}(0,1,2,1,0,1,0,0,1)\, , \quad
&&\mathcal{J}_{64} = \mathcal{I}_{\text{PLAx123}}(0,1,1,1,0,1,0,0,1)\, , \\
&\mathcal{J}_{65} = \mathcal{I}_{\text{PLAx123}}(0,1,2,1,0,1,0,0,1)\, , \quad
&&\mathcal{J}_{66} = \mathcal{I}_{\text{PLA}}(0,1,2,1,0,0,1,0,1)\, , \\
&\mathcal{J}_{67} = \mathcal{I}_{\text{PLA}}(0,1,3,1,0,0,1,0,1)\, , \quad
&&\mathcal{J}_{68} = \mathcal{I}_{\text{PLA}}(0,1,2,1,-1,0,1,0,1)\, , \\
&\mathcal{J}_{69} = \mathcal{I}_{\text{PLAx12}}(0,1,2,1,0,0,1,0,1)\, , \quad
&&\mathcal{J}_{70} = \mathcal{I}_{\text{PLAx12}}(0,1,3,1,0,0,1,0,1)\, , \\
&\mathcal{J}_{71} = \mathcal{I}_{\text{PLAx12}}(0,1,2,1,-1,0,1,0,1)\, , \quad
&&\mathcal{J}_{72} = \mathcal{I}_{\text{PLAx123}}(0,1,2,1,0,0,1,0,1)\, , \\
&\mathcal{J}_{73} = \mathcal{I}_{\text{PLAx123}}(0,1,3,1,0,0,1,0,1)\, , \quad
&&\mathcal{J}_{74} = \mathcal{I}_{\text{PLAx123}}(0,1,2,1,-1,0,1,0,1)\, , \\
&\mathcal{J}_{75} = \mathcal{I}_{\text{PLAx124}}(0,1,2,1,0,0,1,0,1)\, , \quad
&&\mathcal{J}_{76} = \mathcal{I}_{\text{PLAx124}}(0,1,3,1,0,0,1,0,1)\, , \\
&\mathcal{J}_{77} = \mathcal{I}_{\text{PLAx124}}(0,1,2,1,-1,0,1,0,1)\, , \quad
&&\mathcal{J}_{78} = \mathcal{I}_{\text{PLAx1234}}(0,1,2,1,0,0,1,0,1)\, , \\
&\mathcal{J}_{79} = \mathcal{I}_{\text{PLAx1234}}(0,1,3,1,0,0,1,0,1)\, , \quad
&&\mathcal{J}_{80} = \mathcal{I}_{\text{PLAx1234}}(0,1,2,1,-1,0,1,0,1)\, , \\
&\mathcal{J}_{81} = \mathcal{I}_{\text{PLAx1243}}(0,1,2,1,0,0,1,0,1)\, , \quad
&&\mathcal{J}_{82} = \mathcal{I}_{\text{PLAx1243}}(0,1,3,1,0,0,1,0,1)\, , \\
&\mathcal{J}_{83} = \mathcal{I}_{\text{PLAx1243}}(0,1,2,1,-1,0,1,0,1)\, , \quad
&&\mathcal{J}_{84} = \mathcal{I}_{\text{PLA}}(2,1,0,0,0,1,1,0,1)\, , \\
&\mathcal{J}_{85} = \mathcal{I}_{\text{PLAx123}}(2,1,0,0,0,1,1,0,1)\, , \quad
&&\mathcal{J}_{86} = \mathcal{I}_{\text{PLAx124}}(2,1,0,0,0,1,1,0,1)\, , \\
&\mathcal{J}_{87} = \mathcal{I}_{\text{PLA}}(0,1,1,1,0,0,0,1,1)\, , \quad
&&\mathcal{J}_{88} = \mathcal{I}_{\text{PLAx12}}(0,1,1,1,0,0,0,1,1)\, , \\
&\mathcal{J}_{89} = \mathcal{I}_{\text{PLC}}(1,1,1,0,1,1,0,0,0)\, , \quad
&&\mathcal{J}_{90} = \mathcal{I}_{\text{PLC}}(1,1,1,0,1,2,0,0,0)\, , \\
&\mathcal{J}_{91} = \mathcal{I}_{\text{PLC}}(0,1,1,1,1,1,0,0,0)\, , \quad
&&\mathcal{J}_{92} = \mathcal{I}_{\text{PLC}}(1,2,0,1,0,1,1,0,0)\, , \\
&\mathcal{J}_{93} = \mathcal{I}_{\text{PLC}}(1,1,1,0,1,0,0,1,0)\, , \quad
&&\mathcal{J}_{94} = \mathcal{I}_{\text{PLCx12}}(1,1,1,0,1,0,0,1,0)\, , \\
&\mathcal{J}_{95} = \mathcal{I}_{\text{PLC}}(0,1,1,1,1,0,0,1,0)\, , \quad
&&\mathcal{J}_{96} = \mathcal{I}_{\text{PLC}}(0,1,1,1,1,0,0,2,0)\, , \\
&\mathcal{J}_{97} = \mathcal{I}_{\text{PLCx12}}(0,1,1,1,1,0,0,1,0)\, , \quad
&&\mathcal{J}_{98} = \mathcal{I}_{\text{PLCx12}}(0,1,1,1,1,0,0,2,0)\, , \\
&\mathcal{J}_{99} = \mathcal{I}_{\text{PLC}}(1,2,0,1,0,1,0,1,0)\, , \quad
&&\mathcal{J}_{100} = \mathcal{I}_{\text{PLCx12}}(1,2,0,1,0,1,0,1,0)\, , \\
&\mathcal{J}_{101} = \mathcal{I}_{\text{PLCx123}}(1,2,0,1,0,1,0,1,0)\, , \quad
&&\mathcal{J}_{102} = \mathcal{I}_{\text{PLC}}(0,1,1,0,1,1,0,1,0)\, , \\
&\mathcal{J}_{103} = \mathcal{I}_{\text{PLC}}(0,2,1,0,1,1,0,1,0)\, , \quad
&&\mathcal{J}_{104} = \mathcal{I}_{\text{PLC}}(0,1,2,0,1,1,0,1,0)\, , \\
&\mathcal{J}_{105} = \mathcal{I}_{\text{PLCx12}}(0,1,1,0,1,1,0,1,0)\, , \quad
&&\mathcal{J}_{106} = \mathcal{I}_{\text{PLCx12}}(0,2,1,0,1,1,0,1,0)\, , \\
&\mathcal{J}_{107} = \mathcal{I}_{\text{PLCx12}}(0,1,2,0,1,1,0,1,0)\, , \quad
&&\mathcal{J}_{108} = \mathcal{I}_{\text{PLCx123}}(0,1,1,0,1,1,0,1,0)\, , \\
&\mathcal{J}_{109} = \mathcal{I}_{\text{PLCx123}}(0,2,1,0,1,1,0,1,0)\, , \quad
&&\mathcal{J}_{110} = \mathcal{I}_{\text{PLCx123}}(0,1,2,0,1,1,0,1,0)\, , \\
&\mathcal{J}_{111} = \mathcal{I}_{\text{PLC}}(1,0,0,1,2,0,0,1,1)\, , \quad
&&\mathcal{J}_{112} = \mathcal{I}_{\text{PLCx12}}(1,0,0,1,2,0,0,1,1)\, , \\
&\mathcal{J}_{113} = \mathcal{I}_{\text{PLA}}(1,1,1,1,0,1,0,0,1)\, , \quad
&&\mathcal{J}_{114} = \mathcal{I}_{\text{PLAx12}}(1,1,1,1,0,1,0,0,1)\, , \\
&\mathcal{J}_{115} = \mathcal{I}_{\text{PLAx123}}(1,1,1,1,0,1,0,0,1)\, , \quad
&&\mathcal{J}_{116} = \mathcal{I}_{\text{PLA}}(1,1,1,0,0,1,1,0,1)\, , \\
&\mathcal{J}_{117} = \mathcal{I}_{\text{PLAx123}}(1,1,1,0,0,1,1,0,1)\, , \quad
&&\mathcal{J}_{118} = \mathcal{I}_{\text{PLAx124}}(1,1,1,0,0,1,1,0,1)\, , \\
&\mathcal{J}_{119} = \mathcal{I}_{\text{PLA}}(0,1,1,1,0,1,0,1,1)\, , \quad
&&\mathcal{J}_{120} = \mathcal{I}_{\text{PLAx12}}(0,1,1,1,0,1,0,1,1)\, , \\
&\mathcal{J}_{121} = \mathcal{I}_{\text{PLAx123}}(0,1,1,1,0,1,0,1,1)\, , \quad
&&\mathcal{J}_{122} = \mathcal{I}_{\text{PLC}}(1,1,0,1,1,1,1,0,0)\, , \\
&\mathcal{J}_{123} = \mathcal{I}_{\text{PLC}}(1,1,1,0,1,1,0,1,0)\, , \quad
&&\mathcal{J}_{124} = \mathcal{I}_{\text{PLC}}(1,1,1,-1,1,1,0,1,0)\, , \\
&\mathcal{J}_{125} = \mathcal{I}_{\text{PLCx12}}(1,1,1,0,1,1,0,1,0)\, , \quad
&&\mathcal{J}_{126} = \mathcal{I}_{\text{PLCx12}}(1,1,1,-1,1,1,0,1,0)\, , \\
&\mathcal{J}_{127} = \mathcal{I}_{\text{PLCx123}}(1,1,1,0,1,1,0,1,0)\, , \quad
&&\mathcal{J}_{128} = \mathcal{I}_{\text{PLCx123}}(1,1,1,-1,1,1,0,1,0)\, , \\
&\mathcal{J}_{129} = \mathcal{I}_{\text{PLC}}(0,1,1,1,1,1,0,1,0)\, , \quad
&&\mathcal{J}_{130} = \mathcal{I}_{\text{PLC}}(-1,1,1,1,1,1,0,1,0)\, , \\
&\mathcal{J}_{131} = \mathcal{I}_{\text{PLCx12}}(0,1,1,1,1,1,0,1,0)\, , \quad
&&\mathcal{J}_{132} = \mathcal{I}_{\text{PLCx12}}(-1,1,1,1,1,1,0,1,0)\, , \\
&\mathcal{J}_{133} = \mathcal{I}_{\text{PLCx123}}(0,1,1,1,1,1,0,1,0)\, , \quad
&&\mathcal{J}_{134} = \mathcal{I}_{\text{PLCx123}}(-1,1,1,1,1,1,0,1,0)\, , \\
&\mathcal{J}_{135} = \mathcal{I}_{\text{PLC}}(1,1,0,1,1,0,0,1,1)\, , \quad
&&\mathcal{J}_{136} = \mathcal{I}_{\text{PLCx12}}(1,1,0,1,1,0,0,1,1)\, , \\
&\mathcal{J}_{137} = \mathcal{I}_{\text{NPA}}(1,1,1,1,0,0,1,1,0)\, , \quad
&&\mathcal{J}_{138} = \mathcal{I}_{\text{NPA}}(1,2,1,1,0,0,1,1,0)\, , \\
&\mathcal{J}_{139} = \mathcal{I}_{\text{NPAx123}}(1,1,1,1,0,0,1,1,0)\, , \quad
&&\mathcal{J}_{140} = \mathcal{I}_{\text{NPAx123}}(1,2,1,1,0,0,1,1,0)\, , \\
&\mathcal{J}_{141} = \mathcal{I}_{\text{NPAx124}}(1,1,1,1,0,0,1,1,0)\, , \quad
&&\mathcal{J}_{142} = \mathcal{I}_{\text{NPAx124}}(1,2,1,1,0,0,1,1,0)\, , \\
&\mathcal{J}_{143} = \mathcal{I}_{\text{NPA}}(1,1,1,0,0,1,1,1,0)\, , \quad
&&\mathcal{J}_{144} = \mathcal{I}_{\text{NPA}}(1,2,1,0,0,1,1,1,0)\, , \\
&\mathcal{J}_{145} = \mathcal{I}_{\text{NPAx123}}(1,1,1,0,0,1,1,1,0)\, , \quad
&&\mathcal{J}_{146} = \mathcal{I}_{\text{NPAx123}}(1,2,1,0,0,1,1,1,0)\, , \\
&\mathcal{J}_{147} = \mathcal{I}_{\text{NPAx124}}(1,1,1,0,0,1,1,1,0)\, , \quad
&&\mathcal{J}_{148} = \mathcal{I}_{\text{NPAx124}}(1,2,1,0,0,1,1,1,0)\, , \\
&\mathcal{J}_{149} = \mathcal{I}_{\text{NPB}}(1,1,1,0,0,1,1,1,0)\, , \quad
&&\mathcal{J}_{150} = \mathcal{I}_{\text{NPBx123}}(1,1,1,0,0,1,1,1,0)\, , \\
&\mathcal{J}_{151} = \mathcal{I}_{\text{NPBx124}}(1,1,1,0,0,1,1,1,0)\, , \quad
&&\mathcal{J}_{152} = \mathcal{I}_{\text{PLA}}(1,1,1,1,0,1,1,0,1)\, , \\
&\mathcal{J}_{153} = \mathcal{I}_{\text{PLA}}(1,1,1,1,-1,1,1,0,1)\, , \quad
&&\mathcal{J}_{154} = \mathcal{I}_{\text{PLA}}(1,1,1,1,0,1,1,-1,1)\, , \\
&\mathcal{J}_{155} = \mathcal{I}_{\text{PLA}}(1,1,1,1,-1,1,1,-1,1)\, , \quad
&&\mathcal{J}_{156} = \mathcal{I}_{\text{PLAx12}}(1,1,1,1,0,1,1,0,1)\, , \\
&\mathcal{J}_{157} = \mathcal{I}_{\text{PLAx12}}(1,1,1,1,-1,1,1,0,1)\, , \quad
&&\mathcal{J}_{158} = \mathcal{I}_{\text{PLAx12}}(1,1,1,1,0,1,1,-1,1)\, , \\
&\mathcal{J}_{159} = \mathcal{I}_{\text{PLAx12}}(1,1,1,1,-1,1,1,-1,1)\, , \quad
&&\mathcal{J}_{160} = \mathcal{I}_{\text{PLAx123}}(1,1,1,1,0,1,1,0,1)\, , \\
&\mathcal{J}_{161} = \mathcal{I}_{\text{PLAx123}}(1,1,1,1,-1,1,1,0,1)\, , \quad
&&\mathcal{J}_{162} = \mathcal{I}_{\text{PLAx123}}(1,1,1,1,0,1,1,-1,1)\, , \\
&\mathcal{J}_{163} = \mathcal{I}_{\text{PLAx123}}(1,1,1,1,-1,1,1,-1,1)\, , \quad
&&\mathcal{J}_{164} = \mathcal{I}_{\text{PLAx124}}(1,1,1,1,0,1,1,0,1)\, , \\
&\mathcal{J}_{165} = \mathcal{I}_{\text{PLAx124}}(1,1,1,1,-1,1,1,0,1)\, , \quad
&&\mathcal{J}_{166} = \mathcal{I}_{\text{PLAx124}}(1,1,1,1,0,1,1,-1,1)\, , \\
&\mathcal{J}_{167} = \mathcal{I}_{\text{PLAx124}}(1,1,1,1,-1,1,1,-1,1)\, , \quad
&&\mathcal{J}_{168} = \mathcal{I}_{\text{PLAx1234}}(1,1,1,1,0,1,1,0,1)\, , \\
&\mathcal{J}_{169} = \mathcal{I}_{\text{PLAx1234}}(1,1,1,1,-1,1,1,0,1)\, , \quad
&&\mathcal{J}_{170} = \mathcal{I}_{\text{PLAx1234}}(1,1,1,1,0,1,1,-1,1)\, , \\
&\mathcal{J}_{171} = \mathcal{I}_{\text{PLAx1234}}(1,1,1,1,-1,1,1,-1,1)\, , \quad
&&\mathcal{J}_{172} = \mathcal{I}_{\text{PLAx1243}}(1,1,1,1,0,1,1,0,1)\, , \\
&\mathcal{J}_{173} = \mathcal{I}_{\text{PLAx1243}}(1,1,1,1,-1,1,1,0,1)\, , \quad
&&\mathcal{J}_{174} = \mathcal{I}_{\text{PLAx1243}}(1,1,1,1,0,1,1,-1,1)\, , \\
&\mathcal{J}_{175} = \mathcal{I}_{\text{PLAx1243}}(1,1,1,1,-1,1,1,-1,1)\, , \quad
&&\mathcal{J}_{176} = \mathcal{I}_{\text{PLC}}(1,1,1,0,1,1,1,1,0)\, , \\
&\mathcal{J}_{177} = \mathcal{I}_{\text{PLC}}(1,1,1,-1,1,1,1,1,0)\, , \quad
&&\mathcal{J}_{178} = \mathcal{I}_{\text{PLC}}(1,1,1,-1,1,1,1,1,-1)\, , \\
&\mathcal{J}_{179} = \mathcal{I}_{\text{PLCx12}}(1,1,1,0,1,1,1,1,0)\, , \quad
&&\mathcal{J}_{180} = \mathcal{I}_{\text{PLCx12}}(1,1,1,-1,1,1,1,1,0)\, , \\
&\mathcal{J}_{181} = \mathcal{I}_{\text{PLCx12}}(1,1,1,-1,1,1,1,1,-1)\, , \quad
&&\mathcal{J}_{182} = \mathcal{I}_{\text{PLCx123}}(1,1,1,0,1,1,1,1,0)\, , \\
&\mathcal{J}_{183} = \mathcal{I}_{\text{PLCx123}}(1,1,1,-1,1,1,1,1,0)\, , \quad
&&\mathcal{J}_{184} = \mathcal{I}_{\text{PLCx123}}(1,1,1,-1,1,1,1,1,-1)\, , \\
&\mathcal{J}_{185} = \mathcal{I}_{\text{PLC}}(0,1,1,1,1,1,0,1,1)\, , \quad
&&\mathcal{J}_{186} = \mathcal{I}_{\text{PLC}}(-1,1,1,1,1,1,0,1,1)\, , \\
&\mathcal{J}_{187} = \mathcal{I}_{\text{PLC}}(-1,1,1,1,1,1,-1,1,1)\, , \quad
&&\mathcal{J}_{188} = \mathcal{I}_{\text{PLCx12}}(0,1,1,1,1,1,0,1,1)\, , \\
&\mathcal{J}_{189} = \mathcal{I}_{\text{PLCx12}}(-1,1,1,1,1,1,0,1,1)\, , \quad
&&\mathcal{J}_{190} = \mathcal{I}_{\text{PLCx12}}(-1,1,1,1,1,1,-1,1,1)\, , \\
&\mathcal{J}_{191} = \mathcal{I}_{\text{PLCx123}}(0,1,1,1,1,1,0,1,1)\, , \quad
&&\mathcal{J}_{192} = \mathcal{I}_{\text{PLCx123}}(-1,1,1,1,1,1,0,1,1)\, , \\
&\mathcal{J}_{193} = \mathcal{I}_{\text{PLCx123}}(-1,1,1,1,1,1,-1,1,1)\, , \quad
&&\mathcal{J}_{194} = \mathcal{I}_{\text{NPA}}(1, 1, 1, 1, 0, 1, 1, 1, -1)\, , \\
&\mathcal{J}_{195} = \mathcal{I}_{\text{NPA}}(1, 2, 1, 1, 0, 1, 1, 1, -1)\, , \quad
&&\mathcal{J}_{196} = \mathcal{I}_{\text{NPA}}(1, 1, 1, 1, 0, 1, 1, 1, -2)\, , \\
&\mathcal{J}_{197} = \mathcal{I}_{\text{NPA}}(1, 1, 1, 1, 0, 1, 1, 1, 0)\, , \quad
&&\mathcal{J}_{198} = \mathcal{I}_{\text{NPAx123}}(1, 1, 1, 1, 0, 1, 1, 1, -1)\, , \\
&\mathcal{J}_{199} = \mathcal{I}_{\text{NPAx123}}(1, 2, 1, 1, 0, 1, 1, 1, -1)\, , \quad
&&\mathcal{J}_{200} = \mathcal{I}_{\text{NPAx123}}(1, 1, 1, 1, 0, 1, 1, 1, -2)\, , \\
&\mathcal{J}_{201} = \mathcal{I}_{\text{NPAx123}}(1, 1, 1, 1, 0, 1, 1, 1, 0)\, , \quad
&&\mathcal{J}_{202} = \mathcal{I}_{\text{NPAx123}}(1, 1, 1, 1, 0, 1, 1, 1, -1)\, , \\
&\mathcal{J}_{203} = \mathcal{I}_{\text{NPAx123}}(1, 2, 1, 1, 0, 1, 1, 1, -1)\, , \quad
&&\mathcal{J}_{204} = \mathcal{I}_{\text{NPAx123}}(1, 1, 1, 1, 0, 1, 1, 1, -2)\, , \\
&\mathcal{J}_{205} = \mathcal{I}_{\text{NPAx123}}(1, 1, 1, 1, 0, 1, 1, 1, 0)\, , \quad
&&\mathcal{J}_{206} = \mathcal{I}_{\text{NPB}}(1, 1, 1, 1, 0, 1, 1, 1, 0)\, , \\
&\mathcal{J}_{207} = \mathcal{I}_{\text{NPB}}(1, 1, 1, 1, 0, 1, 1, 1, -1)\, , \quad
&&\mathcal{J}_{208} = \mathcal{I}_{\text{NPB}}(1, 1, 1, 1, -1, 1, 1, 1, 0)\, , \\
&\mathcal{J}_{209} = \mathcal{I}_{\text{NPB}}(1, 1, 1, 1, 0, 1, 1, 1, -2)\, , \quad
&&\mathcal{J}_{210} = \mathcal{I}_{\text{NPB}}(1, 1, 1, 1, -1, 1, 1, 1, -1)\, , \\
&\mathcal{J}_{211} = \mathcal{I}_{\text{NPBx123}}(1, 1, 1, 1, 0, 1, 1, 1, 0)\, , \quad
&&\mathcal{J}_{212} = \mathcal{I}_{\text{NPBx123}}(1, 1, 1, 1, 0, 1, 1, 1, -1)\, , \\
&\mathcal{J}_{213} = \mathcal{I}_{\text{NPBx123}}(1, 1, 1, 1, -1, 1, 1, 1, 0)\, , \quad
&&\mathcal{J}_{214} = \mathcal{I}_{\text{NPBx123}}(1, 1, 1, 1, 0, 1, 1, 1, -2)\, , \\
&\mathcal{J}_{215} = \mathcal{I}_{\text{NPBx123}}(1, 1, 1, 1, -1, 1, 1, 1, -1)\, , \quad
&&\mathcal{J}_{216} = \mathcal{I}_{\text{NPBx124}}(1, 1, 1, 1, 0, 1, 1, 1, 0)\, , \\
&\mathcal{J}_{217} = \mathcal{I}_{\text{NPBx124}}(1, 1, 1, 1, 0, 1, 1, 1, -1)\, , \quad
&&\mathcal{J}_{218} = \mathcal{I}_{\text{NPBx124}}(1, 1, 1, 1, -1, 1, 1, 1, 0)\, , \\
&\mathcal{J}_{219} = \mathcal{I}_{\text{NPBx124}}(1, 1, 1, 1, 0, 1, 1, 1, -2)\, , \quad
&&\mathcal{J}_{220} = \mathcal{I}_{\text{NPBx124}}(1, 1, 1, 1, -1, 1, 1, 1, -1)\, .
\end{align*}

\begin{figure}[H]
    \centering
\includegraphics[width=0.96\textwidth]{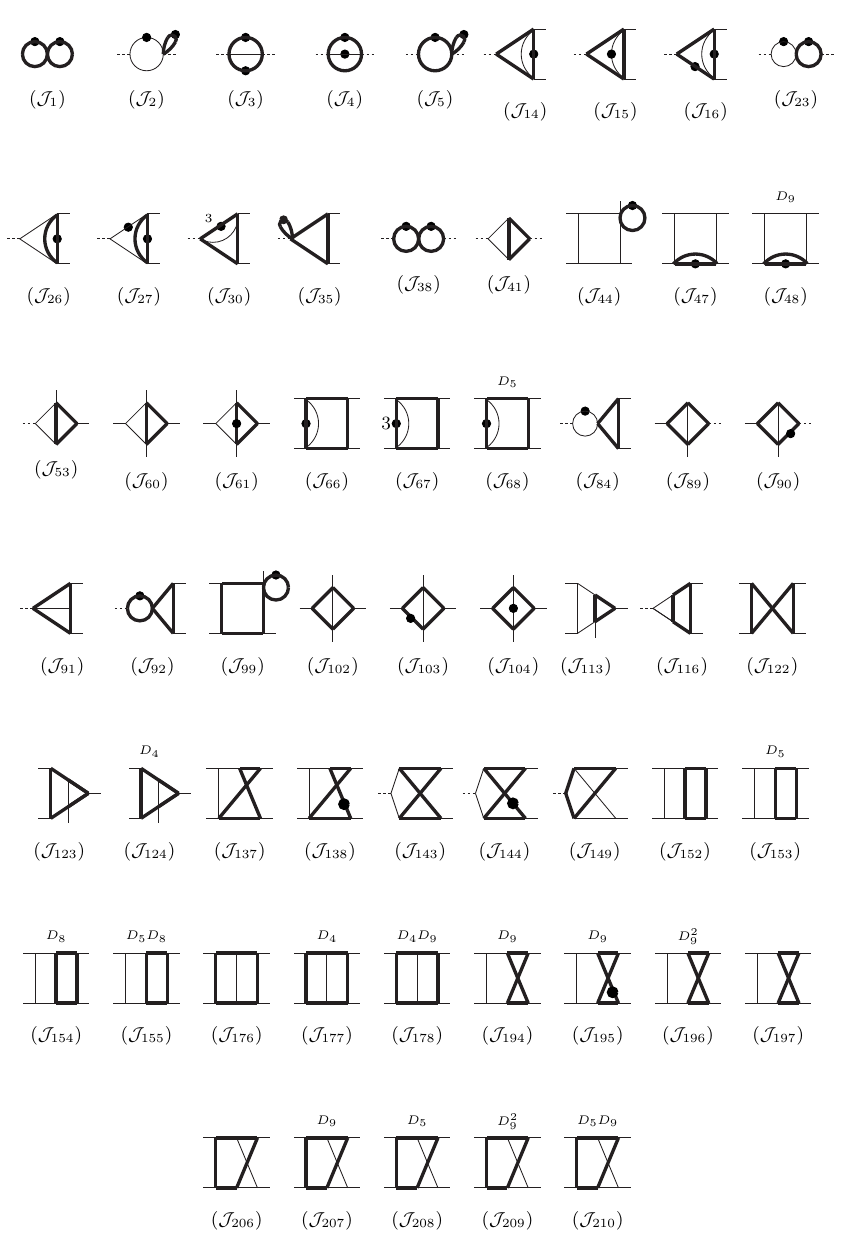}
    \caption{Graphical representations of a minimal subset of master integrals with independent graphs under crossings ($59$ in total). Note that $\mathcal{J}_{137},\mathcal{J}_{138},\mathcal{J}_{143},\mathcal{J}_{144},\mathcal{J}_{149}$ and all integrals after and including $\mathcal{J}_{194}$ are non-planar, i.e., the $X$-shaped crossing of two internal does \emph{not} represent a vertex.}
    \label{fig:axodraw}
\end{figure}

To the two four-gauge-boson amplitudes, all $220$ contribute, with the exception of $\mathcal{J}_{44},\mathcal{J}_{45}$ and $\mathcal{J}_{46}$ in the case, where one of the gauge bosons is a photon. For the three amplitudes involving at least one quark-antiquark pair, the sets of relevant master integrals are given by
\begin{equation}
\begin{aligned}
    \vec{\mathcal{J}}_{q\bar{q}g\gamma} =& \left\{
    \mathcal{J}_1,\mathcal{J}_2,
    \mathcal{J}_3,\mathcal{J}_4,
    \mathcal{J}_5,\mathcal{J}_8,
    \mathcal{J}_9,\mathcal{J}_{10},\mathcal{J}_{11},\mathcal{J}_{12},\mathcal{J}_{13},\mathcal{J}_{14},\mathcal{J}_{15},\mathcal{J}_{16},\mathcal{J}_{23},\mathcal{J}_{26},\mathcal{J}_{27},\right. \\
    & \ \ \mathcal{J}_{28}, \mathcal{J}_{29},\mathcal{J}_{30},\mathcal{J}_{31},\mathcal{J}_{32},\mathcal{J}_{33},\mathcal{J}_{34},\mathcal{J}_{35},\mathcal{J}_{41},\mathcal{J}_{44},\mathcal{J}_{45},\mathcal{J}_{46},\mathcal{J}_{51},\mathcal{J}_{52},\mathcal{J}_{53},\mathcal{J}_{54}, \\
    & \ \ \mathcal{J}_{55},\mathcal{J}_{56},\mathcal{J}_{57},
    \mathcal{J}_{58},\mathcal{J}_{59},\mathcal{J}_{60},\mathcal{J}_{61},\mathcal{J}_{62},\mathcal{J}_{63},\mathcal{J}_{64},\mathcal{J}_{65},\mathcal{J}_{66},\mathcal{J}_{67},\mathcal{J}_{68},\mathcal{J}_{69},\mathcal{J}_{70}, \\
    & \ \ \mathcal{J}_{71},\mathcal{J}_{84},\mathcal{J}_{87},\mathcal{J}_{88},\mathcal{J}_{91},\mathcal{J}_{108},\mathcal{J}_{109},\mathcal{J}_{110},\mathcal{J}_{113},\mathcal{J}_{114},\mathcal{J}_{115},\mathcal{J}_{116},\mathcal{J}_{121},\mathcal{J}_{137}, \\
    & \ \ \mathcal{J}_{138},\mathcal{J}_{143},\mathcal{J}_{144},\mathcal{J}_{152},\mathcal{J}_{153},\mathcal{J}_{154},\mathcal{J}_{155},\mathcal{J}_{156},\mathcal{J}_{157},\mathcal{J}_{158},\mathcal{J}_{159},\mathcal{J}_{194},\mathcal{J}_{195}, \\ 
    & \ \left. \mathcal{J}_{196},\mathcal{J}_{197} \right\} \, ,
\end{aligned}
\end{equation}
\begin{equation}
\begin{aligned}
    \vec{\mathcal{J}}_{q\bar{q}gg} =& \left\{\mathcal{J}_1,\mathcal{J}_2,\mathcal{J}_3,\mathcal{J}_4,\mathcal{J}_5,\mathcal{J}_8,\mathcal{J}_9,\mathcal{J}_{10},\mathcal{J}_{11},\mathcal{J}_{12},\mathcal{J}_{13},\mathcal{J}_{14
   },\mathcal{J}_{15},\mathcal{J}_{16},\mathcal{J}_{23},\mathcal{J}_{26},\mathcal{J}_{27}, \right. \\
   & \ \ \mathcal{J}_{28},\mathcal{J}_{29},\mathcal{J}_{30},\mathcal{J}_{31},\mathcal{J}_{32},\mathcal{J}
   _{33},\mathcal{J}_{34},\mathcal{J}_{35},\mathcal{J}_{38},\mathcal{J}_{41},\mathcal{J}_{44},\mathcal{J}_{45},\mathcal{J}_{46},\mathcal{J}_{47},\mathcal{J}_{48},\mathcal{J}_{49}, \\
   & \ \ \mathcal{J}_{50},\mathcal{J}_{51},\mathcal{J}_{52},\mathcal{J}_{53},\mathcal{J}_{54},\mathcal{J}_{55},\mathcal{J}_{56},\mathcal{J}_{57},\mathcal{J}_{58},\mathcal{J}_{59},\mathcal{J}_{60},\mathcal{J}_{61},\mathcal{J}_{62},\mathcal{J}_{63},\mathcal{J}_{64},\mathcal{J}_{65}, \\
   & \ \ \mathcal{J}_{66},\mathcal{J}_{67},\mathcal{J}_{68},\mathcal{J}_{69},\mathcal{J}_{70},\mathcal{J}_{7
   1},\mathcal{J}_{84},\mathcal{J}_{87},\mathcal{J}_{88},\mathcal{J}_{89},\mathcal{J}_{90},\mathcal{J}_{91},\mathcal{J}_{92},\mathcal{J}_{108},\mathcal{J}_{109}, \\
   & \ \mathcal{J}_{110},\mathcal{J}_{113},\mathcal{J}_{114},\mathcal{J}_{115},\mathcal{J}_{116},\mathcal{J}_{119},\mathcal{J}_{120},\mathcal{J}_{121},\mathcal{J}_{137},\mathcal{J}_{138},\mathcal{J}_{143},\mathcal{J}_{144},\mathcal{J}_{152}, \\
   & \ \left . \mathcal{J}_{153},\mathcal{J}_{154},\mathcal{J}_{155},\mathcal{J}_{156},\mathcal{J}_{157},\mathcal{J}_{158},\mathcal{J}_{159},\mathcal{J}_{194},\mathcal{J}_{195},\mathcal{J}_{196},\mathcal{J}_{197}\right\} \, ,
\end{aligned}
\end{equation}
\begin{equation}
\begin{aligned}
    \vec{\mathcal{J}}_{q\bar{q}\bar{Q}Q} =& \left\{\mathcal{J}_1,\mathcal{J}_2,\mathcal{J}_3,\mathcal{J}_4,\mathcal{J}_5,\mathcal{J}_8,\mathcal{J}_9,\mathcal{J}_{14},\mathcal{J}_{15},\mathcal{J}_{16},\mathcal{J}_{23},\mathcal{J}_{26},\mathcal{J}_{27},\mathcal{J}_{28},\mathcal{J}_{29},\mathcal{J}_{32},\mathcal{J}_{33}, \right. \\
    & \ \left. \mathcal{J}_{38},\mathcal{J}_{41},\mathcal{J}_{44},\mathcal{J}_{46},\mathcal{J}_{47},\mathcal{J}
   _{48},\mathcal{J}_{49},\mathcal{J}_{50}\right\} \, .
\end{aligned}
\end{equation}

\section{Relations between bare and renormalized helicity coefficients}
\label{app:renhelcoeffs}
The explicit relations between the bare and normalized helicity coefficients are given by
\begin{align}
    \boldsymbol{\Omega}_{q\bar{q}g\gamma}^{(0,\rm UV)} &= S_{\epsilon}^{-\frac{1}{2}} \, \boldsymbol{\Omega}_{q\bar{q}g\gamma}^{(0,b)}\, ,  \\
    \boldsymbol{\Omega}_{q\bar{q}g\gamma}^{(1,\rm UV)} &= S_{\epsilon}^{-\frac{1}{2}} \left[ \left(\frac{C_{\epsilon}}{S_\epsilon}\right) \boldsymbol{\Omega}_{q\bar{q}g\gamma}^{(1,b)} - \frac{\beta_0}{2\epsilon} \boldsymbol{\Omega}_{q\bar{q}g\gamma}^{(0,b)}  \right] \, ,  \\
    \boldsymbol{\Omega}_{q\bar{q}g\gamma}^{(2,\rm UV)} &= S_{\epsilon}^{-\frac{1}{2}} \left[ \left(\frac{C_{\epsilon}}{S_\epsilon}\right)^2 \boldsymbol{\Omega}_{q\bar{q}g\gamma}^{(2,b)} - \left(\frac{C_{\epsilon}}{S_\epsilon}\right) \left(\frac{3\beta_0}{2\epsilon} + \delta_w \right) \boldsymbol{\Omega}_{q\bar{q}g\gamma}^{(1,b)}  \right. \\
    &\quad \left. + \left(\frac{C_{\epsilon}}{S_\epsilon}\right)  (2 m^2) \, \delta_{m} \frac{\partial \boldsymbol{\Omega}_{q\bar{q}g\gamma}^{(1,b)}}{\partial m^2} + \left(\frac{3\beta_0^2}{8\epsilon^2} - \frac{\beta_1}{4\epsilon}-\frac{\delta_A}{2}+ \delta_q \right) \boldsymbol{\Omega}_{q\bar{q}g\gamma}^{(0,b)} \right] \, , \\
    \boldsymbol{\Omega}_{ggg\gamma}^{(1,\rm UV)} &= S_{\epsilon}^{-\frac{1}{2}} \, \left(\frac{C_{\epsilon}}{S_\epsilon}\right)  \, \boldsymbol{\Omega}_{ggg\gamma}^{(1,b)}\, ,  \\
    \boldsymbol{\Omega}_{ggg\gamma}^{(2,\rm UV)} &= S_{\epsilon}^{-\frac{1}{2}} \left(\frac{C_{\epsilon}}{S_\epsilon}\right)  \left[ \left(\frac{C_{\epsilon}}{S_\epsilon}\right) \boldsymbol{\Omega}_{ggg\gamma}^{(2,b)} - \frac{3\beta_0}{2\epsilon} \boldsymbol{\Omega}_{ggg\gamma}^{(1,b)} + (2 m^2) \, \delta_{m} \frac{\partial \boldsymbol{\Omega}_{ggg\gamma}^{(1,b)}}{\partial m^2} \right] \, ,  \\
    \boldsymbol{\Omega}_{q\bar{q}\bar{Q}Q}^{(0,\rm UV)} &= S_{\epsilon}^{-1} \, \boldsymbol{\Omega}_{q\bar{q}\bar{Q}Q}^{(0,b)}\, ,  \\
    \boldsymbol{\Omega}_{q\bar{q}\bar{Q}Q}^{(1,\rm UV)} &= S_{\epsilon}^{-1} \left[ \left(\frac{C_{\epsilon}}{S_\epsilon}\right) \boldsymbol{\Omega}_{q\bar{q}\bar{Q}Q}^{(1,b)} - \left( \frac{\beta_0}{\epsilon} + \delta_w \right)\boldsymbol{\Omega}_{q\bar{q}\bar{Q}Q}^{(0,b)}  \right] \, ,  \\
    \boldsymbol{\Omega}_{q\bar{q}\bar{Q}Q}^{(2,\rm UV)} &= S_{\epsilon}^{-1} \left[ \left(\frac{C_{\epsilon}}{S_\epsilon}\right)^2 \boldsymbol{\Omega}_{q\bar{q}\bar{Q}Q}^{(2,b)} - \left(\frac{C_{\epsilon}}{S_\epsilon}\right) \left(\frac{2\beta_0}{\epsilon} + 2 \delta_w \right) \boldsymbol{\Omega}_{q\bar{q}\bar{Q}Q}^{(1,b)}  \right. \\
    &\quad \left. + \left(\frac{C_{\epsilon}}{S_\epsilon}\right)  (2 m^2) \, \delta_{m} \frac{\partial \boldsymbol{\Omega}_{q\bar{q}\bar{Q}Q}^{(1,b)}}{\partial m^2} + \left( \left( \frac{\beta_0}{\epsilon} +\delta_w \right)^2 - \frac{\beta_1}{2\epsilon}+\Delta + 2\delta_q \right) \boldsymbol{\Omega}_{q\bar{q}\bar{Q}Q}^{(0,b)} \right] \, , \\
    \boldsymbol{\Omega}_{q\bar{q}gg}^{(0,\rm UV)} &= S_{\epsilon}^{-1} \, \boldsymbol{\Omega}_{q\bar{q}gg}^{(0,b)}\, ,  \\
    \boldsymbol{\Omega}_{q\bar{q}gg}^{(1,\rm UV)} &= S_{\epsilon}^{-1} \left[ \left(\frac{C_{\epsilon}}{S_\epsilon}\right) \boldsymbol{\Omega}_{q\bar{q}gg}^{(1,b)} - \frac{\beta_0}{\epsilon} \boldsymbol{\Omega}_{q\bar{q}gg}^{(0,b)}  \right] \, ,  \\
    \boldsymbol{\Omega}_{q\bar{q}gg}^{(2,\rm UV)} &= S_{\epsilon}^{-1} \left[ \left(\frac{C_{\epsilon}}{S_\epsilon}\right)^2 \boldsymbol{\Omega}_{q\bar{q}gg}^{(2,b)} - \left(\frac{C_{\epsilon}}{S_\epsilon}\right) \left(\frac{2\beta_0}{\epsilon} + \delta_w \right) \boldsymbol{\Omega}_{q\bar{q}gg}^{(1,b)}  \right. \\
    &\quad \left. + \left(\frac{C_{\epsilon}}{S_\epsilon}\right)  (2 m^2) \, \delta_{m} \frac{\partial \boldsymbol{\Omega}_{q\bar{q}gg}^{(1,b)}}{\partial m^2} + \left(\frac{\beta_0^2}{\epsilon^2} - \frac{\beta_1}{2\epsilon}-\delta_A+ \delta_q \right) \boldsymbol{\Omega}_{q\bar{q}gg}^{(0,b)} \right] \, , \\
    \boldsymbol{\Omega}_{gggg}^{(0,\rm UV)} &= S_{\epsilon}^{-1} \, \boldsymbol{\Omega}_{gggg}^{(0,b)}\, ,  \\
    \boldsymbol{\Omega}_{gggg}^{(1,\rm UV)} &= S_{\epsilon}^{-1} \left[ \left(\frac{C_{\epsilon}}{S_\epsilon}\right) \boldsymbol{\Omega}_{gggg}^{(1,b)} + \left(- \frac{\beta_0}{\epsilon} +\delta_w \right) \boldsymbol{\Omega}_{gggg}^{(0,b)}  \right] \, ,  \\
    \boldsymbol{\Omega}_{gggg}^{(2,\rm UV)} &= S_{\epsilon}^{-1} \left[ \left(\frac{C_{\epsilon}}{S_\epsilon}\right)^2 \boldsymbol{\Omega}_{gggg}^{(2,b)} - \left(\frac{C_{\epsilon}}{S_\epsilon}\right) \frac{2\beta_0}{\epsilon} \boldsymbol{\Omega}_{gggg}^{(1,b)}  \right. \\
    &\quad \left. + \left(\frac{C_{\epsilon}}{S_\epsilon}\right)  (2 m^2) \, \delta_{m} \frac{\partial \boldsymbol{\Omega}_{gggg}^{(1,b)}}{\partial m^2} + \left(\frac{\beta_0^2}{\epsilon^2} -\frac{2\beta_0 \delta_w}{\epsilon} - \frac{\beta_1}{2\epsilon} - \Delta - 2\delta_A \right) \boldsymbol{\Omega}_{gggg}^{(0,b)} \right] \, .
\end{align}
We thereby stress that the bare amplitude is defined using the unphysical normalization of the integration measure given in~\cref{eq:norm}. Note that while we focus on the diagrams containing loops of massive quarks at the two-loop level, the knowledge of the full tree level and one-loop amplitudes is necessary for the cancellation of the UV-poles.

\section{Details on the structure of the IR divergences}
\label{app:IRstructure}
In this appendix, we discuss in more detail the structure of the IR divergences of our amplitudes.\footnote{See for example~\cite{Becher:2009qa} for a detailed discussion.}
On general grounds, it is possible to define finite remainders of our amplitudes as
\begin{equation}
    \boldsymbol{\Omega}^{\rm fin}_{Z} = \lim_{\epsilon \rightarrow 0} \left( \mathbfcal{Z}_{Z}^{-1} \, \boldsymbol{\Omega}^{\rm UV}_{Z} \right) \, ,
\end{equation}
where $\mathbfcal{Z}_{Z}$ is a non-diagonal operator in colour space, that can be written in the form
\begin{equation}
    \mathbfcal{Z}_{Z} = \mathbb{P} \exp{ \int\limits_{\mu}^{\infty} \frac{\mathrm{d}\mu'}{\mu'} \mathbf{\Gamma}_{Z}(\mu')} \, . \label{eq:Zopexponential}
\end{equation}
Here, $\mathbf{\Gamma}_{Z}$ denotes the anomalous dimension operator and $\mathbb{P}$ the path-ordering of multiple $\mathbf{\Gamma}_{Z}$ by increasing values of $\mu'$ from left to right. Up to the order we are interested in, we only need to consider colour-dipole contributions to $\mathbf{\Gamma}_{Z}$, but it is well-known that at higher-loop order, colour-quadrupole corrections become relevant~\cite{Almelid:2015jia}. The dipole contributions take the form 
\begin{equation}
    \mathbf{\Gamma}_{Z}\left(\alpha_s(\mu),\mu \right) = \sum_{(n_1,n_2) \in Z} \mathbf{T}_{n_1}^a \mathbf{T}_{n_2}^a \,\gamma^{\text{cusp}} \left(\alpha_s (\mu) \right) \log{\left( \frac{\mu^2}{-s_{n_1 n_2}-i 0^+} \right)} + \sum_{n \in Z} \gamma^{[n]} \left(\alpha_s (\mu) \right)  \, ,
    \label{eq:anomdimop}
\end{equation}
where the second sum runs over all partons ((anti)quarks and gluons) involved in the process, while the first sum runs over all \emph{pairs} of partons involved in the process. The colour operators $\mathbf{T}_n^a$ in the dipole operator are defined as
\begin{align}
    &(\mathbf{T}_n^a)_{bc} = -i f^{a}_{bc} &&\text{for parton} \ n \ \text{a gluon,} \\
    &(\mathbf{T}_n^a)_{ij} = T^{a}_{ij} &&\text{for parton} \ n \ \text{a final state quark or an initial state antiquark,} \\
    &(\mathbf{T}_n^a)_{ij} = - T^{a}_{ji} &&\text{for parton} \ n \ \text{an initial state quark or a final state antiquark.}
\end{align}
Their action on the colour structures $\mathcal{C}_k$ of the amplitude is defined as 
\begin{align}
    \mathbf{T}_n^c \, \mathcal{C}_k^{i_1,\dots,i_{n_q},a_1,\dots,a_n,\dots,a_{n_g}} \equiv \left(\mathbf{T}_n^c\right)_{a_n b} \, \mathcal{C}_k^{i_1,\dots,i_{n_q},a_1,\dots,b,\dots,a_{n_g}} 
\end{align}
for parton $n$ a gluon with colour index $a_n$ and
\begin{align}
    \mathbf{T}_n^c \, \mathcal{C}_k^{i_1,\dots,i_n,\dots,i_{n_q},a_1,\dots,a_{n_g}} \equiv \left(\mathbf{T}_n^c\right)_{i_n j} \, \mathcal{C}_k^{i_1,\dots,j,\dots,i_{n_q},a_1,\dots,a_{n_g}}
\end{align}
for parton $n$ a (anti)quark with colour index $i_n$. Further, in~\cref{eq:anomdimop}, $\gamma^{\text{cusp}}$ denotes the cusp anomalous dimension and $\gamma^{[n]}$ the anomalous dimension of parton $n$. These quantities can be computed in a perturbative expansion in the strong coupling. For the purpose of this work, it suffices to know their leading orders, which are respectively given by
\begin{align}
    \gamma^{\text{cusp}} (\alpha_s) &= \left(\frac{\alpha_s}{2\pi}\right) 2 + \mathcal{O}\left( \alpha_s^2 \right) \, ,  \\ 
    \gamma^{[g]} (\alpha_s) &= \left(\frac{\alpha_s}{2\pi}\right)\left(- \frac{3C_F}{2}\right)+ \mathcal{O}\left( \alpha_s^2 \right) \, , \\ 
    \gamma^{[q]} (\alpha_s) &= \gamma^{[\bar{q}]} (\alpha_s) = \left(\frac{\alpha_s}{2\pi}\right)\left(- \beta_0 \right)+ \mathcal{O}\left( \alpha_s^2 \right) \, .
\end{align}
Changing variables in the integral in~\cref{eq:Zopexponential} from $\mu$ to $\alpha_s$ via the renormalization group equation for the QCD beta-function 
\begin{equation}
    \frac{\mathrm{d}\alpha_s}{\mathrm{d}\log \mu} = -2 \alpha_s \left[\epsilon + \left(\frac{\alpha_s}{2\pi}\right) \beta_0 + \mathcal{O}\left( \alpha_s^2 \right) \right] \, ,
\end{equation}
the integral can be performed as a series in $\alpha_s$ up to the desired order. It is then easy to obtain an expression for $\mathbfcal{Z}_{Z}$ by expanding the exponential using $[ \mathbf{\Gamma}_{Z}(\mu_1) , \mathbf{\Gamma}_{Z}(\mu_2) ] =0$, which is valid for the colour-dipole term. After inverting, the result can be written as
\begin{equation}
    \mathbfcal{Z}_{Z}^{-1} = 1 - \left(\frac{\alpha_s}{2\pi}\right) \mathbfcal{I}_{Z}^{(1)} + \mathcal{O}\left( \alpha_s^2 \right) \, ,
\end{equation}
which provides us with a subtraction operator $\mathbfcal{I}_{Z}^{(1)}$ that we can use to check the IR divergences of our helicity amplitudes, see~\cref{eq:IRsubtraction}.